%% file: main.tex
\documentclass[12pt]{article}
\usepackage{xr-hyper}
\input{Preambles/header}

\input{Preambles/notation}

\usepackage[T1]{fontenc}
\newtheorem{innercustomass}{Assumption}
\newenvironment{customass}[1] {
	\renewcommand\theinnercustomass{#1}
	\innercustomass 
} {\endinnercustomass}

\allowdisplaybreaks
\usepackage{etoolbox}
\newcommand{\zerodisplayskips}{%
	\setlength{\abovedisplayskip}{5pt}%
	\setlength{\belowdisplayskip}{5pt}%
	\setlength{\abovedisplayshortskip}{5pt}%
	\setlength{\belowdisplayshortskip}{5pt}}
\appto{\normalsize}{\zerodisplayskips}
\appto{\small}{\zerodisplayskips}
\appto{\footnotesize}{\zerodisplayskips}

\crefname{namedtheorem}{Theorem}{Theorems}

\crefname{assumption}{Assumption}{Assumptions}
\crefname{customass}{Assumption}{Assumptions}
\crefname{lemma}{Lemma}{Lemmas}

\title{ Difference-in-Differences Designs: A Practitioner's Guide}

\author{ {Andrew Baker}\thanks{University of California, Berkeley} \and {Brantly Callaway}\thanks{University of Georgia} \and {Scott Cunningham}\thanks{Baylor University} \and {Andrew Goodman-Bacon}\thanks{Opportunity and Inclusive Growth Institute, Federal Reserve Bank of Minneapolis} \and {Pedro H. C. Sant'Anna}\thanks{Emory University}}

\date{ \vspace{.2cm} \normalsize{\today}}

\begin{document}
\begin{bibunit}
\maketitle
\begin{abstract}
Difference-in-differences (DiD) is arguably the most popular quasi-experimental research design. Its canonical form, with two groups and two periods, is well-understood. However, empirical practices can be ad hoc when researchers go beyond that simple case. This article provides an organizing framework for discussing different types of DiD designs and their associated DiD estimators. It discusses covariates, weights, handling multiple periods, and staggered treatments. The organizational framework, however, applies to other extensions of DiD methods as well.
\end{abstract}

\section{Introduction}
Dating to the 1840s, difference-in-differences (DiD) is now the most common research design for estimating causal effects in the social sciences.\footnote{\citet{Currie2020} find that almost 25\% of all NBER empirical working papers and 17\% of empirical articles in five leading general-interest economics journals in 2018 mention DiD. The earliest DiD applications we are aware of are from Ignaz Semmelweis from the 1840s \citep{Semmelweis1983} and \citet{Snow1855}. For a brief overview of the long history of DiD in economics, see Section 2 of \citet{Lechner2011}.} A basic DiD design requires two time periods, one before and one after some treatment begins, and two groups, one that receives a treatment and one that does not. The DiD estimate equals the change in outcomes for the treated group minus the change in outcomes for the untreated group: the difference of two differences. If the average change in the outcomes would have been the same in the two groups had the treatment not occurred, which is referred to as a ``parallel trends'' assumption, this comparison estimates the average treatment effect among treated units.

In practice, however, researchers apply DiD methods to situations that are more complicated than the classic two-period and two-group ($2\times2$) setup. 
Most datasets cover multiple periods, and units may enter (or exit) treatment at different times. Treatment might also vary in its amount or intensity. Other variables are often used to make treated and untreated units more comparable. Today's typical DiD study includes at least one of these deviations from the canonical $2\times2$ setup. 

For many years, the common practice in applied research was to estimate complex DiD designs using linear regressions with unit and time fixed effects (two-way fixed effects, henceforth TWFE). Their identifying assumptions and interpretation were informally traced to the fact that, in the $2\times2$ case, a TWFE estimator gives the same estimate as a  DiD estimator calculated directly from sample means, and thus inherits a clear causal interpretation under a specific parallel trends identification assumption. This appeared to justify the use of a single technique for any type of design or specification. Recent research, however, has shown that simple regressions can fail to estimate meaningful causal parameters when DiD designs are complex and treatment effects vary, producing estimates that are not only misleading in their magnitudes but potentially of the wrong \textit{sign}. The significance of these findings is substantial; given the prevalence of DiD analysis in modern applied econometrics work, common empirical practices have almost certainly yielded misleading results in several concrete cases \citep{baker2022much}.

So, what should applied researchers do instead? This paper proposes a unified framework for discussing and conducting DiD studies that is rooted in the principles of causal inference in the presence of treatment effect heterogeneity. The central conclusion of recent methodological research is that even complex DiD studies can be understood as aggregations of $2\times2$ comparisons between one set of units for which treatment changes and another set for which it does not. This fact links a wide variety of DiD designs used in practice and guides methodological choices about estimating them. Viewing DiD studies through the lens of $2\times2$ ``building blocks'' aids in interpretability by clarifying that they yield causal quantities that aggregate the treatment effects identified by each $2\times2$ component. It also means that identification comes from the simple parallel trends assumptions required for each $2\times2$ building block. Practically, this framework suggests first estimating each $2\times2$ building block and then aggregating them. 
As long as the effective sample size is large, this approach allows for asymptotically valid inference using standard techniques.  

This framework is a ``forward-engineering'' approach to DiD that embraces treatment effect heterogeneity and constructs estimators that recover well-motivated causal parameters under explicitly stated assumptions. By fixing the goals of the study (the target parameters) and deriving analytical techniques, forward engineering provides clear benefits over ``reverse-engineering'' approaches that begin with a familiar regression specification and derive the assumptions under which it has \textit{some} causal interpretation. The methods we describe in this paper combine familiar techniques with some newer ones, but expressly avoid the difficulties of interpretation inherent in common regression estimators \citep{Goodman2021, deChaisemartin2020, Sun2021, Borusyak2023}. Moreover, the interpretation of common regression estimators changes across specifications, which makes it hard to understand the difference between non-robustness and a shifting target parameter. In contrast, our proposed framework naturally leads to estimation procedures that target the same parameter under different transparent identification assumptions. Thus, two estimates can be distinguished easily by their identifying assumptions. Finally, the principles of the forward-engineering approach provide guidance to good econometric practices even in settings without well-established methodological findings. 

This paper is not designed to be a comprehensive literature review; its goal is to provide guidelines for practitioners who want to better understand DiD and its various forms. Because of the tremendous variations in design, data, and specification that practitioners encounter, we opt to focus on three of the most common aspects of modern DiD studies: the use of weights, covariates, and staggered treatment timing. Table \ref{tab:acronym} includes a list of the acronyms that we, and the econometrics literature on DiD, use to distinguish different methods. We apply techniques to address these issues to a specific example: the causal effect of recent public health insurance expansions in the US on county-level mortality. Our replication materials include data as well as R and Stata code that can serve as a template for any DiD study using these methods. In an appendix, we briefly discuss related DiD designs with different treatment variables (ones that turn on and off or take many values), additional comparisons (i.e., triple-difference designs), distributional target parameters, or different data structures (repeated cross-sections or unbalanced panels). Several recent reviews follow the logic laid out here and cover additional DiD-related topics and technical details: \citet{Roth2023a, deChaisemartin2023-survey, Callaway2023_review}.

The rest of the paper is structured as follows. Section \ref{sec:ACA} introduces the Medicaid example. Section \ref{sec:2x2} discusses the canonical $2\times2$ DiD setups with and without weights, and Section \ref{sec:covariates} discusses threats to the identification assumptions, how to assess them, and how to incorporate covariates. Section \ref{sec:binary-multiple-t} extends the $2\times2$ setup to multiple periods with potentially staggered treatment adoption. Section \ref{sec:conclusion} concludes and briefly discusses some extensions that involve more complex DiD designs.

\section{Medicaid and mortality: The running example} \label{sec:ACA}

To make our methodological discussion concrete, we revisit a timely and important causal question: How did the expansion of public health insurance (Medicaid) under the Affordable Care Act (ACA) affect mortality? 

Medicaid expansion is a great example of a staggered treatment adoption. The ACA originally mandated that in 2014 all states expand Medicaid eligibility to adults with incomes up to 138\% of the federal poverty threshold. In upholding the law's constitutionality in a 2012 decision, however, the Supreme Court made Medicaid expansion optional. As a result, many states expanded Medicaid after 2014, but several have not done so as of 2024. 

Columns 1 and 2 of Table \ref{tab:adoptions} illustrate the variation in Medicaid expansion dates. 

\input{Tables/adoptions}

States expanded Medicaid largely because of economic and political considerations \citep{sommers2013us}, which created observable differences between expansion and non-expansion states. For instance, just four out of the 22 states that expanded Medicaid in 2014 are in the southern Census region; conversely, seven out of 10 non-expansion states are in the South. This suggests a potential role for covariates when analyzing Medicaid expansion. 

Finally, mortality is measured in jurisdictions like states and counties, which are of very different sizes. Choices about (population) weights determine not only how different estimation approaches average the units within a given expansion group but also how a given estimation technique averages estimated effects across those groups. California, for example, represented 4.5\% of the states that expanded Medicaid in 2014, 5.4\% of the counties, but 27.7\% of the adults ages 20-64; its contribution to ``the'' average outcome for the 2014 expansion group is very different with weights than without. The final three columns of Table \ref{tab:adoptions} show that, in our data the entire 2014 expansion group contains 44\% of the states, 36\% of the counties, but 45\% of all adults. Weighting will therefore change how important the estimated treatment effects are for the 2014 group. 

Several recent papers study the effect of ACA Medicaid expansion on mortality rates for lower-income adults, who are most likely to gain insurance through Medicaid. \citet{miller2021medicaid} and \citet{wyse2024} use simple DiD methods to provide evidence that Medicaid reduced adult mortality rates for targeted sub-populations. Unfortunately, their analyses require restricted links between income and mortality data, which are important for overcoming the low statistical power in studies using aggregate mortality data \citep{black2022simulated}. Our goal is to pursue a replicable and shareable example based on a related analysis by \citet{Borgschulte2020}. They use a sophisticated strategy to select and use covariates in a weighted TWFE regression using restricted access data, and find that Medicaid expansion reduced aggregate county-level mortality rates. We use publicly available data, which we include in a fully-reproducible replication package, and consider only a handful of intuitive demographic and economic covariates sufficient to illustrate several practical challenges that can arise with DiD. This empirical exercise is meant solely to illustrate how to tackle several common features of DiD designs. The results are pedagogical in spirit and do not represent the best possible estimates of Medicaid’s effect on adult mortality.

Our outcome variable is the crude adult mortality rate, $Y_{i,t}$, for people ages 20-64 (measured per 100,000) by county ($i$) from 2009 to 2019 released by the \citet{CDC_Vital_Statistics}.\footnote{It is common to adjust mortality rates by the county age distribution. Unfortunately, the CDC measurements of age-specific deaths are restricted for counties with fewer than 10 annual deaths. We aim to use publicly available and shareable data for pedagogical purposes; we follow \citet{Borgschulte2020} and use the crude mortality rate.} We denote county $i$'s adult population in 2013 by $W_i$ and its socioeconomic covariates in year $t$ (discussed below) by $X_{i,t}$. The information in Table \ref{tab:adoptions} defines the treatment group variable $G_i$ that equals the year in which county $i$'s state expanded Medicaid; $G_i=\infty$ for the non-expansion states. Our final sample contains 2,604 counties in states with complete data on mortality rates from 2009 to 2019 and covariates for 2013 and 2014.

Faced with a setup such as this, researchers need to make a range of tightly related choices. Which treatment groups in Table \ref{tab:adoptions} should be compared with each other and over what time horizons? What must be true for those comparisons to identify causal effects, and how should one empirically evaluate their plausibility? How can other information, such as covariates or pre-period outcomes, be used to improve the credibility of the design? How do these methodological choices affect the causal interpretation of a given analysis? The aim of this review is to demonstrate to practitioners using DiD in realistic scenarios why and how to use state-of-the-art econometric tools to answer these questions.

\section{\texorpdfstring{2$\times$2}{2x2} DiD designs} \label{sec:2x2}
We begin our discussion by focusing on the canonical $2\times2$ DiD setup, which has two time periods---one before and one after treatment---and two groups---one that remains untreated in both periods and one that becomes treated in the second period. In our Medicaid example, we focus on comparisons in 2014 and 2013 between the 2014 expansion group (978 counties) and the group that had not expanded by 2019 (1,222 counties). When we consider more complex designs, this kind of comparison will still play a role: it will be one $2 \times 2$ ``building block'' among many.

Using these basic ingredients, we can now define a $2\times2$ DiD \textit{design}, composed of a causal target parameter, a treatment variable, an assumption under which it is identified, and an estimation approach, which will be the classic difference of two differences. This may be familiar territory in the simple case, but it is a crucial framework for building up appropriate techniques in more complex cases.  

We first define a treatment group dummy $D_{i}$ that equals one for the treated units (expansion states, $G_i=2014$) and zero for the untreated units (states that had not expanded by 2019, $G_i>2019$). The treatment \emph{status} dummy, $D_{i,t} = D_i \times 1\{t\geq2014\}$, then equals one for counties in 2014 expansion states during post-expansion years.
To highlight how weights enter different kinds of DiD analyses, we use the following notation for expected values. For generic random variables $A$ and  $C$, for a given set of non-negative weights $\omega$, define $\E_\omega [A | C] = \E [\omega A | C] \big/ \E[\omega |C]$ as the $\omega$-weighted population expectation of $A$ given $C$. When $\omega = 1$ for all units in the population, we simply write $\E_\omega [A | C] = \E [ A | C] $. Henceforth, unless otherwise noted, we assume that we have a balanced panel data random sample of $(Y_{t=1},\dots, Y_{t=T}, G, X)$.

\subsection{Causal effects and target parameters: The ATT} \label{sec:target_param}
The first step of any causal analysis is to define the causal quantity of interest, also called the {target parameter}. We use the potential outcomes framework of \citet{Rubin1974} and \citet{Robins1986} to do so. Let $Y_{i,t}(0,0)$ denote unit $i$'s potential outcome at time $t$ if it remained untreated in both periods. Analogously, let $Y_{i,t}(0,1)$ denote unit $i$'s potential outcome at time $t$ if untreated in the first period but exposed to treatment by the second period. In our example, $Y_{i,t}(0,0)$ is county $i$'s mortality rate in period $t$ in a world in which Medicaid did not expand in its state, and $Y_{i,t}(0,1)$ is its mortality rate in a world in which Medicaid did expand in 2014.\footnote{We have implicitly introduced the stable unit treatment value assumption, which holds if the only treatment determining county $i$'s mortality rate is its own. In other words, Medicaid expansion in neighboring counties must not affect deaths in county $i$. If this fails, then there are effectively many different treatment variables and counterfactuals. The potential outcomes notation and ensuing analysis would then need to account for this.} To simplify notation, we will write $Y_{i,t}(0) = Y_{i,t}(0,0)$ and $Y_{i,t}(1) = Y_{i,t}(0,1)$, as the potential outcomes are defined by treatment exposure in period two (Medicaid expansion status by 2014). Nonetheless, it will be useful for later discussions that these potential outcomes correspond to treatment sequences.

In practice, we never observe $Y_{i,t}(1)$ and $Y_{i,t}(0)$ for the the same unit. Instead, the data we observe, $Y_{i,t}$, are treated outcomes $Y_{i,t}(1)$ for treated units, and untreated outcomes $Y_{i,t}(0)$ for untreated units, as in the following equation:
\begin{equation}\label{eqn:switching}
Y_{i,t} = (1 - D_i)Y_{i,t}(0) + D_i Y_{i,t}(1) .
\end{equation}
We additionally assume that county mortality rates were not affected by the Medicaid expansion \textit{before} Medicaid expanded, which is crucial to the validity of the DiD estimator (see, e.g., \citealp{Abbring2003}, \citealp{Malani2015}, and \citealp{Roth2023a}). This standard ``no anticipation'' assumption ensures that we observe untreated potential outcomes before Medicaid expansion takes effect: $Y_{i,2013} = Y_{i,2013}(0)$. It also helps us define effective treatment dates. For instance, if the announcement of Medicaid expansion affected mortality before its actual expansion, ``treatment'' would begin when the policy was announced rather than implemented. We formally state this assumption for completeness and maintain it throughout the paper. 

\begin{customass}{NA}[No-Anticipation]\label{ass:NA}
For all treated units $i$ and all pre-treatment periods $t$, $Y_{i,t}(1) = Y_{i,t}(0)$. 
\end{customass}

The potential outcomes define a causal effect for every unit in every time period, $Y_{i,t}(1)-Y_{i,t}(0)$. These describe what Medicaid expansion did to mortality rates in a specific treated county or what it would have done in a specific untreated county. This framework allows for arbitrary heterogeneity in the effects across units and time; that is, it allows the effect of Medicaid expansion to be different in every county and year. But it is hard to learn about this degree of rich heterogeneity without additional strong conditions holding. Instead, DiD analyses typically seek to estimate (weighted) averages of heterogeneous treatment effects. In particular, most DiD designs target the average treatment effect on the treated at time $t$, or $ATT(t)$:
\begin{align}\label{eqn:ATT(t)}
ATT(t) &= \E_\omega[Y_{i,t}(1)-Y_{i,t}(0)|D_i=1] \nonumber \\
&=\E_\omega[Y_{i,t}|D_i=1]-\textcolor{black}{\E_\omega[Y_{i,t}(0)|D_i=1]}.
\end{align}
Equation \eqref{eqn:ATT(t)} shows that $ATT(t)$ compares (weighted) average \textit{observed} post-expansion mortality rates among treated counties ($\E_\omega[Y_{i,t}|D_i=1]$) with the (weighted) average untreated mortality rates for the same treated counties ($\textcolor{black}{\E_\omega[Y_{i,t}(0)|D_i=1]}$). The second quantity is counterfactual because untreated outcomes are never observed for treated counties. Note that by the no-anticipation assumption, $ATT(t)=0$ for all pre-treatment periods; that is, $ATT(2013)=0$ in our two-period Medicaid example. This ensures that cross-group outcome comparisons before treatment begins reflect \textit{untreated} potential outcome gaps, which is central to the logic of DiD. Note that we abuse notation and omit the weight index when defining $ATT$'s; we do that to unclutter notation throughout the paper.

Equation \eqref{eqn:ATT(t)} shows that weighting enters the analysis early on, as part of the definition of the causal parameter. In the Medicaid context, the unweighted $ATT(2014)$ answers the question, ``What was the average causal effect of Medicaid expansion on 2014 mortality rates among the 2014 expansion state counties?'' The weighted $ATT(2014)$ answers the question, ``What was the average causal effect of Medicaid expansion on 2014 mortality rates among adults in counties in states that expanded Medicaid in 2014?'' This point interacts with other justifications for weighting, such as improving precision. With heterogeneous treatment effects, adopting a weighting scheme designed to improve precision in the presence of heteroskedasticity in a constant-coefficient regression model will also change the target parameter, potentially by a lot when treatment effects are correlated with the weights \citep{solon2015}. Comparing weighted and unweighted estimates, therefore, does not show whether weighting matters for estimation or inference; these reflect different target parameters. In our example, the population-weighted parameter is probably more policy-relevant, but we conduct some of our empirical exercises both ways to show how weighting can affect a given DiD result. Which parameter is ``of interest'' is an argument about theoretical importance, policy relevance, and the use to which it will be put.  

Other target parameters are also possible. Designs other than DiD identify different kinds of average treatment effects, and some DiD methods use quantile regression \citep{Athey2006, Callaway2019} or distribution regression \citep{Fernandez-Val2024} approaches to target features of the marginal distributions of $Y_{i,t}(1)$ and $Y_{i,t}(0)$ among treated units. We focus on identification and estimation strategies that target ATT parameters but emphasize that the $2\times2$ building block framework applies to DiD methods more broadly; see our appendix for more discussions about distributional parameters.

\subsection{Identifying assumptions: Parallel trends} \label{sec:PT}
A research design is a strategy---a set of assumptions---to identify and estimate specific target parameters. Many different assumptions can identify the missing counterfactual for $ATT(2014)$ in the Medicaid example. For example, mean independence between $Y_{i,2014}(0)$ and $D_i$ implies that the counterfactual equals average 2014 mortality rates in non-expansion counties ($\E_\omega[Y_{i,2014}(0)|D_i=0]$). Under this assumption, which essentially entails assuming that Medicaid expansion is as good as random, the cross-sectional mortality gap in 2014 between expansion and non-expansion counties is the $ATT(2014)$. Similarly, time invariance of $Y_{i,t}(0)$ among expansion counties (plus the fact that we ruled out anticipatory behavior) implies that the counterfactual equals 2013 mortality rates in expansion counties ($\E_\omega[Y_{i,2013}(0)|D_i=1]$). Under this assumption, which essentially rules out non-treatment-related changes in the outcome variable, the ``time trend'' in average mortality in expansion counties is the $ATT(2014)$. 

DiD comes from an alternative assumption that identifies the relevant counterfactual even when the average untreated potential outcome differs across treatment groups (which violates mean independence) and changes over time (which violates time invariance). The {parallel trends} assumption states that in the absence of treatment, the average outcome evolution is the same among treated and comparison groups. For general assumptions and results, we denote time periods by $t=1,2$, but continue to be explicit about which years are being used when we reference the Medicaid example.

\begin{customass}{PT}[$2\times2$ Parallel Trends]\label{ass:parallel-trends}
The (weighted) average change of $Y_{i,t=2}(0)$ from $Y_{i,t=1}(0)$ is the same between treated and comparison groups; that is, 
\begin{align}
	\textcolor{black}{\E_\omega[Y_{i,t=2}(0)|D_i=1]}-\E_\omega[Y_{i,t=1}(0)|D_i=1] = \E_\omega[Y_{i,t=2}(0)|D_i=0]-\E_\omega[Y_{i,t=1}(0)|D_i=0].\label{eqn:parallel-trends}
\end{align}
\end{customass}
If the parallel trends assumption holds, then it is easy to construct \textcolor{black}{$\E_\omega[Y_{i,t=2}(0)|D_i=1]$} from observable quantities---that is, to identify it:
\begin{align}
\textcolor{black}{\E_\omega[Y_{i,t=2}(0)|D_i=1]} = \E_\omega[Y_{i,t=1}|D_i=1] + \big(\E_\omega[Y_{i,t=2}|D_i=0]-\E_\omega[Y_{i,t=1}|D_i=0]\big). \label{eqn:parallel-trends-counterfactual}
\end{align}
In the Medicaid example, assumption \ref{ass:parallel-trends} says that to calculate expansion counties' average 2014 mortality rate in a counterfactual world without Medicaid expansion, start with their average 2013 mortality rate and add the observed change in average mortality rates in non-expansion counties. Substituting \eqref{eqn:parallel-trends-counterfactual} into the definition of $ATT(2014)$ and replacing potential outcomes with observed outcomes using \eqref{eqn:switching} gives the $2\times2$ DiD estimand, an expression for the target parameter in terms of four estimable \textit{population} averages:
\begin{small}\begin{align}
	ATT(2014) =&~ \overbrace{\E_\omega[Y_{i,2014}|D_i=1]}^{=\E_\omega[Y_{i,2014}(1)|D_i=1]} - \overbrace{\big( \E_\omega[Y_{i,2013}|D_i=1]+\big(\E_\omega[Y_{i,2014}|D_i=0]-\E_\omega[Y_{i,2013}|D_i=0]\big)\big)}^{\textcolor{black}{=\E_\omega[Y_{i,2014}(0)|D_i=1]}} \nonumber \\
	=&~ \underbrace{(\E_\omega[Y_{i,2014}|D_i=1]-\E_\omega[Y_{i,2013}|D_i=1])}_{\text{(weighted) average change for } D_i=1} - \underbrace{(\E_\omega[Y_{i,2014}|D_i=0]-\E_\omega[Y_{i,2013}|D_i=0])} _{\text{(weighted) average change for } D_i=0}.\label{eqn:2x2estimand} 
\end{align}     \end{small}

Equation \eqref{eqn:2x2estimand} highlights what makes DiD so attractive. It is intuitive, it has very mild data requirements (just four means), it answers \textit{ex post} questions like ``what did the treatment do?'', and its identifying assumption can be stated precisely. 

Parallel trends makes DiD distinct from causal designs that are based on statistical independence between treatment and potential outcomes. 
In designs like randomized trials or instrumental variables, the conditions---mean equalities across groups, for instance---that identify counterfactuals are often a statistical \textit{consequence} of the randomness induced externally \citep{Heckman2000}. In contrast, parallel trends is just a restriction on untreated potential outcome trends. It does not necessarily come from exogenous variation ``outside the model.'' In fact, because treatment adoption is often chosen by economic actors or policymakers ``inside the model,'' parallel trends need not hold. For this reason, DiD analyses (correctly) devote significant attention to evaluating parallel trends. 

We discuss how to generate empirical evidence about the plausibility of parallel trends below, but new theoretical findings about treatment choice behaviors or selection mechanisms also inform the plausibility of parallel trends.\footnote{In fact, some of the earliest economic research on DiD methods examined exactly these questions \citep{Ashenfelter1985, Heckman1985}.} These results explicitly connect DiD to behaviors, they provide grounding for stories about why a given DiD analysis is internally valid (or not), and they discipline empirical tests of parallel trends. A full review of this literature is outside the scope of this paper, but some helpful broad themes emerge.\footnote{For details on selection models and outcome models that are consistent with parallel trends, see \citet{Ghanem2023} and \citet{Marx2024}; \citet{Chabeferret2015} applies these arguments to earnings dynamics.} For instance, a trade-off exists between the information the agents who choose treatment know and how they act on it, and the time-series properties of untreated potential outcomes. At one extreme, consider someone who knows $Y_{i,t}(0)$ before and after treatment and can opt into or out of treatment based on it. Parallel trends can only hold in this case if, other than common shifts for every unit, $Y_{i,t}(0)$ is constant \citep{Ghanem2023}. In our Medicaid example, neither of these conditions---that state legislatures in 2013 knew their 2014 untreated mortality rates or that untreated mortality rates would all have shifted in parallel---is plausible. The opposite extreme is when treatment timing is random, in which case parallel trends holds without any time-series restrictions. Yet, in this case, DiD is not necessary, and more efficient estimators exist \citep{Roth2023c}. Against the political and economic backdrop of the early 2010s, it is clear that Medicaid expansion decisions were not random.

Therefore, in realistic scenarios, parallel trends only holds if \textit{some} restrictions on the way untreated outcomes enter the treatment selection mechanism hold as well. As one example, imagine that treatment selection depends on the permanent component of $Y_{i,t}(0)$ (fixed effects) but not on shorter-term fluctuations (``shocks''). For instance, if state legislatures knew and considered their long-run mortality levels only when making their expansion decision, they would be following this kind of selection mechanism. Expansion and non-expansion states might then have large differences in the permanent part of untreated outcomes, which cancel in equation \eqref{eqn:parallel-trends}, and so parallel trends would hold if shocks to $Y_{i,t}(0)$ had a time invariant mean conditional on the fixed effects.\footnote{Some researchers may find easier to understand these as ``parallel changes'' rather than ``parallel trends.'' However, the use of ``parallel trends'' is now firmly established in the literature, and other influential work has used ``changes-in-changes'' to refer to an alternative estimator to classical DiD estimation \citep{Athey2006}. To avoid confusion, we use ``parallel trends'' throughout this paper.} State legislatures, however, may have also known whether their 2013 mortality rates were especially high or low when considering expanding Medicaid. If the expansion choice is related to these 2013 mortality shocks as well, parallel trends would hold only if stronger time-series restrictions on $Y_{i,t}(0)$ hold. \citet{Ghanem2023} provide a fuller discussion of the selection/time-series trade-off and theory-driven templates to assess parallel trends, while \citet{Marx2024} discuss economic models that are and are not compatible with parallel trends. 

Another implication of the fact that DiD does not rely on statistical independence between $Y_{i,t}(0)$ and treatment status is that there is no guarantee that parallel trends holds across different transformations of $Y_{i,t}(0)$. As stated, it is simply an assumption about averages for a particular $Y_{i,t}(0)$. \citet{Roth2023b} show that parallel trends is insensitive to functional form if and only if it holds between groups and across the distribution of $Y_{i,t}(0)$. This can only be true if Medicaid adoption is random, the mortality distribution is constant between 2013 and 2014, or a combination of the two cases. As these conditions are arguably ex-ante implausible, our DiD analysis may depend on our choices to measure $Y_{i,t}$ in rates (deaths per 100,000) as opposed to logs, for example. One approach to this measurement choice is to propose and evaluate a theory that delivers it, though we recognize that this is not always possible. To assess whether cases where parallel trends holding for one functional form come at the cost of ruling other transformations, we recommend that researchers use the \citeauthor{Roth2023b}'s (\citeyear{Roth2023b}) falsification tests for the null that parallel trends are insensitive to functional form. In our application, we do not reject the null that parallel trends is insensitive to functional form, with p-values above 0.80.

The interplay between treatment selection and the properties of the outcome variable characterize the structural basis for a DiD analysis (see \citealp{DiNardo2011}) and engaging with them is essential to any DiD application. While every study will have its own institutions, choices, and outcomes to consider, a rigorous DiD analysis must provide a transparent discussion about the reliability of the underlying identification assumptions. If parallel trends requires implausible behavioral restrictions, one may be better off using an alternative research design. 

\subsection{Estimation and inference: Four means or one regression?} \label{sec:estimation}
Mapping the DiD estimand in equation \eqref{eqn:2x2estimand} to the canonical {${2\times2}$ DiD estimator} follows immediately from replacing population expectations with their sample analogs:
\begin{align}
\widehat{ATT}(2014) = (\overline{Y}_{\omega, D=1, t=2014} - \overline{Y}_{\omega,D=1, t=2013}) - (\overline{Y}_{\omega,D=0, t=2014} - \overline{Y}_{\omega,D=0, t=2013}),\label{eqn:2x2_ATT}
\end{align}
\noindent where $\overline{Y}_{\omega,D=g, t=t'} = \dfrac{\sum_{i=1}^n\1\{D_i=g, t=t'\}\omega_iY_{i,t'}}{\sum_{i=1}^n \omega_i\1\{D_i=g, t=t'\}}$ is the $\omega$-weighted sample mean of $Y$ for treatment group $g$ in period $t'$. Equation \eqref{eqn:2x2_ATT} is the classic difference of two differences written in terms of sample means. It is a direct recipe for actually estimating $ATT(t)$ and can be read directly from the following table of average mortality rates in 2013 and 2014 by expansion group.

\input{Tables/two_by_two_ex}

The two across-time changes in equation \eqref{eqn:2x2_ATT} are in the third row of the table. Without weighting, average county-level mortality rates in expansion states rose by 9.3 deaths per 100,000 and 9.1 deaths in non-expansion states, so after rounding, the DiD estimate of $ATT(2014)$ is 0.1 deaths per 100,000. This result implies that the average treatment effect of Medicaid expansion on mortality in 2014 among \emph{counties} that are part of an expansion state was an increase of 0.1 deaths per 100,000. In contrast, the DiD result using population weights suggests that Medicaid expansion caused a reduction of 2.6 deaths per 100,000 for the average \textit{adult} in expansion states.\footnote{Columns 3 and 6 show cross-group gaps in average mortality in each year. These can also be used to construct the DiD estimate by rearranging equation \eqref{eqn:2x2_ATT}: $(\overline{Y}_{D=1, t=2014} - \overline{Y}_{D=0, t=2014}) - (\overline{Y}_{D=1, t=2013} - \overline{Y}_{D=0, t=2013})$.}

The same result can be obtained as the (weighted) least squares estimate of $\beta^{2\times2}$ in the following linear regression specification (which only uses data from $t=2013$ and $t=2014$):\begin{equation}
Y_{i,t} = \beta_0 + \beta_1 \indicator{D_i=1} + \beta_2 \1\{t=2014\}  + \beta^{2\times2} (\indicator{D_i=1} \times \1\{t=2014\}) + \varepsilon_{i,t}, \label{eqn:twfe_2_by_2}
\end{equation}
where $\beta$'s are unknown coefficients and $\varepsilon_{i,t}$ is an idiosyncratic term uncorrelated with $D_i$. To see why, let's focus on the unweighted case ($\omega_i=1$) and write each of the four means in $\widehat{ATT}(2)$ in terms of the estimated coefficients from \eqref{eqn:twfe_2_by_2}:
\begin{itemize}
\item Sample average of $Y_{i,t}$ in post-period for treatment group is $\overline{Y}_{D=1,t=2014} = \widehat{\beta_0} + \widehat{\beta_1} + \widehat{\beta_2} + \widehat{\beta}^{2\times2}$.
\item Sample average of $Y_{i,t}$ in pre-period for treatment group is $\overline{Y}_{D=1,t=2013} = \widehat{\beta_0} + \widehat{\beta_1} $.
\item Sample average of $Y_{i,t}$ in post-period for comparison group is $\overline{Y}_{D=0,t=2014} = \widehat{\beta_0} + \widehat{\beta_2} $.
\item Sample average of $Y_{i,t}$ in pre-period for comparison group is $\overline{Y}_{D=0,t=2013} = \widehat{\beta_0}$.
\end{itemize}

Substituting these expressions into the definition of $\widehat{ATT}(2)$ yields:
\begin{align}
\widehat{ATT}(2014) &= \bigg [( \widehat{\beta_0} + \widehat{\beta_1} + \widehat{\beta_2} + \widehat{\beta}^{2\times2} ) -  (\widehat{\beta_0} + \widehat{\beta_1}) \bigg ] - \bigg [ (\widehat{\beta_0} + \widehat{\beta_2}) -\widehat{\beta_0} \bigg ]  = \widehat{\beta}^{2\times2} \nonumber.
\end{align}
Table \ref{tab:regdid_2x2} demonstrates this equivalence for both unweighted (column 1) and weighted (column 4) regressions. In fact, with balanced panel data, the estimate of $\beta^{2\times2}$ is numerically the same if the regression instead contains fixed effects for each unit (columns 2 and 5) or if one regresses outcome changes on a constant and the treatment group dummy $D_i$ (columns 3 and 6).\footnote{When population weights vary over time, the equivalence between a by-hand DiD estimate and one that comes from a regression with unit fixed effects no longer holds.} Provided that one adopts the same notion of uncertainty, standard errors also coincide; such as when one clusters the standard errors at the county level.

\input{Tables/regdid_2x2}

The equivalence between calculating a $2\times2$ DiD by hand or with a regression has appealing features. Regressions are simple to run, and they do the averaging and differencing behind the scenes. They also allow the use of statistical inference tools from ordinary least squares (OLS), which are themselves the subject of a large econometrics literature that is particularly important when it comes to standard error estimation \citep{Wooldridge2003, Bertrand2004, Donald2007, Cameron2008, Conley2011, Abadie2020, Abadie2023}.

Many inference procedures exist for DiD-type analyses, arising from a combination of choices about the target parameter, details of the data structure and sampling process, and maintained assumptions about the structure of outcomes. In practice, one needs to determine and discuss the forms of uncertainty the standard errors are designed to capture---that is, what is (conceptually) being resampled and what may or may not vary across those resamples. As discussed in \citet{Abadie2020}, these details come from the nature of the parameter of interest---whether the focus is on sample-specific average treatment effects or population-level average treatment effects---and the stochastic elements of the model that make the estimator random. Heuristically, this involves a thought experiment (or stochastic process) hypothesized to generate the random components of the model (or the data-generating process). Different inferential frameworks highlight different sources of uncertainty by resampling distinct model components and treating other components as fixed (non-random).

Inferential frameworks on two extremes help cement these concepts. Design-based frameworks treat potential outcomes and covariates as non-random, focus on finite-population parameters (e.g., sample average treatment effects), and consider the allocation of treatment as the only source of the randomness in the model \citep{imbensrubin2015}.\footnote{Traditionally, design-based inference procedures are justified when treatment assignment is fully random, which is a much stronger requirement than parallel trends. See \citet{Rambachan2022} for a discussion on design-based inference for quasi-experimental designs, including a discussion of the Medicaid expansion.} The only thing that is random and thus varies across the hypothetical resamples from this point of view is the treatment allocation. On the other hand, a traditional sampling-based approach to inference presumes that we independently sample units from a superpopulation. In this case, it is customary to focus on population parameters (like $ATT(2)$), treat all variables in the model as random variables, and compute standard errors by clustering at the level in which the (hypothesized) sampling was conducted. In this framework, every variable in the analysis---outcomes, covariates, and treatment---is randomly redrawn across the hypothetical resamples. A drawback of the sampling approach is that sometimes, it is unnatural to think of the data as a random sample from a well-defined population. 

A third popular approach to inference---the model-based approach---is more structural and involves taking a stand on the structure of the error component of the model (e.g., imposing a putative model for how shocks affect outcomes and their relationship with treatment and other variables in the model). The uncertainty reflected in this model-based setting entails a thought experiment in which different values of these shocks and the other random variables in the model are drawn from their joint distribution \citep{Abadie2023}. This model-based approach is common in econometrics, and it almost always takes the linear regression specification (or model, in this case) as the starting point of the analysis. Although this is often convenient, it is important to note that imposing model restrictions on the error component of the model necessarily imposes restrictions on treatment effect heterogeneity and on the relationship between potential outcomes; see Section 5 and Appendix A of \citet{Roth2023a} for a discussion. Another challenge with the model-based approach is that it is hard to use this framework when adopting estimation strategies other than linear regressions---for example, inverse probability weighting or doubly robust procedures---which we will discuss in Session \ref{sec:did_covariates}.

Ultimately, as the discussion above highlights, each approach has pros and cons, and discussions about the best way to compute standard errors are complex and intrinsically context-specific. A detailed treatment of the topic is outside the scope of this paper. It requires information about the sampling process, the research design and target parameters, what is treated as fixed and random, and the structure of the error components of their models (e.g., presence of spatial or serial correlation), among other factors. We refer interested readers to \citet{Abadie2020, Abadie2023} and Section 5 of \citet{Roth2023a} for discussions on these topics, though we also emphasize that further methodological research in this area is warranted. For the remainder of this article, we adopt a sampling perspective for uncertainty and cluster our standard errors at the county level. In our context, this is compatible with treating all variables as random, including treatment groups and potential outcomes. It also allows us to avoid (a) making time-series dependence restrictions on potential (and realized) outcomes---as we are in a short-panel framework with a large number of units and a fixed number of time periods---and (b) taking an explicit stand on the structure of error components of the model, which is particularly appealing as the starting point of our analysis is potential outcomes rather than regression models. It is also worth mentioning that as our treatment in the empirical example is assigned at the state level, clustering at the county level would also be compatible with treating state-specific shocks as fixed (or conditioned on) and assessing if they lead to violations of parallel trends \citep[Section 5.1]{Roth2023a}. Clustering at the state level would be justified if we were using a design-based perspective \citep{Rambachan2022}, though that would require us to treat potential outcomes as fixed (which we do not do in this paper). Our choice of inference procedure is not without controversies, and other inferential approaches may also be rationalized.

We conclude this section by stressing that the appeal of using regressions like \eqref{eqn:twfe_2_by_2} to estimate ATT in DiD designs comes from the fact that their numerical equivalence to the ``by-hand'' DiD estimator \eqref{eqn:2x2_ATT}, which was explicitly derived from the $ATT(t)$ and the parallel trends assumption. This ensures that the regression specification respects the underlying identifying assumptions and estimates the desired target parameter. Unfortunately, the tight connection between (TWFE) regressions and DiD designs breaks under more complex setups that are ubiquitous in practice. We now turn to some of these issues and how approaching them from the point of view of $2\times2$ building blocks can guide good econometric practices.

\section{Incorporating covariates into \texorpdfstring{2$\times$2}{2x2} DiD}\label{sec:covariates}
So far, we have focused on $2\times2$ DiD designs that do not leverage any information about covariates, but researchers frequently incorporate them into DiD analyses in one of three ways: checking for balance in variables thought to influence $Y_{i,t}(0)$, controlling for those variables in the main estimates, and estimating treatment effect heterogeneity. For example, in the absence of Medicaid expansion, mortality rates likely would have evolved differently in poorer and richer counties; they certainly did before 2014 \citep{Currie2016}. Therefore, parallel trends may fail if poverty rates differ between expansion and non-expansion counties. If they do, then one may want to ``control for'' poverty rates when estimating $ATT$ parameters. Finally, because Medicaid expansion reached more people in higher-poverty counties, its average effects on overall mortality may be larger there. This heterogeneity may be of interest in its own right or may be used to assess the plausibility of the overall DiD design. 

This section discusses how to use auxiliary information on covariates to evaluate parallel trends, identify $ATT$ parameters under potentially weaker assumptions, and study heterogeneity. These approaches stem from viewing a DiD with covariates in terms of $2\times2$ building blocks that themselves condition on those variables, which creates a clear link to the assumptions, estimators, and interpretation of unconditional $2\times2$ designs. We also discuss how regression estimators that control for covariates impose extra assumptions and can fail to identify $ATT$ parameters. 

\subsection{Covariate balance: Is unconditional parallel trends plausible?}\label{sec:covariate-balance}
Assumption \ref{ass:parallel-trends} is fundamentally untestable, as it contains an unobserved counterfactual component. Therefore, ``tests'' of these assumptions are necessarily indirect and rely on other \textit{observed} variables thought to be related to untreated potential outcome trends. For example, during the 2010s, demographics and economic conditions were strongly correlated with mortality levels and trends. If these relationships would have held in the absence of Medicaid expansion, and if expansion and non-expansion counties differ in those demographic and economic characteristics, then the parallel trends assumption \eqref{eqn:parallel-trends} may fail to hold. Checking balance in observable determinants of changes in $Y_{i,t}(0)$ is thus a common and sensible way to evaluate parallel trends.

Most DiD analyses check for balance across groups in baseline covariate levels ($\E_\omega[X_{i,t=1}|D_i=1]-\E_\omega[X_{i,t=1}|D_i=0]$) or covariate trends before and after treatment ($\E_\omega[\Delta X_{i,t=2}|D_i=1]-\E_\omega[\Delta X_{i,t=2}|D_i=0]$, where $\Delta X_{i,t=2} = X_{i,t=2}-X_{i,t=1}$). We consider the following variables, $X_{i,t}$: the percentages of a county's population that are female, white, or Hispanic; the unemployment rate; the poverty rate; and county-level median income (in thousands of dollars).\footnote{We focus on these variables for convenience. \citet{Borgschulte2020} use a LASSO procedure that selects more and different covariates to include in their analysis. We replicate their findings when we follow their methodology but diverge here for the sake of brevity.}  The top panel of Table \ref{tab:cov_balance} reports averages of these variables by group in 2013 with and without population weights. 
We also report a measure of imbalance that is comparable across variables: the normalized difference in means between the treatment and comparison groups \citep[Chapter 14]{imbensrubin2015},
$$\text{Norm. Diff}_\omega = \frac{\overline{X}_{\omega,T} - \overline{X}_{\omega,C}}{\sqrt{(S_{\omega,T}^2 +S_{\omega,C}^2)/2}} ,$$
\noindent where $\overline{X}_{\omega,T}$ and $\overline{X}_{\omega,C}$ are the sample weighted or unweighted averages for the treatment and comparison groups, respectively, and $S_{\omega,T}^2$ and $S_{\omega,C}^2$ are the sample weighted or unweighted variances of the covariates for the treatment and comparison group. As a general rule of thumb, values of the normalized difference in excess of 0.25 in absolute value indicate a potentially problematic imbalance between the two groups \citep[page 277]{imbensrubin2015}.\footnote{As discussed in \citet[Section 3.2]{Austin2009_Balance}, the normalized difference was initially proposed in the psychological literature and is sometimes referred to as Cohen’s effect size index. In this context, normalized differences of 0.2, 0.5, and 0.8 are sometimes used to represent small, medium, and large imbalances; however, normalized differences as small as 0.1 are sometimes considered worrisome, depending on how important the covariate in question is. Ultimately, there is no universally accepted threshold for what value indicates important imbalances. See \citet{Ho_Imai_King_Stuart_2007_matching} and \citet{Austin2009_Balance} for additional discussions.}

\input{Tables/cov_balance}

We find meaningful imbalance in several baseline measures. Expansion counties in 2013 were whiter and had higher unemployment rates despite lower poverty and higher median income than non-expansion counties. Because DiD uses \textit{changes} in outcomes, researchers sometimes argue that the effect of pre-treatment variables is differenced out. This logic does not hold, though, if baseline covariates are related to untreated potential outcome trends themselves. The imbalance in the top panel of Table \ref{tab:cov_balance} will lead to violations of parallel trends to the extent that counties with different racial composition or income distributions would have had different mortality changes even without Medicaid expansion.

Nevertheless, checks of balance in covariate changes can be informative about parallel trends as well. The bottom panel of Table \ref{tab:cov_balance} reports average changes by group between 2013 and 2014 as well as normalized differences. Many of the imbalances evident in baseline levels change, or even flip signs, when measured in changes. Unemployment, for example, was higher in expansion states in 2013 but fell faster. To the extent that these changes are important determinants of $\Delta Y_{i, t}(0)$, then these results could suggest that Assumption \ref{ass:parallel-trends} is violated.  

Why do we say ``could''? A major challenge in interpreting cross-group gaps in $\Delta X_{i,t}$ involves deciding which variables are truly covariates and which are mechanisms/outcomes. If an element of $X_{i,t}$ cannot be affected by the treatment, it is a (strictly exogenous) covariate, and differential changes in exogenous covariates may indicate a PT violation. Since the treatment cannot have caused $X_{i,t}$ to change (by assumption), something else that differs across groups and over time must have. Since little research suggests an effect of Medicaid expansion on unemployment, this may be a good assumption. On the other hand, if Medicaid expansion can change the demographic and economic composition of its counties, then differential changes in these variables may actually be a consequence of the expansion itself.\footnote{In fact, comparing mean covariate changes in expansion and non-expansion is the same as using $X_{i,t}$ as the outcome in a $2\times2$ DiD estimator.} If so, then differential post-treatment changes in them would not necessarily indicate a parallel trends violation; they could partially reflect a causal effect. As with the plausibility of Assumption \ref{ass:parallel-trends} itself, whether something is a covariate or a mechanism is not a data question per se. It requires context-specific knowledge about how the treatment works. 

\subsection{DiD with covariates: Identification under conditional parallel trends}\label{sec:2x2_DiD_with_X}
Having detected covariate imbalance that casts doubt on Assumption \ref{ass:parallel-trends}, how should we proceed to estimate $ATT(2)$? Because the imbalance documented in Table \ref{tab:cov_balance} suggested that unconditional parallel trends may not hold, our goal is to develop a DiD identification strategy based on an assumption that accounts for this imbalance. Working from a conditional parallel trends assumption shows how to construct $ATT(2)$ from $2\times2$ comparisons that are each conditioned on specific covariate values, thus addressing the imbalance problem.

Let $X_i$ be a vector of observed determinants of changes in $Y_{i,t}(0)$. Here, we purposefully omit the time subscript on $X_i$ because the covariates in this section can be time-invariant, such as fixed variations or baseline values ($X_{i,t=1}$), or time-varying in the sense of including values from the second period, $X_{i,t=2}$. The empirical content of a ``new'' identification assumption that incorporates $X_i$, henceforth conditional parallel trends (CPT) assumption, is formalized as follows.

\begin{customass}{CPT}[$2\times2$ Conditional Parallel Trends] \label{ass:cond-parallel-trends}
The (weighted) average change of $Y_{i,t=2}(0)$ from $Y_{i,t=1}(0)$ is the same between treated and comparison units that share the same covariate values,
\begin{align}
	\E_\omega[Y_{i,t=2}(0) - Y_{i,t=1}(0) | X_i, D_i=1] = \E_\omega[ Y_{i,t=2}(0)- Y_{i,t=1}(0) | X_i, D_i=0].\label{eqn:cond-parallel-trends}
\end{align}
\end{customass}

Assumption \ref{ass:cond-parallel-trends} has the same structure as Assumption \ref{ass:parallel-trends} but states that PT holds within each covariate-specific stratum rather than across the whole population. With respect to the baseline covariates in Table \ref{tab:cov_balance}, this amounts to assuming parallel trends in $Y_{i,t}(0)$ between expansion and non-expansion counties that in 2013 had the same female, white, and Hispanic shares, as well as the same unemployment and poverty rates and median income. This does not restrict \textit{how} $Y_{i,t}(0)$ changes in different covariate strata, nor is not necessary to estimate these trends. Assumption \ref{ass:cond-parallel-trends} only requires that covariate-specific trends are common between expansion and non-expansion counties.

For both expectations in Assumption \ref{ass:cond-parallel-trends} to be well-defined for all values of $X_i$, there must be both untreated and treated units in the population at each covariate value. If, for some covariate values, there are only treated units, for example, then the right-hand side of \eqref{eqn:cond-parallel-trends} is undefined. A formal statement of this assumption, called ``common support'' or ``strong overlap,'' is as follows.\footnote{$ATT(2)$ is still identified under conditional parallel trends if some values of $X_i$ have only untreated observations. We require $P_\omega[D_i=1|X_i]$ to be bounded away from zero and one to avoid irregular inference procedures; see \citet{Khan2010} for additional details.}

\begin{customass}{SO}[Strong overlap] \label{ass:overlap}
The conditional (weighted) probability of belonging to the treatment group, given observed covariates $X_i$, which are determinants of untreated potential outcome growth, is uniformly bounded away from zero and one. That is, for some $\epsilon>0$, $\epsilon < P_\omega[D_i=1|X_i] < 1-\epsilon$.
\end{customass}

Under assumptions \ref{ass:cond-parallel-trends} and \ref{ass:overlap} the $ATT(2)$ is identified: 
\begin{align}
ATT(2) &= \E_\omega[Y_{i,t=2}(1) | D_i=1] - \textcolor{black}{\E_\omega[Y_{i,t=2}(0) | D_i=1]} \nonumber \\
&= \E_\omega[Y_{i,t=2} | D_i=1] - \textcolor{black}{\E_\omega\Big[ \E_\omega[Y_{i,t=2}(0) | X_i, D_i=1] \Big| D_i=1\Big]} \nonumber \\
&= \E_\omega[Y_{i,t=2} | D_i=1\Big] - \E_\omega\Big[ \E_\omega[ Y_{i,t=1}(0) | X_i, D_i=1] +\E_\omega[\Delta Y_{i,t=2}(0) | X_i, D_i=0]  \Big| D_i=1\Big] \nonumber \\
&= \E_\omega[\Delta Y_{i,t=2} | D_i=1\Big] - \E_\omega\Big[ \E_\omega[\Delta Y_{i,t=2} | X_i, D_i=0] \Big| D_i=1\Big]  .\label{eqn:ATT_OR}
\end{align}
The first line restates the definition of $ATT(2)$, and the second line uses the law of iterated expectations to write the counterfactual mean for the whole treatment group as an average of counterfactual means conditional on $X_i$. These quantities are exactly the ones that appear in the conditional parallel trends assumption (and the overlap condition). Therefore, as in section \ref{sec:2x2}, the third line uses Assumption \ref{ass:cond-parallel-trends} to rewrite the conditional counterfactuals in terms of observable population quantities under no anticipation (Assumption \ref{ass:NA}). Equation \eqref{eqn:ATT_OR} then uses the law of iterated expectations again to group terms and express $ATT(2)$ in terms of observed variables $(Y_{i,t=2}, Y_{i,t=1}, G_i, X_i)$; that is, it establishes that $ATT(2)$ is nonparametrically identified under our assumptions.  This expression has a clear intuition: the $ATT(2)$ is equal to the path of outcomes experienced by the treated group (the term on the left) minus the average path of outcomes in the comparison group for each value of the covariates, averaged over the treated group's distribution of covariates (the term on the right).

\subsection{DiD estimation with covariates: TWFE}
Moving from the population identification result in equation \eqref{eqn:ATT_OR} to sample analogs is a challenge unless the covariates are discrete and the conditional expectations themselves are easily calculable. This difficulty does not arise in an unconditonal DiD. With continuous covariates, or many discrete ones, it may not be feasible to construct $\E_\omega[\Delta Y_{i,t=2} | X_i, D_i=0]$. Conditional DiD estimation, therefore, uses additional econometric techniques to bridge this gap. We begin, however, by discussing how regression DiD estimators that include covariates relate to the assumptions used for identification in \eqref{eqn:ATT_OR}.

Because the TWFE specification in \eqref{eqn:twfe_2_by_2} recovers the $ATT(2)$ in $2\times 2$ DiD setups without covariates, it is natural to extend this logic to regressions with covariates. Indeed, this is by far the most popular approach adopted by practitioners, arguably because it is both easy and familiar. A typical regression specification is
\begin{align}
Y_{i,t} = \theta_t + \eta_i + \beta_{treat} D_{i,t} + X_{i,t}'\beta_{covs} + e_{i,t}. \label{eq:reg_did_covs}
\end{align}
where the unit and time fixed effects, treatment status, and covariates have already been defined, $e_{i,t}$ is an error term, and $\beta_{treat}$ is interpreted as the parameter of interest.
A related specification explicitly controls for baseline covariates by replacing $X_{i,t}$ with interactions of the pre-treatment covariates $X_{i,t=1}$ and a post-treatment dummy,
\begin{align}
Y_{i,t} = \theta_t + \eta_i + \beta_{treat,2} D_{i,t} + (\1\{t=2\}X_{i,t=1})\beta_{covs,2} + e_{i,t}, \label{eq:reg_did_covs_interact}
\end{align}
In Table \ref{tab:regdid_2x2_covs}, we report the OLS and weighted least squares estimates of the unconditional $2\times2$ DiD estimate, $\beta_{covs}$ from \eqref{eq:reg_did_covs_interact}, $\beta_{covs,2}$ from \eqref{eq:reg_did_covs}, and their cluster-robust standard errors, using the covariates from Table \ref{tab:cov_balance}.

\input{Tables/regdid_2x2_covs}

Although only one covariate-adjusted estimate in Table \ref{tab:regdid_2x2_covs} is (marginally) statistically significant, the point estimates differ noticeably. In the unweighted case, adjusting for the 2013 levels of the covariates decreases the estimated effect of Medicaid expansion on short-run mortality rates from a point estimate of roughly 0.12 to -2.35. However, if we include their time-varying values instead, we estimate an effect of -0.49, a large difference. We find a similar result when using weighted regressions; while the coefficient remains constant (-2.56) when using 2013 values of the covariates, it attenuates to -1.37 if we use \eqref{eq:reg_did_covs}. 

The jump from the conditional DiD identification result in \eqref{eqn:ATT_OR} to the TWFE \textit{estimators} in \eqref{eq:reg_did_covs} and \eqref{eq:reg_did_covs_interact} skips a crucial question about $\beta_{treat}$ or $\beta_{treat,2}$: do they equal the target parameter $ATT(2)$ under the conditional parallel trends assumption? It turns out that the close relationship between regression DiD, $ATT(2)$, and parallel trends in a design without covariates does not hold with covariates. The issues come from exactly what kinds of covariates are effectively being ``controlled for'' in these specifications and how the regression estimator combines outcome trends for covariate sub-groups. 

Note that in our two-period setup, \eqref{eq:reg_did_covs} and \eqref{eq:reg_did_covs_interact} are respectively equivalent to (with some abuse of notation),
\begin{align*}
\Delta Y_{i,t=2} &= \alpha + \beta_{treat} D_{i} + \Delta X_{i,t=2}'\beta_{covs} + \Delta e_{i,t=2},\\
\Delta Y_{i,t=2} &= \alpha + \beta_{treat,2} D_{i} + X_{i,t=1}'\beta_{covs,2} + \Delta e_{i,t=2}.
\end{align*}
The first thing that is clear from these representations is that because time-invariant variables drop out of equation \eqref{eq:reg_did_covs}, a TWFE specification can account for differential trends related to baseline covariate levels only if they enter as interactions with the post-treatment dummy as in equation \eqref{eq:reg_did_covs_interact}. The exact regression specification, therefore, determines the implied conditional parallel trends assumption. Controlling for annual poverty rates really means controlling for poverty changes, and areas that are poor are not the same as areas that are becoming poor.  

Another limitation evident in \eqref{eq:reg_did_covs} relates to ``bad controls.'' Whenever $X_{i,t=2}$ is affected by the treatment, then conditioning on it (in any way) can bias estimates of the $ATT(2)$. If Medicaid expansion lowered poverty rates, for example, then including 2014 poverty rates or the 2013-2014 change in poverty rates as a covariate is problematic. This echoes our discussion about testing balance in $\Delta X_{i,t}$ in the sense that time-varying covariates must be unaffected by treatment in order to interpret imbalance in their trends as a source of bias, and to be able to control for them to address that bias. See \citet{Caetano2022} for a discussion.

Suppose we have decided on which variables to include in a conditional parallel trends assumption and whether to measure them in levels or changes. If Assumptions \ref{ass:cond-parallel-trends} and \ref{ass:overlap} hold with respect to this set of covariates, does $\beta_{treat}$ recover the $ATT(2)$? In the DiD context, \citet{Caetano2024} tackle exactly this question. They show that $\beta_{treat}$ equals a weighted average of conditional average treatment effects, defined as $ATT_{x_{k}}(2) \equiv \E_\omega\big[Y_{i,t=2}(1)-Y_{i,t=2}(0)| D_i=1, X_i = x_k\big]$, with weights that may not be convex, plus three bias terms reflecting misspecification either in the set of control variables or the fact that they are included linearly. These conclusions relate to recent findings about the properties of regression estimators in the presence of heterogeneity in other contexts, including instrumental variables \citep{Mogstad_Torgovitsky_2024}, cross-sectional designs \citep{Angrist_1998_ECMA, Aronow_Samii_2015, Sloczyski2022, Goldsmith-Pinkham_Hull_Kolesar_2024_AER}, and panel data \citep{Goodman2021, deChaisemartin2020, Sun2021, Poirier_Sloczyski_2024}. 

The weighting results suggest that even when the covariates are correctly selected, measured, and added with the correct functional form, $\beta_{treat}$ could be negative even when $ATT_{X_{i}}(2)$ is positive for all values of covariates. Short of this extreme sign-reversal case, $\beta_{treat}$'s weighting scheme generally does not yield the $ATT(2)$ target parameter and instead 
puts too much weight on $ATT_{X_{i}}(2)$ for $X_i$'s that are relatively uncommon among the treated group relative to the untreated group and puts too little weight on $ATT_{X_{i}}(2)$ for $X_i$'s that are relatively common among the treated group relative to the untreated group \citep{Sloczyski2022, Caetano2024}. 

Taken together, these results imply that $\beta_{treat}$ identifies $ATT(2)$ under the \textit{additional} assumption that treatment effects across covariate strata are constant. To see why, write the conditional ATT in period two given $\Delta X_{i,t=2}$ as
\begin{align*}
ATT_{\Delta X_{i,t=2}}(2)= \E_\omega[Y_{i,t=2}(1) - Y_{i,t=2}(0)|D_i=1, \Delta X_{i,t=2} ],
\end{align*}
and note that, under Assumptions \ref{ass:cond-parallel-trends} and \ref{ass:overlap}, it is identified by 
\begin{align*}
ATT_{\Delta X_{i,t=2}}(2)= \E_\omega[\Delta Y_{i,t=2}|D_i=1, \Delta X_{i,t=2} ] - \E_\omega[\Delta Y_{i,t=2}|D_i=0, \Delta X_{i,t=2} ].
\end{align*}
If we take \eqref{eq:reg_did_covs} to be a correctly specified regression, then
\begin{align*}
ATT_{\Delta X_{i,t=2}}(2)= \big(\beta_{treat} + \Delta X_{i,t=2}'\beta_{covs}\big) - \big( \Delta X_{i,t=2}'\beta_{covs}\big) =\beta_{treat}.
\end{align*}
In other words, \eqref{eq:reg_did_covs} implicitly rules out that treatment effects can vary across covariate-strata, which makes the weighting issues identified by \citet{Caetano2024} irrelevant to the interpretation of $\beta_{treat}$. Research on the Medicaid expansion using data on mortality rates by income, however, shows clear evidence of heterogeneous effects \citep{miller2021medicaid, wyse2024}.

One way to avoid these limitations would be to make \eqref{eq:reg_did_covs_interact} (or \eqref{eq:reg_did_covs}) more flexible by including interactions of the covariates with treatment group, time, and treatment-group-by-time. An alternative possibility is to adopt a ``forward-engineering'' perspective \citep{Mogstad_Torgovitsky_2024} and derive an estimator for $ATT(2)$ that directly leverages Assumptions \ref{ass:NA}, \ref{ass:cond-parallel-trends} and \ref{ass:overlap}. In some situations, the forward-engineering approach will also use regressions. Still, the target parameter and the identifying assumptions will guide the specification we should use (and not the other way around). 



\subsection{DiD estimators with covariates that target the ATT(2)}\label{sec:did_covariates}

Fortunately, TWFE is not the only way to bring covariates into DiD estimation. We now discuss alternative ways to use covariates in a DiD analysis that start from the identification result in \eqref{eqn:ATT_OR}:
\begin{align*}
ATT(2) &= \E_\omega[\Delta Y_{i,t=2} | D_i=1] - \E_\omega\Big[ \E_\omega[\Delta Y_{i,t=2} | X_i, D_i=0] \Big| D_i=1\Big].
\end{align*}
This equation provides an intuitive recipe for estimating $ATT(2)$. The first term in $ATT(2)$ is the same as in an unconditional $2\times2$ design and can be replaced by its sample analog as in equation \eqref{eqn:2x2_ATT}: $(\overline{Y}_{\omega, D=1, t=2} - \overline{Y}_{\omega,D=1, t=1})$. 

One way to obtain the second term is to first estimate the inner conditional expectation, $\E_\omega[\Delta Y_{i,t=2} | X_i, D_i=0]$. This object is just an (unknown) function that relates average outcome trends for untreated units to their covariates. The most common way to proceed, especially in cases where $X_i$ contains many variables or continuous ones, is to specify a working model, ${\mu}_{\omega,\Delta,D=0}(X_i)$, with parameters that are simple to estimate. A natural and empirically friendly choice is a linear model, ${\mu}_{\omega,\Delta,D=0}(X_i) = X_i'\beta_{D=0}$, whose parameters come from a (weighted) regression of $\Delta Y_{i,t=2}$ on $X_i$ in the sample of untreated units. The results of this regression describe \textit{untreated} outcome trends as a function of $X_i$. The fitted model then generates predicted values, $\widehat{\mu}_{\omega,\Delta,D=0}(X_i) =  X_i'\widehat{\beta}_{D=0}$, for all units in the sample, including treated units. With these fitted values we can estimate $\E_\omega\Big[ \E_\omega[\Delta Y_{i,t=2} | X_i, D_i=0] \Big| D_i=1\Big]$ using the plug-in principle; that is, by replacing $\E_\omega[\Delta Y_{i,t=2} | X_i, D_i=0]$ with its fitted value $\widehat{\mu}_{\omega,\Delta,D=0}(X_i)$, and replacing population (weighted) expectations with their sample analogs: $\frac{\sum_{i=1}^n D_i ~\omega_i~  \widehat{\mu}_{\omega,\Delta,D=0}(X_i)}{\sum_{i=1}^n D_i~ \omega_i}$.

Putting the pieces together gives the following estimator for $ATT(2)$:
\begin{align}
\widehat{ATT}_{ra}(2) &= \frac{\sum_{i=1}^n D_i ~\omega_i~  \big(\Delta Y_{i,t=2} - \widehat{\mu}_{\omega,\Delta,D=0}(X_i)\big)}{\sum_{i=1}^n D_i~ \omega_i}.\label{eqn:att_OR_estimator}
\end{align}
This strategy is often referred to as the regression-adjustment (RA) or outcome regression approach to DiD; see, for example, \citet{Heckman1997}.

We apply this strategy in our application using only the baseline covariates from Table \ref{tab:cov_balance}, and report the parameters $\widehat{\beta}_{D=0}$ of our working model in columns (1) and (3) of Table \ref{tab:reg_pscore_cs}. In practice, mortality changes in non-expansion states are only weakly related to our baseline covariates. Multiplying the weighted coefficients in Table \ref{tab:reg_pscore_cs} times the weighted treatment group 2013 means in Table \ref{tab:cov_balance} gives a predicted change in untreated mortality rates for the average treated county of 7.2 deaths per 100,000, the second term in equation \eqref{eqn:att_OR_estimator}, compared to the observed (weighted) change among untreated counties of 6.3 deaths. The observed weighted trend in mortality for expansion counties from Table \ref{tab:two_by_two_ex} is 3.7 deaths. Together, these results imply that this approach, based only on assumptions \ref{ass:cond-parallel-trends}, \ref{ass:overlap}, \ref{ass:NA}, the choice of baseline covariates, and a linear model for $\E_\omega[\Delta Y_{i,t=2} | X_i, D_i=0]$, yields an estimated $ATT(2014)$ of -3.5. This matches the formal RA DiD estimate we report in column 4 of Table \ref{tab:2x2_csdid} (labeled as ``Regression''). Column 1, however, gives an unweighted estimate from the same procedure of -1.62.

\input{Tables/reg_pscore_cs}

To compute standard errors, we need to account for the fact that \eqref{eqn:att_OR_estimator} is a two-step estimation procedure and take into account the uncertainty associated with estimating the working model $\mu_{\omega,\Delta,D=0}(X_i)$. This is standard, though, and most statistical software automates this process; see, for example, \citet{SantAnna2020} and \citet{Callaway2021}.

\input{Tables/2x2_csdid}

A linear working model for $\E_\omega[\Delta Y_{i,t=2} | X_i, D_i=0]$ is a familiar choice, but not the only one. More flexible, or even fully nonparametric, working models are possible, and the procedure is the same: estimate the model on untreated units, get its fitted values for the covariate values of the treated units, and then estimate the $ATT(2)$ using \eqref{eqn:att_OR_estimator}. An important consideration when choosing the working models is sample size. Large samples permit more flexible estimators that do not sacrifice too much precision. In smaller samples, a parametric linear model may be more appealing. Ultimately, the reliability of DiD RA estimators for $ATT(2)$ depends on how well $\widehat{\mu}_{\omega,\Delta,D=0}(X_i)$ approximates $\E_\omega[\Delta Y_{i,t=2} | X_i, D_i=0]$. If the working model is misspecified---for example by omitting relevant nonlinear terms---the resulting DiD RA estimator will be biased.

Table \ref{tab:reg_pscore_cs} showed that our covariates did not strongly predict untreated mortality trends and thus do little to change our potential biased unconditional DiD estimates. However, $ATT(2)$ can be estimated conditional on covariates in a different way without needing to specify which variables determine outcome trends. Instead, one can improve the comparability of the comparison group directly by selecting a model for the conditional probability of being treated and applying an inverse probability weighted (IPW) DiD procedure \citep{Abadie2005}. The logic of IPW builds on the balance checks we conducted in Table \ref{tab:cov_balance}: if imbalance in covariates is the source of parallel trends violations, then adjusting the comparison group to be balanced on covariates can address that bias. The adjustment takes the form of re-weighting the observed changes in adult mortality rates for non-expansion counties to ensure that the expansion and non-expansion counties are similar on covariates, thus addressing the compositional source of bias.

To implement the IPW DiD procedure and construct these ``balancing weights,'' we need to model $p_\omega(X_i) = P_\omega(D_i=1|X_i)$, the (weighted) conditional probability of belonging to the treatment group, known as the ``propensity score.'' IPW weights are a function of $p_\omega(X_i)$, and for estimating $ATT(2)$, they take a form that forces the underlying weighted distribution of covariates for comparison units to match the distribution for treatment units \citep{Rosenbaum1983}. 
Intuitively, these weights are formed so that if we find some units that were likely to be observed in the treatment groups (based on their covariate values) but ended up in the comparison group, we give these untreated observations ``extra'' weight. 

Formally, we can show that, under Assumptions \ref{ass:cond-parallel-trends} and \ref{ass:overlap}, when panel data are available, 
\begin{align}
ATT(2) &= 
\E\left[ \Big(w_{\omega,D=1}(D_i) - w_{\omega,D=0}(D_i,X_i)\Big) \Delta Y_{i, t=2} \right],\label{eqn:ATT_IPW}
\end{align}
where 
\begin{align}
w_{\omega, D=1}(D) = D \omega \Big/{\E[ \omega D]}, ~ \text{and } ~ w_{\omega, D=0}(D,X) = \dfrac{\omega (1 - D) p_\omega(X)}{1-p_\omega(X)}\Bigg/\E\left[\dfrac{\omega (1 - D) p_\omega(X)}{1-p_\omega(X)}\right]; \label{eqn:weights_ATT}
\end{align}
see, for example, \citet{Abadie2005} and \citet{SantAnna2020}. The structure of the weights in equation \eqref{eqn:weights_ATT} and the way they enter equation \eqref{eqn:ATT_IPW} highlight several intuitive features of how the IPW estimator works. First, $w_{\omega, D=1}(D)$ is only non-zero for treated units, and $w_{\omega, D=0}(D,X)$ is only non-zero for untreated units, which means the estimator subtracts a particular mean of outcome trends for untreated units from a particular mean of outcome trends for treated units. Second, the $w_{\omega, D=1}(D)$ weights do not involve $X_i$ and simply lead simply to a ($\omega$-weighted) mean for the treatment group. Third, the $w_{\omega, D=0}(D,X)$ weights are functions of $X_i$ that, as \eqref{eqn:weights_ATT} shows, give increasingly more weight to untreated units with high propensity scores. This specific IPW weighting function builds a comparison group whose covariate distribution matches the treatment group. Fourth, equation \eqref{eqn:weights_ATT} clarifies that two types of weights both enter an IPW analysis. We already discussed how the $\omega$ weights shape the parameter of interest, parallel trends assumptions, and estimation. These simply multiply the IPW weights which act to address imbalance (and their product is rescaled to integrate to one within group). Finally, the appeal of \eqref{eqn:ATT_IPW} is that if we have a better sense of how units sort into treatment than of the factors that shape outcome trends, we may be more comfortable modeling $p_\omega(X_i)$ than $\E_\omega[\Delta Y_{i,t=2} | X_i, D_i=0]$. 

To leverage the characterization of the $ATT(2)$ in \eqref{eqn:ATT_IPW} for estimation and inference purposes, we need a working model for the true propensity score $p_\omega(X_i)$. When $X$'s are all discrete and low dimensional, this is simple and does not involve functional form restrictions: create covariate-specific strata and then, within each strata, compute the proportion of treated units, and call these estimates $\widehat{\pi}(x_k)$, $x_k$ being a strata-indicator. When $X$'s have continuous components, or when there are too many strata relative to the available sample size, one can adopt a flexible working model, $\pi_\omega(X_i)$, for the propensity score. A common choice for $\pi_\omega(X_i)$ is a (weighted) logistic model whose parameters can be estimated using maximum likelihood---or an alternative estimation procedure such as inverse probability tilting \citep{Graham2012_IPT}. We follow this strategy in our Medicaid application and report in columns (2) and (4) of Table \ref{tab:reg_pscore_cs} the unweighted and weighted maximum likelihood logit coefficients from our propensity score model. Our covariates appear to explain expansion decisions better than untreated outcome trends, suggesting that an estimation approach based on propensity scores may change our $ATT(2014)$ estimate more than the RA approach did. We then use these logit coefficients to get the fitted values for all observations, $\widehat{\pi}_\omega(X_i)$, that will serve as our estimates of $p_\omega(X_i)$.

From these fitted values, we can estimate $ATT(2)$ by
\begin{align}
\widehat{ATT}_{ipw}(2) &= \dfrac{1}{n}\sum_{i=1}^n\Big(\widehat{w}_{\omega,D=1}(D_i) - \widehat{w}_{\omega,D=0}(D_i,X_i)\Big) \Delta Y_{i, t=2},\label{eqn:ATT_IPW_estimator}
\end{align}
where 
\begin{small}
\begin{align}
	\widehat{w}_{\omega,D=1}(D) &= D \omega \Big/\frac{1}{n}\sum_{i=1}^n \omega_i D_i,~~~~ \widehat{w}_{\omega, D=0}(D,X) = \dfrac{\omega (1 -D) \widehat{\pi}_\omega(X)}{1-\widehat{\pi}_\omega(X)}\Bigg/\frac{1}{n}\sum_{i=1}^n \dfrac{\omega_i (1 - D_i)\widehat{\pi}_\omega(X_i)}{1-\widehat{\pi}_\omega(X_i)}. \label{eqn:weights_ATT_estimated}
\end{align}
\end{small}

We report the $ATT(2014)$ estimates and their standard errors using this IPW DiD procedure in Table \ref{tab:2x2_csdid} (labeled as ``IPW'') using both unweighted and population-weighted procedures, where we use a logistic regression that is linear in covariates as our propensity score working model. We use the delta method to account for the estimation uncertainty inherited in this two-step estimation procedure when computing standard errors. The IPW DiD estimates are generally similar to the RA DiD estimates in that we obtain negative estimates that are larger in magnitude when using population weights. However, despite neither being statistically significant, the unweighted IPW estimate is less than half the size of the RA estimate. This is consistent with the broad conclusion from Table \ref{tab:reg_pscore_cs} that our covariates explain Medicaid expansion better than mortality trends.

\begin{figure}[!ht]
\includegraphics[width=7in]{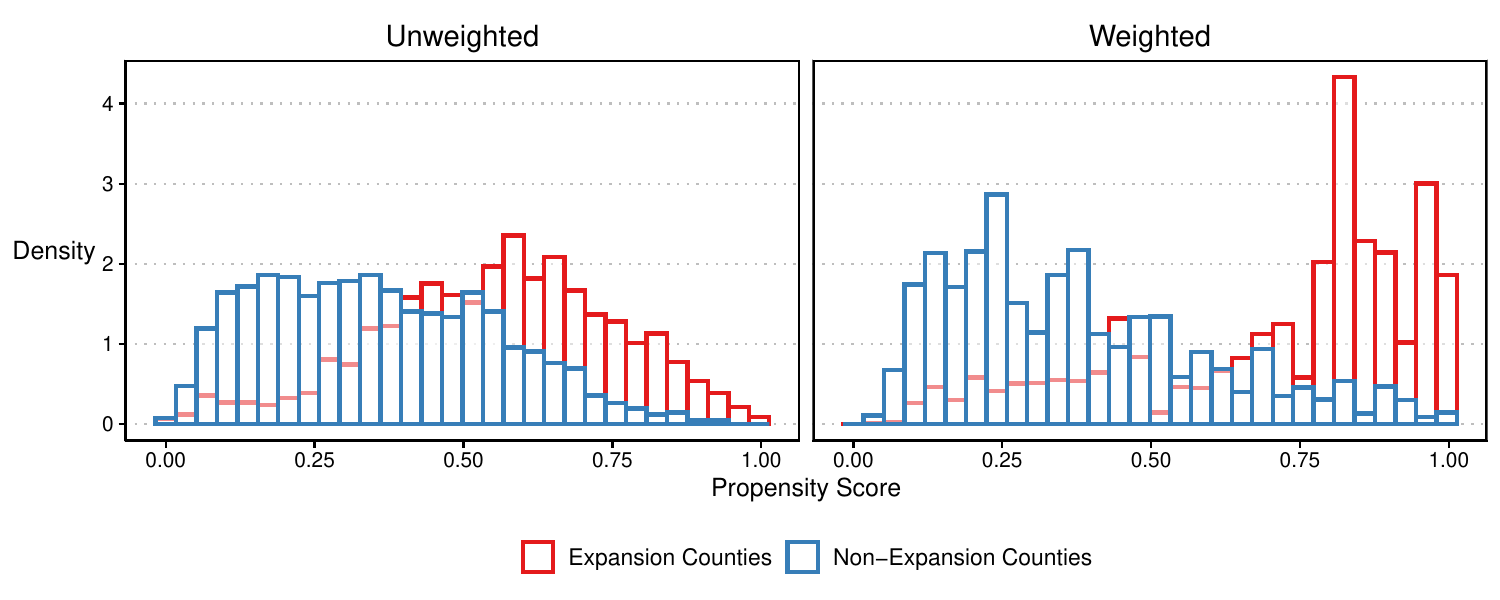}
\caption{\textbf{Distribution of Propensity Scores}}\label{fig:propensity_distribution}
\justifying
\vspace{-.1cm} \noindent\scriptsize{Notes: This figure shows the distribution of propensity scores using both the weighted and unweighted logit propensity score estimates in our $2 \times 2$ Medicaid example.}
\end{figure}

IPW estimators for $ATT(2)$ tend to be noisy when fitted propensity score estimates are too close to 1 among untreated units, a condition related to the Assumption \ref{ass:overlap}. The reason for this is simple: when $\widehat{\pi}_\omega(X_i)$ is close to one among units with $D_i=0$, the (estimated) inverse probability weights $\widehat{w}_{\omega, D=0}(D,X)$ become more volatile, as one is essentially dividing by zero. A good practice when using IPW estimators is to check the plausibility of strong overlap using estimated propensity scores. Figure \ref{fig:propensity_distribution} plots the propensity scores that come from the logit models in Table \ref{tab:reg_pscore_cs} for the two groups of counties. Few untreated units have very high estimated propensity scores, so extreme weighting is not a significant concern.\footnote{By default, the \texttt{did} and \texttt{DRDID R} packages from \citet{Callaway2021} and \citet{Santanna2018} trim untreated units with propensity scores in excess of 0.995.} In addition, propensity scores of non-expansion counties seem to lie within the support of the expansion counties' propensity scores, supporting strong overlap. Trimming high- or low-propensity score observations from the sample may be warranted when overlap is weak; for a discussion, see, for example, \citet{Crump_et_al_2009_overlap}, \citet{Sasaki2022}, and \citet{Ma2023}.

The reliability of these two approaches to estimating conditional DiD designs centers on selecting good models for two different functions: the conditional expectation of untreated outcome changes and the true, unknown propensity score ${p}_\omega(X_i)$. So which approach should one pick? In practice, there are major advantages to combining both in a way that leads to estimators for the $ATT(2)$ that are more robust against model misspecification \citep{SantAnna2020}. This is the so-called {doubly robust (DR)} approach, sometimes referred to as the augmented inverse probability weighting approach; see, for instance, \citet{SantAnna2020} and \citet{Chang2020}.  

The key idea of the DR DiD approach is to express the $ATT(2)$ in terms of both $p(X_i)$ and $\E_\omega[\Delta Y_{i,t=2} | X_i, D_i=0]$ in a way that it gives provides some ``protection'' in case the working models for these functions, $\widehat{\pi}(X_i)$ and $\widehat{\mu}_{\Delta,G=0}(X_i)$, are wrong. The resulting DR estimator for the $ATT(2)$ is consistent when \emph{either} of these nuisance working models is correctly specified. If one is exactly right, it does not matter if the other is wrong. Furthermore, if both working models are only slightly wrong, their errors will multiply, and DR will perform (asymptotically) better than either one alone.\footnote{We usually arrive at such estimands by deriving the efficient influence function; see, for example, \citet{SantAnna2020} for a discussion in DiD setups. See also \citet{Seaman2018} for an overview. When we adopt nonparametric or machine-learning-based estimators for the nuisance functions, we get a different type of double robustness, ``rate double robustness,'' where we can trade-off precision between the different working models. For more details, see, \citet{Kennedy2017} and \citet{Smucler2019}.} Following the steps in \citet{SantAnna2020}, we can express the $ATT(2)$ as
\begin{align}
ATT(2) &= \E\left[ \Big(w_{\omega,D=1}(D_i) - w_{\omega,D=0}(D_i,X_i)\Big) \Big(\Delta Y_{i, t=2} - \E_\omega[\Delta Y_{i,t=2} | X_i, D_i=0]\Big)\right],\label{eqn:ATT_DR}
\end{align}
with the weights as defined in \eqref{eqn:weights_ATT}.  Notice that when we omit the $\E_\omega[\Delta Y_{i,t=2} | X_i, D_i=0]$ term from \eqref{eqn:ATT_DR}, we are back to the IPW estimand \eqref{eqn:ATT_IPW}. When we omit the $w_{\omega,D=0}(D_i,X_i)$ term from \eqref{eqn:ATT_DR}, we are back to the regression adjustment estimand \eqref{eqn:ATT_OR}. 

Constructing a DR DiD estimator for the $ATT(2)$ based on \eqref{eqn:ATT_IPW} is straightforward. Choose a flexible working model for the propensity score and compute its fitted values, $\widehat{\pi}_\omega(X_i)$. Then, a flexible working model for the outcome evolution of untreated units and compute its fitted values for all treated and untreated units, $\widehat{\mu}_{\omega, \Delta, D=0}(X_i)$, and then use the plug-in principle to estimate the $ATT(2)$ by
\begin{align}
\widehat{ATT}_{dr}(2) &= \frac{1}{n}\sum_{i=1}^n  \Big(\widehat{w}_{\omega,D=1}(D_i) - \widehat{w}_{\omega,D=0}(D_i,X_i)\Big) \Big(\Delta Y_{i, t=2} - \widehat{\mu}_{\omega, \Delta, D=0}(X_i)\Big)\label{eqn:ATT_DR_estimator},
\end{align}
with the estimated weights as defined in \eqref{eqn:weights_ATT_estimated}. 

We report the $ATT(2014)$ estimates and standard errors using the DR DiD procedure with a linear in covariates working model for ${\mu}_{\omega, \Delta, D=0}(X_i)$ and (weighted) logistic regression working model that is also linear in covariates for ${p}_\omega(X_i)$ in Table \ref{tab:2x2_csdid} (labeled as ``Doubly Robust''). Population-weighted results are fairly similar across our methods, but the unweighted DR DiD estimates are about halfway between the RA and IPW results. However, an advantage of the DR DiD procedures is that model misspecifications are arguably less of a concern.

Overall, in this section, we discussed different ``forward-engineered'' estimators for the $ATT(2)$ that respect conditional parallel trends, correctly incorporate covariates, and do not restrict treatment effect heterogeneity. Between RA, IPW, and DR DiD estimators, we recommend practitioners favor the \emph{doubly robust} one compared with the RA and IPW appraoches on their own, as this procedure adds additional ``protection.'' When strong overlap fails, though, RA DiD estimators can still work because they extrapolate the outcome model to obtain predicted counterfactual outcome changes even for treated units with covariate values not observed in the comparison group. The credibility of this extrapolation, however, rests on the accuracy of the working model outside the support of $X_i$ in the untreated group. If this extrapolation is not reliable, one can make the DR approach more robust against weak overlap by trimming ``extreme'' propensity scores and performing a bias correction to ensure that the target parameter (the $ATT(2$) in our case) remains the same. See \citet{Ma2023} for details.

\subsection{Heterogeneity analysis}\label{sec:het_analysis}
A different motivation for using covariates is to estimate heterogeneous treatment effects by the values of $X_{it}$. The identification result in equation \eqref{eqn:ATT_OR} can be altered slightly to show that $ATT(2)$ is simply an aggregation of covariate-specific $2\times2$ DiD estimands:
\begin{align*}
ATT(2) = \E_\omega\Big[ \E_\omega[\Delta Y_{i,t=2} | X_i, D_i=1]-\E_\omega[\Delta Y_{i,t=2} | X_i, D_i=0] \Big| D_i=1\Big] .
\end{align*}
Thus, Assumptions \ref{ass:cond-parallel-trends} and \ref{ass:overlap} also imply that we can identify conditional ATT parameters:
\begin{align*}
ATT_{X_i}(2) &\equiv \E_\omega[Y_{i,t=2}(1) - Y_{i,t=2}(0)|D_i=1, X_i] \nonumber \\
&= \E_\omega[\Delta Y_{i,t=2} | X_i, D_i=1]-\E_\omega[\Delta Y_{i,t=2} | X_i, D_i=0].
\end{align*}
This not only demonstrates the building block structure of conditional designs, but also connects the strategies we discussed for estimating $ATT(2)$ itself to the underlying treatment effect heterogeneity by covariate values. When this heterogeneity is of interest in its own right it can also be targeted, identified, and estimated (at least under some additional conditions). 

When all covariates are discrete, estimating $ATT_{X_i}(2)$ parameters is fairly straightforward. One can form saturated partitions $x_k,~k=1,\dots,K$, and then subset the data to contain only information from the units from the specified partition $k$. At this stage, the analysis is analogous to the unconditional DiD setup (except that you need to repeat this step from all partitions of interest), as each $ATT_{x_k}(2)$ is identified by a (conditional on a discrete-variable) comparison of means' that is,
\begin{align*}
ATT_{x_k}(2) = \E_\omega\big[\Delta Y_{i,t=2} | D_i=1, X_i = x_k\big] - \E_\omega\big[\Delta Y_{i,t=2} | D_i=0, X_i = x_k\big].
\end{align*}
One can estimate these $ ATT_{x_k}(2)$'s using their sample analogs or using two-way fixed effects regression specifications analogous to \eqref{eqn:twfe_2_by_2}. Inference is also standard, provided that each partition is sufficiently large. This type of exercise is commonly used to conduct heterogeneity analysis. For example, suppose one wants to see if the effect of Medicaid expansion on adult mortality rate varies across US census regions. Then, one would partition the data into US regions and run one (unconditional) DiD analysis for each region, provided we have treated and untreated units in each region.

When some covariates are continuous or there are so many partitions that each one contains few observations, one can still \textit{identify} the $ATT_{X_i}(2)$'s, though they are hard to estimate unless auxiliary assumptions hold. In such cases, it is customary to identify and estimate a more aggregated conditional ATT parameter than $ATT_{X_i}(2)$. For instance, one may want to assess if the effect of Medicaid expansion on adult mortality rates is higher (or lower) in counties with an unemployment rate above the median than it is in those below. Similar partitions could be made for any other covariates. To formalize this notion of ``partition specific'' ATT, let $\text{PART}(X_i)$ be some user-specified partition of the covariate space such that $\text{PART}(X_i) \in \{1,2,\dots,K\}$, and define 
\begin{align*}
ATT_{k}(2) &= \E_\omega\big[Y_{i,t=2}(1) - Y_{i,t=2}(0)|D_i=1, PART(X_i) = k].
\end{align*}
Under Assumptions \ref{ass:cond-parallel-trends} and \ref{ass:overlap}, it follows that
\begin{align*}
ATT_{k}(2) &= \E_\omega\big[\Delta Y_{i,t=2} | D_i=1, PART(X_i) = k\big] \\
&~~~~~~~~~~~~~~~~- \E_\omega\Big[\E_\omega\big[\Delta Y_{i,t=2} \big| D_i=0, X_i, PART(X_i) = k\big] \Big| D_i=1, PART(X_i) = k\Big], 
\end{align*}
which implies that 
\begin{align*}
ATT(2) &= \sum_{k=1}^K P_\omega(PART(X_i) = k|D_i=1) ATT_{k}(2).
\end{align*}

Thus, for heterogeneity analysis with covariates, one can partition the data with a user-specified partition map $\text{PART}(X_i)$  and then, within each of these partitions, use arguments like the ones we used to establish \eqref{eqn:ATT_OR} to guarantee that each partition-specific ATT is identified (which is precisely what we did to get $ ATT_{k}(2)$ above). Regarding estimation and inference, one can use regression adjustment, inverse probability weighting, or doubly robust estimators as in Section \ref{sec:did_covariates}; the difference is that these are implemented ``locally'' on each partition. 

Obtaining the overall $ATT(2)$ is just a matter of aggregating these partition-specific ATTs using weights equal to the relative partition size among treated units. All these parameters have clear causal interpretations, can be used to answer different policy questions, and are identified under the same identification assumptions already discussed. Other types of heterogeneity analysis are also possible and even attractive. For instance, \citet{Abadie2005} discusses how one can highlight how the ATT varies across a subset of the covariates required for conditional parallel trends to hold. For example, this would entail checking if the average effect of Medicaid expansion among expansion counties varies with the 2013 poverty rate and/or median income. One can also get the best linear approximation of these conditional ATT curves, which involves estimating fewer parameters. These heterogeneity analyses are generally more granular than the partition-based ones discussed above and do not require discretizing the data. They complement each other well.

We close this section with a remark on the choice of partition and the type of heterogeneity analysis to conduct. Heterogeneity analysis has the potential to offer policymakers and researchers novel insights about the treatment of interest and its mechanisms, opening the door for more informed policy recommendations and targeted expansions. But how should one define the subgroups in which to estimate heterogeneous effects? If researchers knew the relevant partition that policymakers care about, they could aim to estimate these particular partition-specific ATTs. But this is rarely the case. Taking no stand about heterogeneity implies reporting unit-level effects, which requires incredibly strong assumptions and could also lead to noisy estimates. On the other hand, taking a too-coarse partition may mask important types of treatment effect heterogeneity. So, it seems that a balance between these extremes is important for a policy-relevant heterogeneity analysis. Estimating conditional ATTs across one or two covariate dimensions is useful, but it may mix some other interactive effects. Another potential avenue is to go beyond averages and adapt the sorted-effects procedure proposed by \citet{Chernozhukov_FernandezVal_Luo_2018_Sorted} to DiD designs. It would also be interesting and practically relevant to extend the heterogeneity tools for experimental data discussed in \citet{Chernozhukov2023_Fisher-Schultz_Lecture} to DiD setups. In our view, this is an area in which applied econometrics practice would benefit from more thorough methodological guidance. 

\section{DiD designs with multiple time periods}\label{sec:binary-multiple-t}
The previous sections focused on fundamental identification issues and estimation approaches for $2\times2$ building blocks, but generally did not build them into anything because they targeted the only feasible $ATT(t)$, $ATT(2)$. DiD designs with multiple periods are notably different. They can target $ATT$s in each period after treatment to trace out dynamics, and they can produce cross-group outcome trends from \textit{before} treatment starts to evaluate the plausibility of parallel trends. Multiple periods also admit the possibility of more complex treatment variation. We will focus on staggered-timing designs in which treatment ``turns on'' at different times for different units and stays on, as the Medicaid expansion did. It is possible, however, to use DiD methods to study treatments that ``turn off'' or occur more than once, though some modifications are warranted; see, for example, \citet{deChaisemartin2020, deChaisemartin2023b-intertemporal-treatments}.

We need to expand the notation to define the relevant concepts with multiple periods. When treatment can happen only at one point in time (so we still have only two treatment groups with $D_i$ equal to one or zero), the only changes we need to make are to acknowledge that time runs from $t=1,2,\dots,T$, and that the map between potential outcomes and observed outcomes is
$$Y_{i,t} = D_i Y_{i,t}(1) + (1 - D_i)Y_{i,t}(0) .$$ 
By Assumption \ref{ass:NA}, we have that $Y_{i,t} = Y_{i,t}(0)$ for all pre-treatment periods, and that, in post-treatment periods, we observe $Y_{i,t}(0)$ if the group remains untreated by $t=T$, and $Y_{i,t}(1)$ for groups treated at the unique treatment date, $t=g$. 

\subsection{Simple event studies \texorpdfstring{$\mathbf{(2\times T)}$}{2xT}} \label{sec:2_by_T}
\begin{figure}[!ht]
\includegraphics[width=5in]{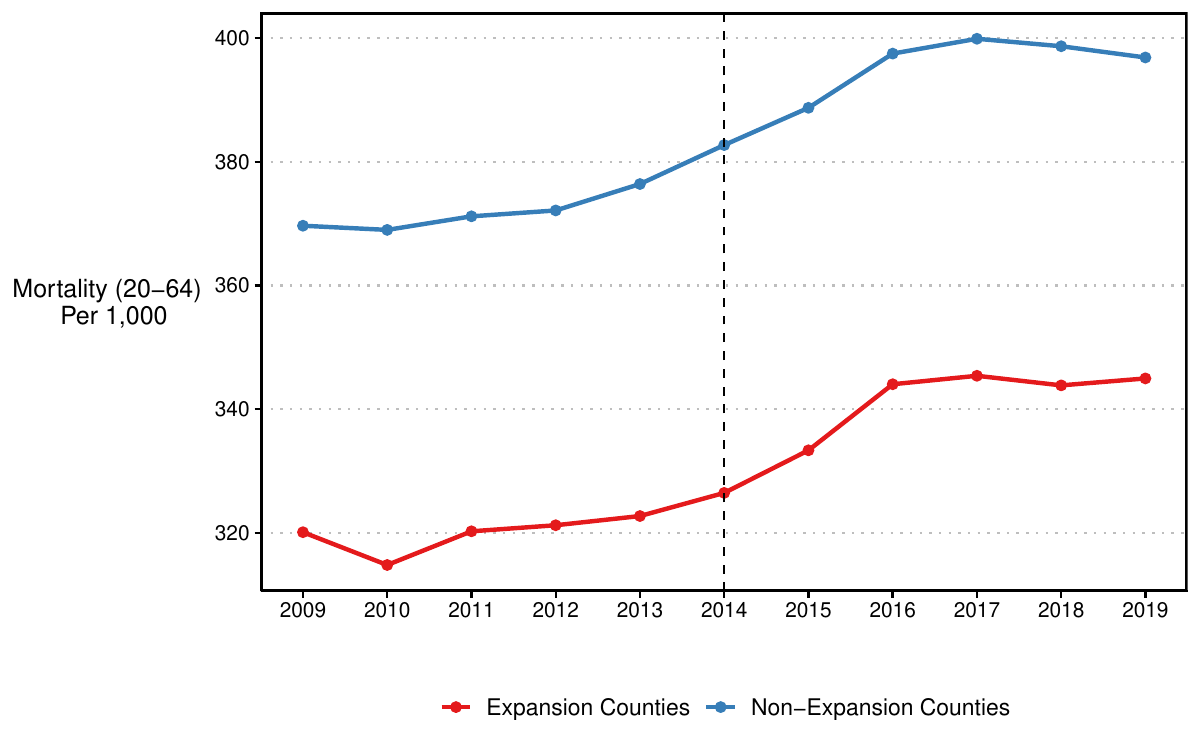}
\caption{\textbf{County Mortality Trends by Expansion Decision}  }\label{fig:trends}
\justifying
\vspace{-.1cm}\noindent\scriptsize{Notes: This figure shows county population-weighted average mortality rates for adults ages 20-64 for 978 counties that expanded Medicaid in 2014 and 1,222 counties that did not expand Medicaid by 2019 from 2009 to 2019.}
\end{figure}

The term ``event study'' refers to estimating and reporting effects across a range of time periods before and after treatment. A design with one treatment timing group and multiple time periods ($2\times T$) is the simplest case in which to discuss event studies. We thus expand our analysis of the 2014 Medicaid expansion group to include data from 2009 to 2019. We report population-weighted results for brevity. Figure \ref{fig:trends} plots the time series of the weighted mortality rates for the 2014 and post-2019 expansion counties. It is the analog of Table \ref{tab:two_by_two_ex} in the sense that it presents the raw data elements necessary to construct event study estimates. The treatment year 2014 naturally divides the x-axis into two windows: post-treatment (2014-2019) and pre-treatment (2009-2013). An event study constructs DiD-type estimates in both windows, but they have different interpretations. 

\subsubsection{Event study estimates in the post-treatment periods}

The target parameters in a simple event study are the $ATT(t)$'s with the same definition as in \eqref{eqn:2x2estimand}; there are just more of them to identify, estimate, and interpret than in a $2\times2$ design. The $ATT(t)$s in the post-treatment period, $t\geq g$, reflect treatment effect dynamics. For example, economic models that view health as a stock imply that $ATT(t)$ may grow if Medicaid stimulates health investments. Furthermore, people and institutions may take time to adjust their behavior after Medicaid expands, suggesting a dynamic treatment effect. Event study parameters answer these kinds of subtle questions. 

Identification of each $ATT(t)$ follows from the same arguments outlined in section \ref{sec:2x2}. Note, however, that an event study analysis requires parallel trends in \textit{every} post-period, as in the following assumption.
\begin{customass}{PT-ES}[$2\times T$ Parallel Trends] \label{ass:parallel-trends-ES}
The average change of $Y_{i,t}(0)$ from $Y_{i,t=g-1}(0)$ is the same between treated and comparison units for all post-treatment periods $t\ge g$; that is, 
\begin{equation}
	\E_\omega[Y_{i,t}(0)-Y_{i,t=g-1}(0)|D_i=1]=\E_\omega[Y_{i,t}(0)-Y_{i,t=g-1}(0)|D_i=0] \; \; \; \forall t \geq g. 
\end{equation}
\end{customass}
Assumption \ref{ass:parallel-trends-ES} suggests that learning about long-run effects requires stronger assumptions than learning about short-run effects. That is, to identify the average effect of Medicaid expansion in 2019 among expansion counties, Assumption \ref{ass:parallel-trends-ES} requires parallel trends to hold in every year from 2014 to 2019. On the other hand, if we are interested in learning only about short-run effects---say effects up until 2015---we would require it to hold from 2014 and 2015 only.

If Assumption \ref{ass:parallel-trends-ES} holds (as well as no-anticipation), then each $ATT(t)$ is identified by a DiD comparison between period $g-1$ and $t$, as in \eqref{eqn:2x2estimand}, and the $ATT(t)$s can be estimated by the familiar comparison of four sample averages:
\begin{align}
\widehat{ATT}(t) = (\overline{Y}_{\omega, D=1, t} - \overline{Y}_{\omega,D=1,t=g-1}) - (\overline{Y}_{\omega,D=0, t} - \overline{Y}_{\omega,D=0, t=g-1}).\label{eqn:2xT_att}
\end{align}

\begin{figure}[!ht]
\includegraphics[width=5in]{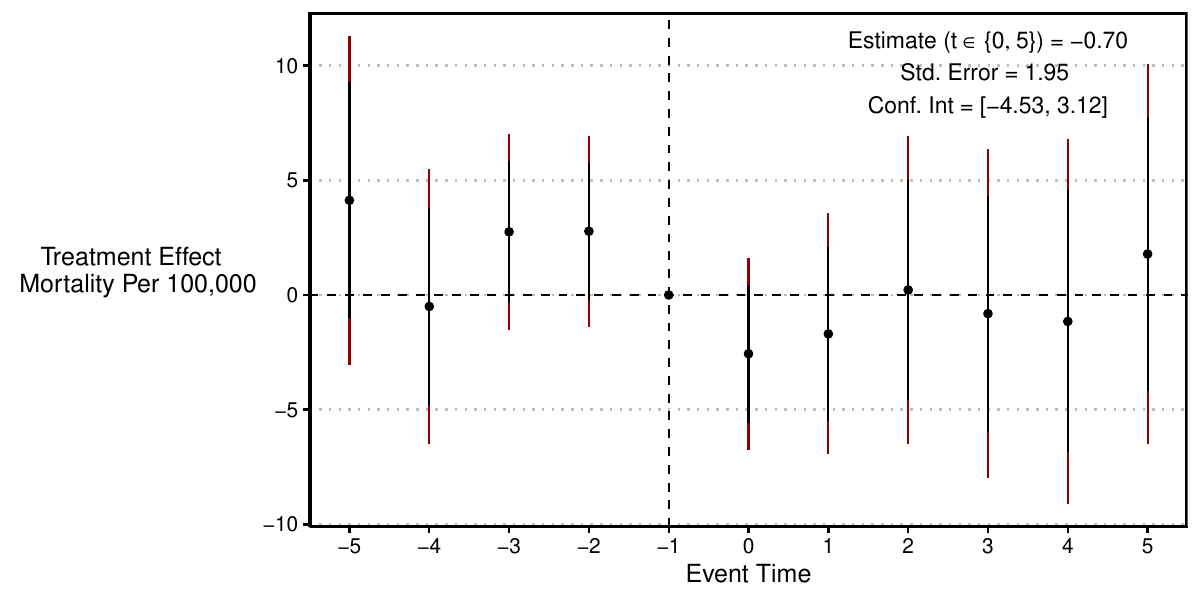}
\caption{\textbf{2 $\times$ T Event Study}}\label{fig:2XT_ES}
\justifying
\vspace{-.1cm}\noindent\scriptsize{Notes: This figure shows the population-weighted event study estimates in the $2 \times T$ case, comparing 978 counties in states that expanded in 2014 to 1,222 counties in states that had not yet expanded by 2019. It uses the unconditional estimator from \citet{Callaway2021}. The outcome variable is the crude mortality rate for adults ages 20-64, and the standard errors are clustered at the county level. The point estimate is reported by the circles, and both point-wise (black) and simultaneous (red) confidence intervals are reported with the vertical lines (see section \ref{sec:estimation_es}).}
\end{figure}

Figure \ref{fig:2XT_ES} plots (weighted) event study estimates for the 2014 Medicaid expansion group. $\widehat{ATT}(t)$'s lie to the right of the vertical dashed line. The estimate for event-time 0 (that is, $t=g=2014$) equals the $2\times2$ results from Table \ref{tab:two_by_two_ex} ($-2.6$). The other estimates have the same DiD form: comparisons of cross-group changes in different post-2014 years ($t$) but always relative to 2013 ($g-1$). The point estimates do not suggest large mortality effects from Medicaid expansion among expansion counties.

\subsubsection{Event study estimates in the pre-periods} \label{sec:pre-trends-ES}
Multiple time periods also allow for falsification/placebo tests based on DiD-type comparisons between \textit{pre}-treatment periods. The no-anticipation assumption implies that all $ATT(t)$'s before time $g$ are equal to zero, which means that a $2\times2$ estimand between periods $t=g-k$ and $t=g-1$, $k>1$, equals a difference in weighted average \textit{untreated} potential outcome trends (as they are all from pre-treatment periods): 
\begin{align}
\tau_{-k} &= \E_\omega[Y_{i,t=g-k}(0)-Y_{i,t=g-1}(0)|D_i=1] - \E_\omega[Y_{i,t=g-k}(0)-Y_{i,t=g-1}(0)|D_i=0] \nonumber \\
&= \E_\omega[Y_{i,t=g-k}-Y_{i,t=g-1}|D_i=1] - \E_\omega[Y_{i,t=g-k}-Y_{i,t=g-1}|D_i=0]. \nonumber
\end{align}
The $\tau_{-k}$ terms are usually called ``differential trends'' or ``pre-trends,'' and they appear to the left of the vertical dashed line in Figure \ref{fig:2XT_ES}. The credibility of event study analyses rests in large part on finding small estimates of $\tau_{-k}$, but Figure \ref{fig:2XT_ES} illustrates how challenging it can be to draw conclusions from informally looking at these pre-trends. No individual $\tau_{-k}$ is statistically significant, so we fail to reject the null hypothesis that the pre-trend estimates equal zero (individually or jointly). This kind of result is often interpreted to mean that parallel trends holds. But the $\tau_{-k}$'s also tend to be positive with a mean of about 2.3, which is larger in magnitude than all post-period point estimates except one. Sometimes, this kind of result is used to argue that parallel trends fails. Do these results support or refute parallel trends? 

Recent methodological work suggests three practical lessons about using pre-trend estimates to assess parallel trends; examples include \citet{Bilinski2018}, \citet{Manski2018}, \citet{KahnLang2019}, \citet{Roth2022}, \citet{Rambachan2023}, \citet{Dette_Schumann_JBES_2024}, and \citet{Freyaldenhoven_etal_2024}. The most fundamental, which will tend to shape language more than practice, is that Assumption \ref{ass:parallel-trends-ES} is not testable, as it only makes restrictions on untreated potential outcomes in post-treatment periods, $t\geq g$. Under no-anticipation, pre-trends do measure differences in untreated outcome trends between treated and untreated units, but they necessarily measure them in the ``wrong'' periods $t<g$. This does not mean parallel pre-trends are not informative; it just means they are not the \textit{same} as the parallel trends assumption in Assumption \ref{ass:parallel-trends-ES}.  

When specifically are observed pre-trends uninformative about the parallel trends assumptions required for identification? One case is when pre-trends are measured too far before treatment starts. The conditions or shocks that jointly shape treatment decisions and untreated outcomes may differ many periods before treatment, but this need not imply that they generate bias in the periods after treatment. A second case, discussed in Section \ref{sec:PT}, is when units select into treatment based on time-varying pre-treatment unobservables. As treatment selection depends on pre-treatment unobservables, non-parallel pre-trends may appear as a consequence of the selection mechanism, though parallel trends may still be plausible. \citet{Ghanem2023} show that in this case, a \emph{necessary} condition for parallel trends is that the untreated potential outcomes satisfy a martingale property. When this martingale property does not hold, parallel trends cannot hold either, \textit{regardless} of the shape of pre-trends. In such situations, it is worthwhile to assess the potential bias components of the DiD estimator using the ``selection-aware''  benchmarking tools presented in \citet{Ghanem2023}, as well as to resort to their theory-based templates for justifying parallel trends based on selection mechanisms. Ultimately, how informative pre-trends are for the plausibility of Assumption \ref{ass:parallel-trends-ES} is case-specific, though we would of course always prefer to have parallel pre-trends.

When Assumption \ref{ass:parallel-trends-ES} holds in all periods (pre- and post-treatment), it is worth noting that one can construct estimators for the $ATT(t)$ that are more precise than those in \eqref{eqn:2xT_att}; see, e.g., \citet{Borusyak2023}, \citet{Gardner2021}, \citet{Harmon2024}, \citet{Marcus_Santanna_2021}, \citet{Wooldridge2021}, and \citet{Chen_SantAnna_Xie_effientDiD_2024}. In such cases, though, one would \emph{require} parallel trends to hold pre-treatment periods, parallel pre-trends becomes directly testable (as the model is overidentified), and $\tau_{-k}$ could be used to assess its plausibility directly. Other types of overidentification tests could also be used in this case to assess the validity of the DiD model directly; see \citet{Marcus_Santanna_2021} and \citet{Chen_SantAnna_Xie_effientDiD_2024} for a discussion. 

The second lesson is that statistical precision shapes the usefulness of pre-trend estimates. The hypothesis tests for parallel pre-trends in Figure \ref{fig:2XT_ES} are low-powered to detect practically important violations (\citealp{Roth2022}, and \citealp{Freyaldenhoven_etal_2024}). The $\tau_{-k}$ estimates fail to rule out flat pre-trends, but as we shall see below, they also fail to rule out large pre-trends that would indicate serious bias in the $\widehat{ATT}(t)$'s. They simply do not say very much.\footnote{Another possibility is that very precisely estimated pre-trends are distinguishable from zero yet also rule out even small violations. The magnitude of the violations also matters: a precisely estimated but small violation of parallel trends in pre-treatment periods is ``better'' than an imprecisely potentially large estimated pre-trend that does not rule out zero.} \citet{Roth2022} discusses how conditioning the analysis on these kinds of low-powered tests can exacerbate biases and should be interpreted with care.

The third lesson is that researchers can make better use of pre-trend estimates by taking a stand on the size of plausible and/or problematic parallel trends violations. For example, \citet{Bilinski2018} propose selecting a value for differential trends that \textit{would} fully explain the estimated treatment effects and then using it instead of zero as the null hypothesis when conducting statistical tests of the estimated pre-trends; see also \citet{Dette_Schumann_JBES_2024}. \citet{Rambachan2023} develop inference methods for an approach that bounds the $ATT(t)$'s under assumptions about the maximum size of parallel trends violations based on pre-treatment periods (see also \citealp{Manski2018}). One can either select a magnitude using contextual knowledge or set it equal to a multiple of the largest period-to-period pre-trend estimate. One then constructs an identified set that contains $ATT(t)$ not under parallel trends but under the weaker assumption that parallel trends is not violated by more than this pre-determined maximum in each post-period.  Importantly, \citet{Rambachan2023} also show how to use the estimated covariance matrix of the event study estimates to construct confidence intervals around the set.\footnote{In cases where a measured variable accounts for the pre-trends, \citet{Freyaldenhoven2019} develop two-stage-least-squares estimators that recover $ATT(t)$ parameters by extrapolating the pre-period relationship between outcomes and the covariate into the post-period.} Given the existence of these methods, which are theoretically grounded in how to consider violations of pre-trends, we caution producers (and consumers) of DiD work against the common practice of using only a simple ``eye-test'' for whether pre-trends differ substantially from zero.

In our application, \citet{Rambachan2023}'s method underscores how little information the pre-trend estimates convey. The largest one-period pre-trend in Figure \ref{fig:2XT_ES} is between event-time -5 and -4, when outcomes fall by roughly four deaths more in the expansion group versus the non-expansion group. If we assume that parallel trend violations are no bigger than this, the identified set for $ATT(2014)$ is $-2.6 \pm 4 = [-6.6, 1.4]$, and given the size of the pre-period standard errors, we obtain a robust confidence interval of $[-11.1, 5.1]$, which spans implausibly large effects in both directions. Assuming smaller parallel trend violations would shrink this interval, but applying the method to subsequent event-times widens it. In general, \citet{Rambachan2023} provide a flexible method to use information about potential parallel trends violations drawn either from external knowledge or the $\tau_{-k}$ estimates, as well as the precision of the pre-trend estimates, to gauge (statistically) the robustness of the $\widehat{ATT}(t)$'s.

When pre-trends suggest that Assumption \ref{ass:parallel-trends-ES} fails, a way forward is to assume that it holds only after conditioning on covariates and proceeding similarly to Section \ref{sec:did_covariates}; we discuss this path in detail in Section \ref{sec:covariates_ES}. Alternatively, one can attempt to parametrically model the violations of parallel trends. Usually this is done by including unit-specific linear trends; see, e.g., \citet{Mora_Reggio_ER_2019}, \citet[Section 7]{Wooldridge2021}, \citet{Lee_Wooldridge_2023}, and \citet{Freyaldenhoven_etal_2024}. We note, however, that the practice of using unit-specific linear trends deviates from the standard DiD procedures: it relies on alternative identification assumptions involving an explicit parametric model for unit-specific trends. We also note that sensitivity analysis procedures that do not rely on such models are available---\citet{Rambachan2023} is one example---and we encourage practitioners to consider them.

\subsubsection{Estimation and aggregating across time in event-studies}\label{sec:estimation_es}
The link between a $2\times T$ event study and a series of $2\times2$ DiD building blocks makes estimation simple. The point estimates in Figure \ref{fig:2XT_ES} are the $ATT(t)$ estimates based on \eqref{eqn:2xT_att}.\footnote{We estimate these with the \texttt{did R} package from \citet{Callaway2021}.} An equivalent way to obtain all the $\widehat{ATT}(t)$'s in one step is to run a TWFE regression with time fixed effects, $\theta_t$, unit fixed effects, $\eta_i$, and a set of interactions between the treatment group dummy and the time dummies. Omitting the treatment interaction for $t=g-1$ avoids multicollinearity and fixes $g-1$ as the baseline period for all $\beta_t$ estimates, which matches Assumption \ref{ass:parallel-trends-ES}. This generalizes the TWFE regression equation for a single $ATT(t)$ in \eqref{eqn:twfe_2_by_2} to
\begin{equation}
Y_{i,t} = \theta_t + \eta_i+ \sum_{k=1}^{g-2}\beta_{k} \left(\1\{G_i=g\}\cdot \1\{t=k\}\right) + \sum_{k=g}^T\beta_{k}\left(\1\{G_i=g\}\cdot \1\{t=k\}\right) +\varepsilon_{i,t}. \label{eqn:2xT_es_twfe}
\end{equation} 

This regression produces identical estimates to those obtained ``by hand'' via \eqref{eqn:2xT_att}: $\widehat{\beta}_t =\widehat{ATT}(t)$. It also generates (point-wise) confidence intervals based on clustered standard errors, as discussed in Section \ref{sec:estimation}. An additional issue in event study inference, however, involves the fact that we are now estimating many treatment effect parameters. Thus, when we compare across event study estimates, we are conducting many hypothesis tests, and the usual normal critical values used to construct confidence intervals do not account for these. Asymptotically correct inferences about the entire event study curve require ``inflating'' critical values to perform a multiple hypothesis test adjustment. In Figure \ref{fig:2XT_ES}, the thick black bars represent the standard pointwise confidence intervals from clustered standard errors at the county level, while the red line shows uniform confidence bands that cover the 95\% confidence interval for the entire treatment path of the event study coefficients after accounting for multiple testing. These are produced by default in the \cite{Callaway2021} statistical packages, using a multiplier bootstrap procedure to compute critical values of the $\sup$-$t$ test statistic. Alternatively, one can construct these using the estimated variance-covariate matrix of all $\widehat{\beta}_t$'s paired with the \citet{Montiel-Olea_Plagborg-Moller-2018-JAE} simulation procedure. Alternative bootstrap procedures, such as the nonparametric bootstrap, the multiplier bootstrap, and the weighted/Bayesian bootstrap, can also be used to compute the $\sup$-$t$ critical values that account for multiple testing.\footnote{The $\sup$-$t$ critical value governs the width of the uniform confidence band that yields simultaneous coverage probabilities for a given confidence level \citep{Montiel-Olea_Plagborg-Moller-2018-JAE}. In our context, its main idea is constructing asymptotically valid critical values for the entire event-study trajectory based on the maximum of all t-statics (one for each event-time considered). This procedure avoids the conservativeness of multiple-testing corrections such as Bonferroni's. See \citet{Montiel-Olea_Plagborg-Moller-2018-JAE} for a general discussion.

}

A final estimation issue arises when targeting aggregations of the $ATT(t)$'s. For example, the average treatment effect in the post-period, $ATT_{\text{avg}} = \frac{1}{T-(g-1)}\sum_{t=g}^T ATT(t)$, is a convenient scalar measure that improves statistical precision, especially when $ATT(t)$ is relatively constant. The easiest way to get an estimate of $\widehat{ATT}_{\text{avg}}$ is just to construct it from the $2\times 2$ $\widehat{ATT}(t)$ building block estimates. Standard post-estimation commands achieve this if the event study estimates come from a regression, and newer DiD packages report this parameter automatically. A common shortcut, however, is to run a second regression that replaces the event study dummies with the treatment status dummy. $D_{i,t}=\1\{D_i=1\}\times \1\{t\geq g\}$:
\begin{equation}
Y_{i,t} = \theta_t + \eta_i + \beta^{OLS}D_{i,t} +\varepsilon_{i,t}.\label{eqn:twfe_2xT_DiD}
\end{equation}
Unfortunately, the (weighted) least squares estimator $\widehat{\beta}^{OLS}$ does not generally equal the $\widehat{ATT}_{\text{avg}}$. The reason is that $\widehat{\beta}^{OLS}$ is equivalent to first collapsing the multiple-periods data to averages in the post- and pre-periods and then estimating a $2\times2$ DiD on the resulting means. This, in turn, is the same as subtracting the average pre-period $\tau_{-k}$ estimate (including the zero at time $g-1$) from the average post-period $\widehat{ATT}(t)$ estimate. This implies that interpreting both $\beta_t$s from \eqref{eqn:2xT_es_twfe} and $\beta^{OLS}$ from \eqref{eqn:twfe_2xT_DiD} requires Assumption \ref{ass:parallel-trends-ES} to hold in every time period, not just in the post-treatment periods.\footnote{\label{fn:gmm}The decision to estimate each $\widehat{ATT}(t)$ relative to period $t=g-1$ comes directly from the choice to define PT that way. If parallel trends holds in every period, one can typically form more efficient estimators than those discussed above. See \citet{Marcus_Santanna_2021} for a discussion} To the extent that the gap in mean outcomes over the whole pre-period differs from the gap in outcomes in period $t=g-1$, the two summary parameters will not be equal. In Figure \ref{fig:2XT_ES}, the two summary parameters are quite different: $\widehat{ATT}_{\text{avg}}$ equals -0.70 while $\widehat{\beta}^{OLS}$ equals -2.53.\footnote{Interestingly, this quantity is almost identical to our estimate of $ATT(2014)$, but rather than representing anything reassuring, it comes from the offsetting effects of positive pre-period estimates and small post-period ones.}

\subsubsection{Covariates in event studies}\label{sec:covariates_ES}
Another advantage of seeing event studies as collections of $2\times2$ building blocks is that all the tools for incorporating covariates from \Cref{sec:covariates} immediately apply to each event study estimate. In fact, the only difference is that instead of using ``short-differences,'' $\Delta Y_{i,t=2}$, one would now use ``long-differences,'' $Y_{i,t} - Y_{i,t=g-1}$. This would imply that using the regression-adjustment procedure would require estimating a working model for $\E_\omega[Y_{i,t} - Y_{i,t=g-1} | D_i=0,X_i]$ for each time $t$. The propensity score working model used to construct the IPW DiD estimate \eqref{eqn:ATT_IPW}, on the other hand, is exactly the same as in a $2\times2$ analysis of the same groups.  Since the DR DiD estimation procedure builds on both RA and IPW procedures, it would involve estimating different outcome-regression working models for each time $t$. We also note that the potential pitfalls of controlling for covariates in a TWFE specification still apply with multiple periods and actually become more complex \citep{Caetano2024}. 

For completeness and ease of access, we list the RA, IPW, and DR estimands for $ATT(t)$:
\begin{align*}
ATT_\text{ra}(t) &= \E_\omega[ Y_{i,t} -  Y_{i,t=g-1} | D_i=1] - \E_\omega\Big[ \E_\omega[ Y_{i,t} -  Y_{i,t=g-1} | X_i, D_i=0] \Big| D_i=1\Big],\\
ATT_\text{ipw}(t) &= \E\left[\Big(w_{\omega,D=1}(D_i) - w_{\omega,D=0}(D_i,X_i)\Big) (Y_{i,t} -  Y_{i,t=g-1}) \right], \\
ATT_\text{dr}(t) &= \E\left[\Big(w_{\omega,D=1}(D_i) - w_{\omega,D=0}(D_i,X_i)\Big) \Big(Y_{i,t} -  Y_{i,t=g-1} - \E_\omega[ Y_{i,t} -  Y_{i,t=g-1} | X_i, D_i=0] \Big) \right],
\end{align*}
where $w_{\omega,D=1}(D_i)$ and $ w_{\omega,D=0}(D_i,X_i)$ are as defined in \eqref{eqn:weights_ATT}.

\begin{figure}[!ht]
\includegraphics[width=7in]{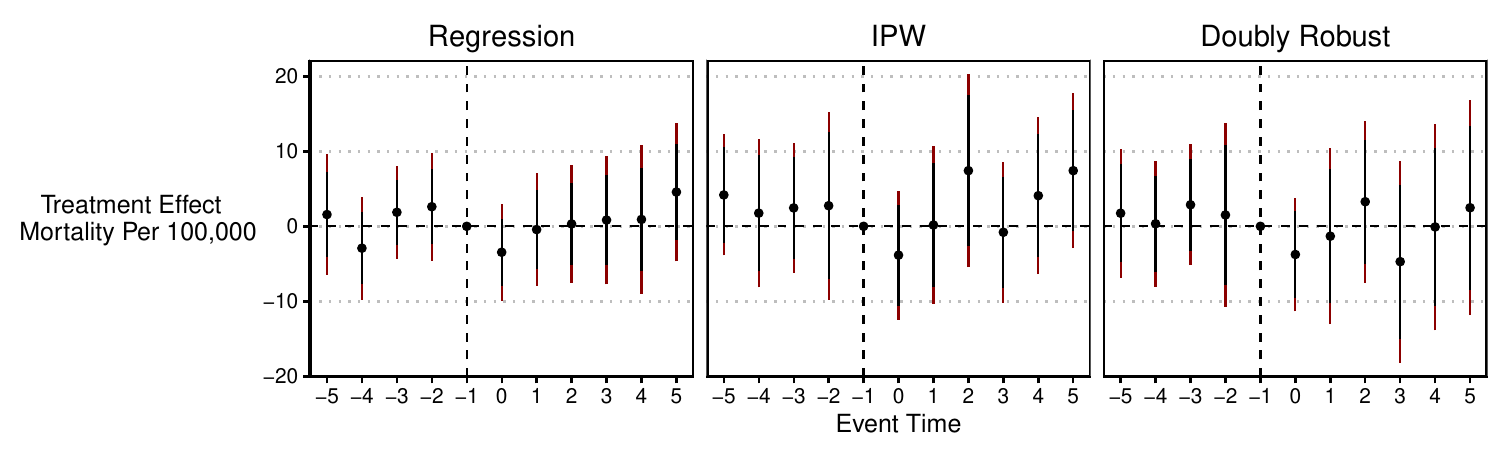}
\caption{\textbf{2 $\times$ T Event Study with Covariates}}\label{fig:2XT_ES_covs}
\justifying
\vspace{-.1cm}\noindent\scriptsize{Notes: This figure shows event study estimates that include covariates. The outcome variable is the crude mortality rate for adults ages 20-64, and the covariates include 2013 values of the percentage of the county population that is female, the percentage of the county population that is white, the percentage of the county population that is Hispanic, the unemployment rate, the poverty rate, and county-level median income. The sample includes 2,200 counties (978 in states that expanded Medicaid by 2014 and 1,222 in states that did not expand Medicaid by 2019). The point estimate is reported by the circles, and both point-wise (black) and simultaneous (red) confidence intervals are reported with the vertical lines. All procedures use population weights.}
\end{figure}

Figure \ref{fig:2XT_ES_covs} shows weighted event study estimates using our three preferred covariate strategies: regression adjustment, inverse propensity weighting, or doubly-robust estimation. In our case, covariates do little to change the unadjusted estimates. Note, however, that \citet{Borgschulte2020} use an IPW estimator with different covariates selected by a lasso procedure and obtain notably stronger evidence of mortality reductions. Evidently, the exact set of covariates one conditions on matters a great deal in this analysis, and, as we have discussed earlier, one should attempt to include in $X_i$ all the determinants of the change in untreated potential outcome or of the treatment assignment. 

\subsection{Staggered treatment adoption (\texorpdfstring{$\mathbf{G_\#\times T}$}{GxT})}\label{sec:staggered}

Viewing $2\times T$ event studies as a collection of $2\times2$ DiD building blocks makes the jump to staggered timing designs straightforward. The key distinction is that with staggered timing, each treatment date defines a distinct treatment group, and each of these has its own set of simple event study parameters. New choices arise about the comparison units used to identify and estimate these group-specific event studies, as well as about how to aggregate the estimates across timing groups. The $2\times2$ structure, as well as all the tools we have developed to evaluate parallel trends (covariate balance and pre-trends) and to estimate (with or without weights and covariates), carry over.

When treatment start dates can vary across units, we need to allow the potential outcomes, and thus the target parameters and identifying assumptions, to reflect this richer notion of treatment. We therefore index potential outcomes by the time treatment begins, $g$: $Y_{i,t}(g)$; and use $Y_{i,t}(\infty)$ to denote never-treated potential outcomes.\footnote{Let $\mathbf{0}_s$ and $\mathbf{1}_s$ be $s$-dimensional vectors of zeros and ones, respectively, and denote the potential outcome for unit $i$ at time $t$ if first exposed to the treatment at time $g$ by $Y_{i,t}(\mathbf{0}_{g-1},\mathbf{1}_{T-g+1})$, and denote by $Y_{i,t}(\mathbf{0}_{T})$ the outcome if untreated by time $t=T$. We discussed the two-period treatment this way in section \ref{sec:target_param} when we defined potential outcomes as a function of the period one and period two treatment. These treatment paths define the potential outcomes we work with: $Y_{i,t}(g) = Y_{i,t}(\mathbf{0}_{g-1},\mathbf{1}_{T-g+1})$ and $Y_{i,t}(\infty) = Y_{i,t}(\mathbf{0}_{T})$. Writing potential outcomes as functions of treatment paths helps with transparency regarding causal parameters of interest and the DiD design we are in. When treatment can turn on and off, writing potential outcomes in terms of the entire path becomes crucial to avoid ``hidden'' assumptions that rule out treatment effect heterogeneity and dynamics. Because of space constraints, we do not cover these cases in this article.} We use $G_i$ to denote each unit's treatment date, and with some abuse of terminology, we call units not exposed to treatment by period $T$ the ``never-treated'' group.\footnote{In practice, ``never-treated'' really means ``not observed to be treated by $t=T$.'' Given more data, units untreated at $T$ could, in many cases, take up the treatment. In fact, this is the case with the Medicaid expansion. We use data through 2019 but include states that expanded Medicaid in 2020, 2021, and 2023 as ``never treated,'' alongside states that have not expanded as of 2024.} Finally, we use $\mathcal{G}$ to represent the set of all treatment times (rows of Table \ref{tab:adoptions} in our example).
With these modifications, we can map potential outcomes to observed outcomes using a generalization of  \eqref{eqn:switching}: $$Y_{i,t} = \sum_{g\in \mathcal{G}} Y_{i,t}(g)\1\{G_i = g\}.$$  

With multiple treatment groups, we also need to extend our notion of no-anticipation (though its empirical content is exactly the same).
\begin{customass}{NA-S}[No-Anticipation with staggered treatment timing]\label{ass:NA_staggered}
For all units $i$ that are eventually treated and all pre-treatment periods $t$, $Y_{i,t}(g) = Y_{i,t}(\infty)$. 
\end{customass}
Like Assumption \ref{ass:NA}, Assumption \ref{ass:NA_staggered} imposes that treatment effects are zero in all pre-treatment periods as a consequence of units not acting on the potential knowledge of future treatment dates before they are actually exposed to treatment. We maintain this assumption throughout this section. 

Finally, we assume that a ``never-treated'' group always exists in our staggered DiD setup. If all units are eventually treated, we drop all the data from when the last cohort is treated, so the last-treated cohort becomes the ``never-treated'' cohort, and $T$ here denotes the number of available periods in the subset of the data that we will use in our analysis. This is essentially without loss of generality, because under standard DiD assumptions, we cannot identify any $ATT$ for periods where all units are treated. We also dropped data from units treated in the first available period, $G_i=1$, as such treatment group does not have any pre-treatment data, preventing us from conducting a DiD analysis.

\begin{figure}[!ht]
\includegraphics[width=6in]{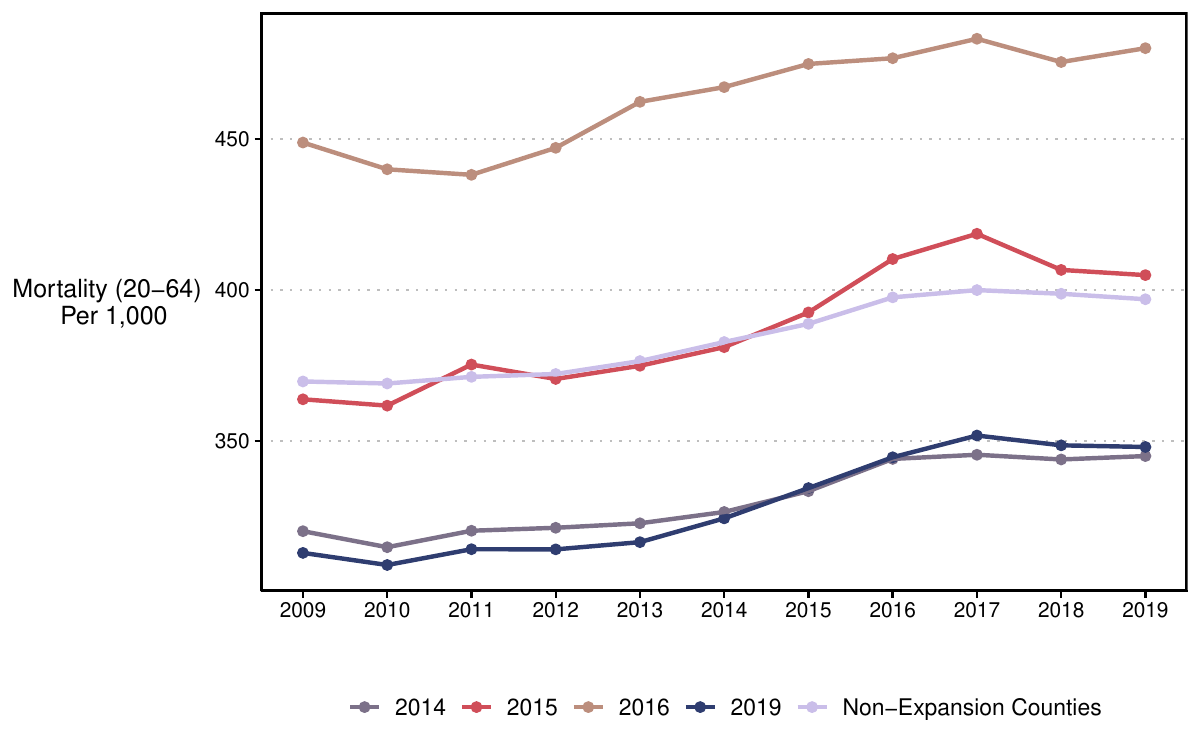}
\caption{\textbf{County Mortality Trends by Expansion Decision with Staggered Timing}}\label{fig:trends_gxt}
\justifying
\vspace{-.1cm}\noindent\scriptsize{Notes: This figure shows county population-weighted average mortality rates for adults ages 20-64 from 2009 to 2019. There are 978 counties in states in the 2014 expansion group, 171 counties in the 2015 expansion group, 93 counties in the 2016 expansion group, 140 counties in the 2019 expansion group, and 1,222 counties that did not expand Medicaid by 2019.}
\end{figure}

Figure \ref{fig:trends_gxt} plots average weighted mortality data by the year of Medicaid expansion ($G_i$) and time. As in Table \ref{tab:two_by_two_ex} and Figure \ref{fig:trends}, these are all the means necessary to calculate a staggered DiD estimate.

\subsubsection{Building block parameters with staggered adoption}\label{sec:building-block}

Staggered treatment timing affects the structure of a DiD analysis because it changes the definition of treatment. Until now, we have used counties in the 2014 expansion states as the only treatment group represented with a single treatment dummy, $D_i$. But Table \ref{tab:adoptions} shows that as of 2019, there were four different groups of expansion states defined by whether they expanded Medicaid in 2014, 2015, 2016, or 2019. Therefore, $D_i$ is not rich enough to capture the relevant definition of treatment groups in staggered setups, because there are many treatment groups, not just two. Fortunately, we can use the treatment timing notation, $G_i$, to define $ATT$ parameters, parallel trends assumptions, and estimators, just as we have done so far. 

A simple $2\times 2$ DiD design had one target parameter ($ATT(2)$), and a $2\times T$ DiD design had $T-1$ of them: $T-(g-1)$ post-treatment parameters, $ATT(t), t\ge g$, and $g-2$ pre-trend parameters. In staggered DiD designs, each treatment group (sometimes referred to as a cohort), defined by its treatment date $g$, has its own set of $T-1$ event study parameters. We call these group-time average treatment effects:
\begin{align*}
ATT(g,t) = \E_\omega[Y_{i,t}(g)-Y_{i,t}(\infty)|G_i=g]. \label{eqn:attgt}
\end{align*}
Each $ATT(g,t)$ is the average treatment effect of starting treatment at period $g$ relative to never-starting it, at time period $t$, among units that actually started treatment in period $g$. It is simply a set of event study parameters for treatment timing group $g$. The unobserved counterfactuals are now untreated potential outcome means for each treatment group in each period, \textcolor{black}{$\E_\omega[Y_{i,t}(\infty)|G_i=g]$}.

\subsubsection{Identification with staggered designs} \label{sec:identification_staggered}
Identifying the $ATT(g,t)$'s works exactly as in the previous sections because they are just group-specific event studies. Under no anticipation, a set of parallel trends assumptions for $t\geq g$ identifies the causal post-treatment parameters. DiD comparisons for $t<g$ represent differential pre-trends in untreated potential outcomes.

The most important way that staggered DiD changes this approach is that having access to multiple treatment groups with different treatment starting dates allows one to use alternative sets of comparison groups. For example, our Medicaid analysis so far has used counties in states that did not expand Medicaid by 2019 as the comparison group. For estimating, say, the $ATT$ for the 2014 expansion group in 2015, counties in states that did not expand until 2016, 2017, or 2018 could also serve as comparison units. Choosing which comparison groups to use to identify an $ATT(g,t)$ is directly tied to which form(s) of parallel trends hold. Relative to simple event-studies and especially $2\times 2$ setups, staggered timing creates many potential parallel trends assumptions. Here, we discuss three types of staggered parallel trends. The first two use either the never-treated units or any not-yet-treated groups as the comparisons for all eventually-treated groups \citep{Callaway2021}, and the third option assumes that parallel trends holds in all periods and in all groups, which is something that several DiD methods require; see \citet{deChaisemartin2020, Sun2021, Wooldridge2021, Borusyak2023, Harmon2024}.\footnote{ While these three types of parallel trends are fairly intuitive choices, others are possible. For instance, \citet{Cengiz2019} use a comparison group of units treated at least $\delta+1$ periods after time $g$. Thus, their parallel trends is tailored to this particular choice. As a result, all of their $ATT(g,t)$ estimates from time $g$ to time $g+\delta$ use the same comparison group. \citet{Marcus_Santanna_2021} also discuss other alternative parallel trends assumptions.}

\begin{customass}{PT-GT-Nev}[Parallel Trends based on never-treated groups] \label{ass:gt-parallel-trends-never}
For every eventually treated group $g$ and post-treatment time period $t\ge g$,
\begin{align*}
	\E_\omega[Y_{i,t}(\infty)-Y_{i,t-1}(\infty)|G_i=g] = \E_\omega[Y_{i,t}(\infty)-Y_{i,t-1}(\infty)|G_i=\infty].
\end{align*}
\end{customass}

\begin{customass}{PT-GT-NYT}[Parallel Trends based on not-yet-treated groups] \label{ass:gt-parallel-trends-nyt}
For every eventually treated group $g$, not-yet-treated group $g'$ and time periods $t$ such that $t\ge g$ and $g' > t$,
\begin{align*}
	\E_\omega[Y_{i,t}(\infty)-Y_{i,t-1}(\infty)|G_i=g] = \E_\omega[Y_{i,t}(\infty)-Y_{i,t-1}(\infty)|G_i=g'].
\end{align*}
\end{customass}

\begin{customass}{PT-GT-all}[Parallel Trends for every period and group] \label{ass:gt-parallel-trends-all}
For every treatment groups $g$ and $g'$ and time periods $t$,
\begin{align*}
	\E_\omega[Y_{i,t}(\infty)-Y_{i,t-1}(\infty)|G_i=g] = \E_\omega[Y_{i,t}(\infty)-Y_{i,t-1}(\infty)|G_i=g'].
\end{align*}
\end{customass}

Assumption \ref{ass:gt-parallel-trends-never} is the analog of the PT assumptions we used in the $2\times T$ design. It uses the never-treated units as the relevant comparison group for all eventually-treated units, and it imposes parallel trends in post-treatment periods only. In our Medicaid application, this would entail using the non-expansion counties as the comparison group for the 2014, 2015, 2016, and 2019 expansion groups. In addition, since Assumption \ref{ass:gt-parallel-trends-never} imposes parallel trends only for the future, the farthest we can go into pre-treatment periods is $t=g-1$, which will serve as the only (justifiable) baseline period. More formally, under Assumption \ref{ass:gt-parallel-trends-never}, it is straightforward to show that for post-treatment periods,
\begin{equation}
ATT(g,t) = \E_\omega[Y_{i,t} - Y_{i,t=g-1} | G_i=g] - \E_\omega[Y_{i,t} - Y_{i,t=g-1} | G_i = \infty], 
\label{eqn:ATTgt-never}
\end{equation}
which shows that $ATT(g,t)$ is identified \citep{Callaway2021, Sun2021}. Note that \eqref{eqn:ATTgt-never} highlights that we are essentially back to a $2\times 2$ design when it comes to learning about $ATT(g,t)$: it leverages data from only two periods, $t$ (post) and $g-1$ (pre), and two treatment groups, $G_i=g$ (treated) and $G_i=\infty$ (comparison). 

Under assumption \ref{ass:gt-parallel-trends-nyt}, one can use not only the never-treated units but any group of units that are not-yet-treated by time $t$. In our Medicaid example, we could now use non-expansion counties and 2016 and 2019 expansion counties as comparison groups when estimating $ATT(2014,2015)$. Using Assumption \ref{ass:gt-parallel-trends-nyt}, \citet{Callaway2021} has shown that we can identify the $ATT(g,t)$ for post-treatment periods $t\ge g$ by\footnote{\citet{deChaisemartin2020} have derived the same results but restrict attention to instantaneous treatment effects---that is, $ATT(g,g)$'s. They do allow for treatment turning on and off, but also impose that being exposed to a treatment today does not affect outcomes tomorrow (a no-carryover assumption). See \citet{deChaisemartin2023b-intertemporal-treatments} for some extensions. }
\begin{align}
ATT(g,t) &= \E_\omega[Y_{i,t} - Y_{i,t=g-1} | G_i=g] - \E_\omega[Y_{i,t} - Y_{i,t=g-1} | G_i >\max\{g,t\}].\label{eqn:ATTgt-nyt}
\end{align}
Like equation \eqref{eqn:ATTgt-never}, this follows a $2 \times 2$ set up: it leverages data from only two periods, $t$ (post) and $g-1$ (pre), and two treatment groups, $G_i=g$ (treated), and $G_i>t$ (comparison).\footnote{We note that more general results are also possible. In fact, for any not-yet-treated group $g'>t$, it is easy to show that under Assumption \ref{ass:gt-parallel-trends-nyt}, for $t\geq g$, $ATT(g,t) = \E_\omega[Y_{i,t} - Y_{i,t=g-1} | G_i=g] - \E_\omega[Y_{i,t} - Y_{i,t=g-1} | G_i = g'].$  One can then flexibly combine the different comparison groups by leveraging user-specified or efficiency-oriented weights. \citet{Chen_SantAnna_Xie_effientDiD_2024} discuss how to efficiently explore all the information implied by the identification assumptions to form semiparametrically efficient DiD estimators---that is, estimators that asymptotically enjoy the shortest possible (theoretically justified) confidence intervals without making strong functional form or model-based assumptions related to error terms such as homoskedasticity and restrictions on serial dependence.}

Finally, under assumption \ref{ass:gt-parallel-trends-all}, one can use of any not-yet-treated units as a comparison group, as well as any pre-treatment period as a baseline period. For instance, to identify the ATT for 2015 expansion counties, we can now use the never-treated group and the 2016 and 2019 expansion groups, and use any or all of the years from 2009 to 2014 as a baseline. Using Assumption \ref{ass:gt-parallel-trends-all}, it is easy to show that, for any pre-treatment period $t_{\text{pre}}<g$ and any not-yet-treated group $g'> t$, we can identify the $ATT(g,t)$ for group $g$'s post-treatment periods $t\geq g$ by
\begin{align}\label{eqn:ATTgt_all_g}
ATT(g,t) &= \E_\omega[Y_{i,t} -Y_{i,t_{\text{pre}}} | G_i=g] - \E_\omega[Y_{i,t} -Y_{i,t_{\text{pre}}} | G_i =g'], 
\end{align}
which again maps back to the $2\times 2$ DiD setup, as it leverages two periods, $t$ (post) and $t_{\text{pre}}$ (pre), and two groups, $G_i=g$ (treated) and $G_i=g'$ (comparison). This representation makes it clear that, in practice, one can use various pre-treatment periods and not-yet-treated comparison groups to characterize the $ATT(g,t)$ under Assumption \ref{ass:gt-parallel-trends-all}. We can also combine several of these to form an $ATT(g,t)$ estimand that uses more data (\citealp{Wooldridge2021, Gardner2021, Liu2022, Borusyak2023, Chen_SantAnna_Xie_effientDiD_2024}). An intuitive estimand that naturally extends \eqref{eqn:ATTgt-nyt} and allows us to use more pre-treatment data is given by
\begin{align}
ATT(g,t) &= \E_\omega[Y_{i,t} - \overline{Y}_{i,t\le g-1} | G_i=g] - \E_\omega[Y_{i,t} - \overline{Y}_{i,t\le g-1} | G_i >\max\{g,t\}],\label{eqn:ATTgt_averages}
\end{align}
where $\overline{Y}_{i,t\le g-1} =  \sum_{s=1}^{g-1} Y_{i,s}\big/ (g-1)$ is the time average of group $g$ pre-treatment periods for each unit $i$; see, for example, \citet[Section 3.2]{Callaway2023_review}, \citet{Lee_Wooldridge_2023}, and the discussion in \citet[Section 3.2.4]{deChaisemartin2023-survey}. Although the estimand does look different from the previous one, it too resembles a $2\times 2$ design: it effectively leverages two periods, $t$ (post) and the average of all period $t<g$ (pre), and two groups, $G_i=g$ (treated) and $G_i>t$ (comparison). In practice, however, there is no general econometrics guarantee that estimators based on \eqref{eqn:ATTgt_averages} will be more precise than estimators based on \eqref{eqn:ATTgt-nyt}; see, for instance, \citet{Harmon2024}. This arises because it is not always optimal to weigh all pre-treatment periods equally when forming estimators for $ATT(g,t)$. In fact, as discussed in \citet{Chen_SantAnna_Xie_effientDiD_2024}, the optimal way (or, more formally, the semiparametric efficient way) to aggregate information across pre-treatment periods and comparison groups under Assumption \ref{ass:gt-parallel-trends-all} depends on the correlation structure of how the outcome changes over time across different comparison groups. And an effective way to leverage this information consists of constructing DiD estimators for $ATT(g,t)$'s that efficiently weigh several $2\times 2$ DiD estimators for the $ATT(g,t)$ that use different comparison groups and different baseline periods.
These efficiency weights do not depend on additional hard-to-motivate assumptions (e.g., homoskedasticity or restriction on the serial correlation), arise as a consequence of the information content of the identification assumptions, and can be transparently visualized using the tools provided \citet{Chen_SantAnna_Xie_effientDiD_2024}. What is perhaps most important here is that even these more complex DiD estimators closely resemble the approach we took in the $2\times 2$ design.

In the end, a natural question arises: Which parallel trends assumption should one use? This context-specific question is hard to answer, as each assumption has pros and cons. For instance, Assumption \ref{ass:gt-parallel-trends-all} leads to more precise $ATT(g,t)$ estimators because it uses data from multiple pre-treatment periods and multiple comparison groups. Given that power is important when conducting causal inference, this is appealing. On the other hand, it imposes parallel pre-trends, an assumption that is not \emph{required} for identification of $ATT(g,t)$ and that we have not imposed in $2\times T$ DiD designs (see our discussion of equation \eqref{eqn:2xT_es_twfe}). If pre-trends are not parallel, then estimates of $ATT(g,t)$ based on Assumption \ref{ass:gt-parallel-trends-all} can be biased. 

The other extreme is to make Assumption \ref{ass:gt-parallel-trends-never} and use only comparison groups made up of never-treated units. This avoids compositional changes in the comparison group over time, does not restrict pre-trends, and identifies all the $ATT(g,t)$'s.\footnote{Without never-treated units, we cannot estimate $ATT(g,t)$ for the last observed treatment date, which shapes the feasible target parameters. For example, in our Medicaid expansion example, without the presence of never-treated (by 2019) counties, we would not be able to estimate the treatment effects in 2019 (i.e., $ATT(g = 2015, t = 2019)$.} It also avoids using as a comparison group units that may have chosen to begin treatment in a given period because of pre-treatment outcomes, which potentially violates parallel trends. For instance, states that expanded Medicaid in 2016 may have done so on the basis of county mortality rates from previous years, likely violating the parallel trends assumption for this group. On the other hand, never-treated units may have remained untreated for reasons related to trends in $Y_{i,t}(0)$. Non-expansion counties may be too different from expansion counties for them to reflect the relevant counterfactual. This could be, in part, justified after examining the differences in covariate levels and trends between treatment and control groups, as we saw in the discussion of Table \ref{tab:cov_balance}. Also, depending on how widespread treatment is, there may be too few never-treated units to obtain precise estimates.

We view Assumption \ref{ass:gt-parallel-trends-nyt} as a middle step that uses all not-yet-treated units as a comparison group without restricting all pre-treatment trends to be parallel. It uses more information than Assumption \ref{ass:gt-parallel-trends-never}, which can lead to gains in precision and helps to incorporate covariates. While it uses less information than Assumption \ref{ass:gt-parallel-trends-all}, it is also less susceptible to bias from violations of parallel pre-trends. For our Medicaid application, we favor Assumption \ref{ass:gt-parallel-trends-nyt}, as we prefer not to impose parallel pre-trends from 2009 to 2014 or to rely exclusively on comparisons to the set of states that have not expanded Medicaid as of 2024. On the other hand, if parallel trends are not plausible for a particular group of eventually-treated units, perhaps owing to selection based on time-varying unobservables \citep{Ghanem2023}, it is important to remove these units from the DiD analysis to retain interpretability. Ultimately, the plausibility of each parallel trends assumption may vary across different contexts. At the very least, we strongly recommend that researchers clearly state the specific parallel trends assumption they are actually imposing in their analysis to allow readers to discuss its plausibility in a scientifically grounded manner.

\subsubsection{Estimators for staggered designs without covariates}

The identification results for $ATT(g,t)$ discussed in Section \ref{sec:identification_staggered} suggest very simple and intuitive estimators for the $ATT(g,t)$. Given the estimand that comes from the chosen parallel trends assumption, the estimators replace the population (weighted) expectations with their sample analogs. The principle is the same as in the $2\times 2$ setup of Section \ref{sec:estimation}. 

For example, under Assumption \ref{ass:gt-parallel-trends-nyt}, we can leverage \eqref{eqn:ATTgt-nyt} and form plug-in estimators for $ATT(g,t)$ using 
\begin{align}
\widehat{ATT}_{\text{nyt}}(g,t) = \frac{\sum_{i=1}^n \indicator{G_i=g} \omega_i (Y_{i,t} - Y_{i,t=g-1})}{\sum_{i=1}^n \indicator{G_i=g} \omega_i} - \frac{\sum_{i=1}^n \indicator{G_i > t} \omega_i (Y_{i,t} - Y_{i,gt=-1})}{\sum_{i=1}^n \indicator{G_i >t}  \omega_i}.\label{eqn:att_gt_nyt_hat}
\end{align}
This simple estimator is what \citet{Callaway2021} propose when one uses the not-yet-treated group as the comparison group.\footnote{ Recently, \citet{Dube_etal_LPDID_2024} show that one can also get $ATT(g,t)$ estimates that are equivalent to $\widehat{ATT}_{\text{nyt}}(g,t)$ by using local projections. One can also get similar estimators using a ``stacked DiD'' procedure akin to what \citet{fadlon2021}, \citet{deshpande2019}, and \citet{Cengiz2019} have implemented.}

When Assumption \ref{ass:gt-parallel-trends-never} holds, it is straightforward to build on \eqref{eqn:ATTgt-never} and estimate $ATT(g,t)$ by 
\begin{align}
\widehat{ATT}_{\text{never}}(g,t) = \frac{\sum_{i=1}^n \indicator{G_i=g}  \omega_i(Y_{i,t} - Y_{i,t=g-1})}{\sum_{i=1}^n \indicator{G_i=g} \omega_i} - \frac{\sum_{i=1}^n \indicator{G_i = \infty} \omega_i (Y_{i,t} - Y_{i,gt=-1})}{\sum_{i=1}^n \indicator{G_i =\infty} \omega_i}.\label{eqn:att_gt_never_hat}
\end{align}
This estimator was proposed by \citet{Callaway2021} and \citet{Sun2021} when using never-treated units as a comparison group, though \citet{Sun2021} arrive at this using a fully saturated regression specification and estimating the regression coefficients $\beta^{SA}_{g,e}$ with (weighted) least squares,
\begin{equation}
Y_{i,t} = \theta_t + \eta_i + \sum_{g\ne \infty} \sum_{e\neq -1} \beta^{SA}_{g,e} \indicator{G_i=g} \indicator{G_i + e = t} + \epsilon_{i,t}.\label{eqn:SA_reg}
\end{equation}
It is straightforward to show that $\beta^{SA}_{g,e} = \widehat{ATT}_{\text{never}}(g,g+e)$, emphasizing that \eqref{eqn:SA_reg} is just a way to contrast sample means across groups and periods that respect Assumption \ref{ass:gt-parallel-trends-never}. 

When Assumption \ref{ass:gt-parallel-trends-all} holds instead, one can construct plug-in estimators for \eqref{eqn:ATTgt_averages}: 
\begin{align*}
\widehat{ATT}_{\text{avg}}(g,t) = \frac{\sum_{i=1}^n \indicator{G_i=g}  \omega_i (Y_{i,t} - \overline{Y}_{i,t\le g-1})}{\sum_{i=1}^n \indicator{G_i=g} \omega_i} - \frac{\sum_{i=1}^n \indicator{G_i >t} \omega_i (Y_{i,t} - \overline{Y}_{i,t\le g-1})}{\sum_{i=1}^n \indicator{G_i>t} \omega_i}.
\end{align*}

Alternatively, \citet{Wooldridge2021} proposed constructing estimators for $ATT(g,t)$ based on Assumption \ref{ass:gt-parallel-trends-all}, using following ``extended'' TWFE specification:
\begin{align}
Y_{i,t} = \theta_t + \eta_i + \sum_{g\ne \infty} \sum_{s=g}^T \beta^{W}_{g,t} \indicator{G_i=g} \indicator{s = t} + \epsilon_{i,t}, \label{eqn:ETWFE}
\end{align}
where the $\beta^{W}_{g,t}$'s are estimated using (weighted) least squares. \citet{Wooldridge2021} shows that $\widehat{\beta}^{W}_{g,t}$ consistently estimate $ATT(g,t)$ under Assumption \ref{ass:gt-parallel-trends-all}, though we do not know the exact way that $\widehat{\beta}^{W}_{g,t}$ combines pre-treatment periods and not-yet-treated units, which is to say that we do not know the statistical estimand associated with ${\beta}^{W}_{g,t}$. \citet{Wooldridge2021} also shows that $\widehat{\beta}^{W}_{g,t}$ is numerically the same as the ``imputation'' estimators proposed by \citet{Gardner2021}, \citet{Liu2022}, and \citet{Borusyak2023} with balanced panel data (and these specifications do not have covariates).\footnote{Imputation procedures work in two-steps. The first step uses all untreated observations to run the regressions $Y_{it} = \theta_t + \eta_i + \epsilon_{it}$ using only data from the $(i,t)$ pairs that satisfy this criterion, and get the fitted values $\widehat{Y}_{i,t}(0) = \widehat{\theta}_t + \widehat{\eta}_i$ for all eventually-treated observations. The second step estimates $ATT(g,t)$ by $\widehat{ATT}_{\text{imp}}(g,t) = \frac{\sum_{i=1}^n \indicator{G_i=g}(Y_{i,t} - \widehat{Y}_{i,t}(0))}{\sum_{i=1}^n \indicator{G_i=g}}$. See \citet{Gardner2021}, \citet{Liu2022}, and \citet{Borusyak2023} for details. Note that the specification in \eqref{eqn:ETWFE} is similar to the \citet{Sun2021}'s specification \eqref{eqn:SA_reg}, but it omits the pre-treatment event-time dummies. This is justified because \eqref{eqn:ETWFE} imposes parallel pre-trends (Assumption \ref{ass:gt-parallel-trends-all}) while \eqref{eqn:SA_reg} does not (it effectively relies on Assumption \ref{ass:gt-parallel-trends-never}). Thus, in general, one should not expect $\widehat{\beta}_{g,t}^W$ to be equal to  $\widehat{\beta}^{SA}_{g,t-g}$.}

\begin{figure}[!ht]
\includegraphics[width=6in]{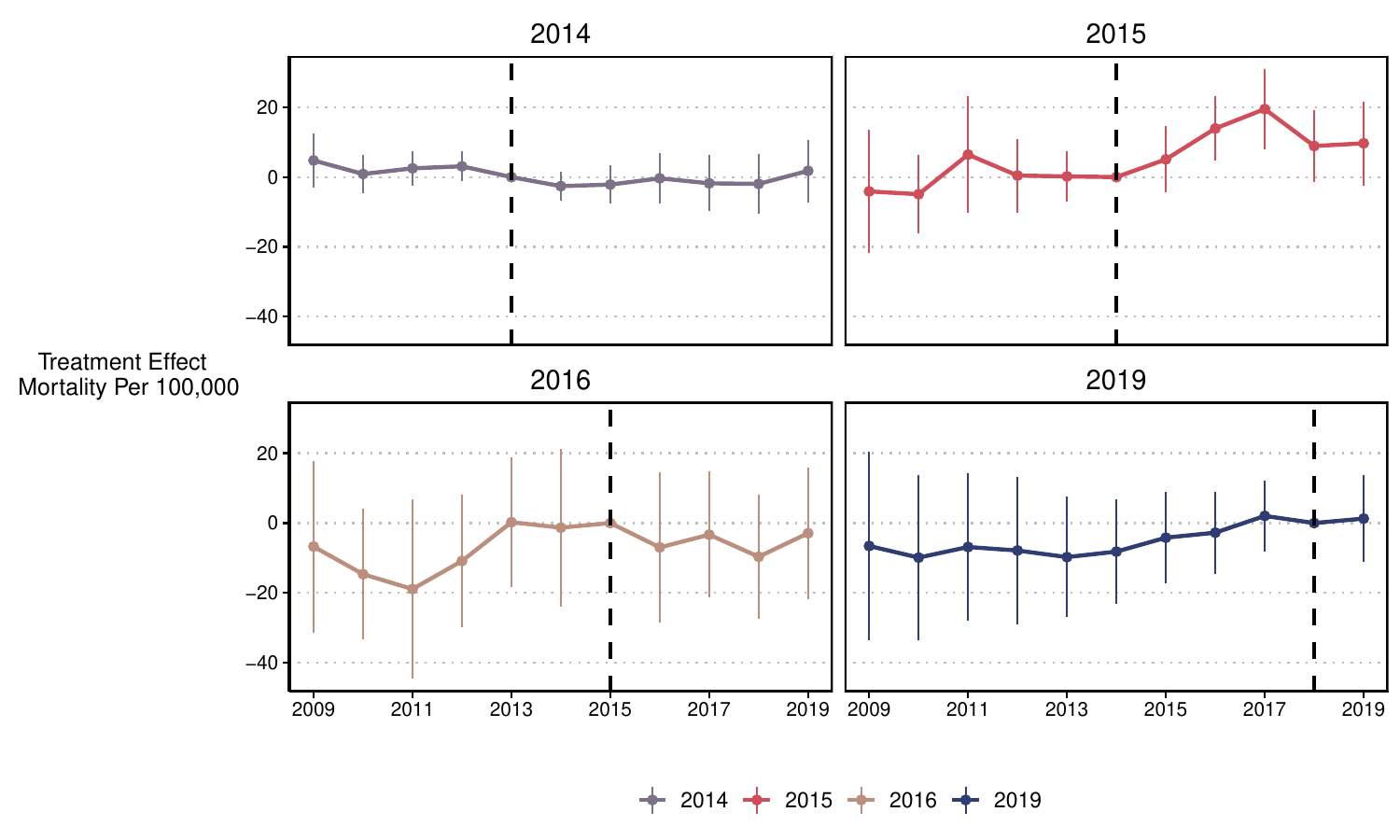}
\caption{\textbf{ATT(g,t)s for Each Expansion Group}}\label{fig:att_gt_calendar}
\justifying
\vspace{-.1cm}\noindent\scriptsize{Notes: This figure shows the group-time ATT estimates ($ATT(g, t)$) in calendar time for the four groups of counties that expanded Medicaid before 2019. Each panel uses 1,222 not-yet-treated counties as the comparison group and shows their uniform confidence intervals at the 95\% significance level. There are 978 counties in states in the 2014 expansion group, 171 counties in the 2015 expansion group, 93 counties in the 2016 expansion group, 140 counties in the 2019 expansion group. The outcome variable is the crude mortality rate for adults ages 20-64, and standard errors are clustered at the county level. The vertical line represents the year before Medicaid expansion (i.e., $g$ - 1) for the timing group.  All results use population weights.}
\end{figure}

Overall, these different $ATT(g,t)$ estimators highlight that we can leverage our DiD expertise built in the $2\times 2$ setup to estimate heterogeneous $ATT(g,t)$ parameters. The exact estimator we can use is shaped by which parallel trends assumption holds. In our Medicaid application, we report in Figure \ref{fig:att_gt_calendar} our $ATT(g,t)$ estimates based on Assumption \ref{ass:gt-parallel-trends-nyt} and \eqref{eqn:att_gt_nyt_hat}. As we have four sets of counties defined by their Medicaid expansion timing, we report four sets of event studies, one for each expansion group. For the 2014, 2016, and 2019 expansion groups, we find that Medicaid did not lead to significant changes in adult mortality rates. For the 2015 expansion group, adult mortality rates rose after expansion.  Regarding pre-trends, Figure \ref{fig:att_gt_calendar} suggests that there may be some non-negligible pre-trends for the 2016 expansion group, though these are not statistically different from zero.  

In Section \ref{sec:covariates}, we highlighted that unconditional assumptiosn such as Assumption \ref{ass:gt-parallel-trends-nyt} may fail in our Medicaid context. One reason is that important determinants of changes in untreated adult mortality rates are imbalanced across treatment and comparison groups. In such cases, one should interpret the results in Figure \ref{fig:att_gt_calendar} with great care. In addition, it is also important to highlight that, as indicated in Table \ref{tab:adoptions}, the 2015, 2016, and 2019 expansion groups are relatively small and represent only 6\%, 2\%, and 3\% of the US population, respectively. Thus, analyzing these groups separately may be ``too noisy'' and not representative of the overall effect for the US population. This does not mean that these $ATT(g,t)$'s are not useful. They are actually an essential part of our DiD analysis, but we may want to aggregate the cohorts to get more informative target parameters for the overall treated population. We now turn to how to aggregate the $ATT(g,t)$'s and then discuss how to incorporate covariates.

\subsubsection{Aggregating group-time average treatment effects}\label{sec:aggregation}

The previous section highlighted that, in many applications, estimating all $ATT(g,t)$'s precisely and attaching policy-relevant interpretations to them may be challenging. Aggregating them into a summary treatment effect measure therefore has clear benefits: it improves precision, reduces the number of results, and yields a parameter that averages over all treated units like the $ATT(2)$ identified in $2\times2$ designs. 

Aggregation in staggered designs involves a notion of time (either calendar time $t$ or event-time $e=t-g$), a length of time (how many periods to aggregate across), and group weights (so larger treatment groups can ``matter more'' than smaller ones). Given some set of weights, it is simple to average the $ATT(g,t)$ building blocks into many kinds of summary parameters:
$$ ATT_{\text{aggte}} = \sum_{g,t} w_{\omega,g,t} ATT(g,t),$$
where $w_{\omega,g,t}$ is a ``generic'' ($\omega$-weighted) group and time-specific (non-negative) weight that sums up to one. Specific choices for the $w_{\omega,g,t}$ weights map to different ways to aggregate and present interpretable causal effects in this kind of complex setting.

As highlighted in Section \ref{sec:2_by_T}, an appealing feature of having access to data from multiple periods is that we can assess how average treatment effects evolve with the time since treatment, or event-time $e= t-g$. Figure \ref{fig:att_gt_calendar} displays event studies for each expansion group, and the aggregation question is how to combine them into a single summary event study.

\begin{figure}[!ht]
\includegraphics[width=5in]{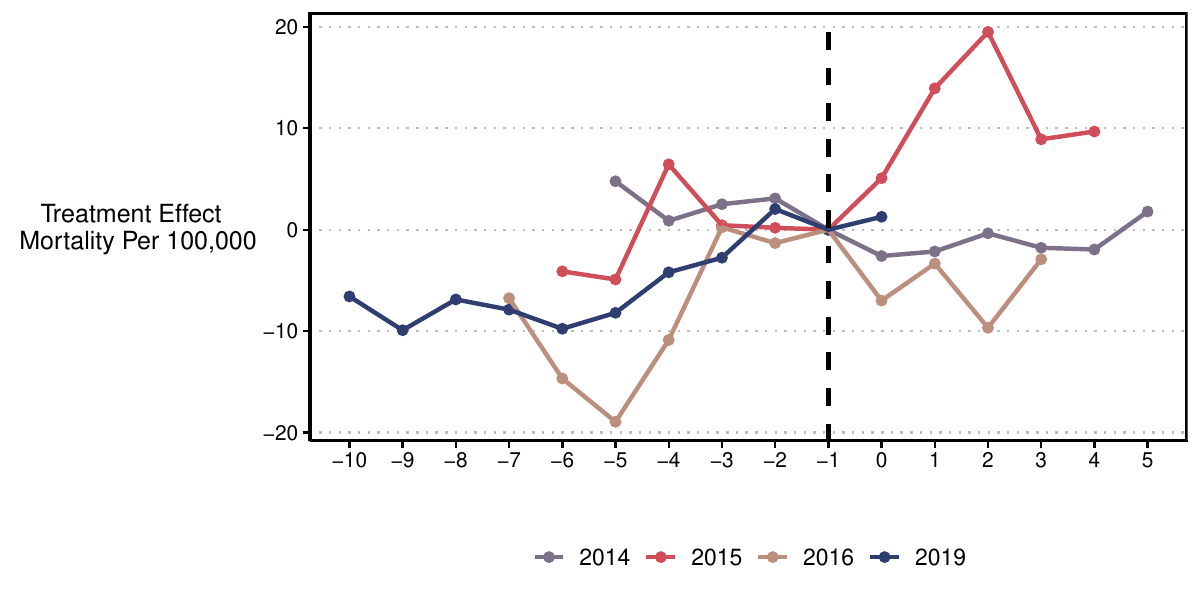}
\caption{\textbf{ATT(g, t) in Event Time}}\label{fig:att_gt_relative}
\justifying
\vspace{-.1cm}\noindent\scriptsize{Notes: This figure shows the group-time ATT estimates ($ATT(g, t)$) in relative event time for the four treatment timing groups of counties that expanded Medicaid before 2019, using not-yet-treated units as the comparison group. The outcome variable is the crude mortality rate for adults ages 20-64.  All estimates use population weights.}
\end{figure}

A basic observation about weighting is that we can give positive weights only to the $ATT(g,t)$'s that we actually identified and estimated. Earlier treated groups have $ATT(g,t)$ estimates for later event-times by definition (and later treated groups have estimates for earlier pre-trends), so it will not be possible to include every group in an aggregated event-study parameter at every event-time. To see which $ATT(g,t)$'s will contribute to our event study, Figure \ref{fig:att_gt_relative} recenters our $ATT(g,t)$ estimates in event time instead of calendar time. That is, we plot $ATT(g,g+e)$ against $e$ for each expansion group.

We will take ``vertical'' weighted averages of each available $ATT(g,g+e)$ for each event time $e$ based on Figure \ref{fig:att_gt_relative}. For instance, to estimate an aggregate event study in event time 0 (a measure of instantaneous treatment effects), we would average estimates of $ATT(2014,2014)$, $ATT(2015,2015)$, $ATT(2016,2016)$ and $ATT(2019,2019)$. When we are interested in event time 1, we would now average $ATT(2014,2015)$, $ATT(2015,2016)$, $ATT(2016,2017)$. The same logic applies to other event times. 

When constructing timing-group weights at a given event-time, it is also important to account for group sizes so that the resulting parameters equal sensible averages of treated units. Table \ref{tab:adoptions} contains all the information necessary for this. The 2014 expansion group accounts for 80\% of treated adults in the groups we consider, while the 2016 expansion group accounts for 3.5\%. If we would like our aggregate event study to be a representative summary of the dynamic effects among treated counties, we should choose weights that are proportional to the treatment group size.

Putting these pieces together, we can formally state the exact summarized causal parameter that highlights treatment effect dynamics in terms of event time: 
\begin{align}
ATT_{\text{es}}(e) &= \E_\omega \bigg[ATT(G,G+e) \bigg|G+e\in[1,T], G\leq T\bigg]\nonumber \\
&=\sum_{g <\infty }w_{\omega,g,e}^{es} ATT(g,g+e),\label{eqn:es_gt}
\end{align} 
where each weight $w_{g,e}^{es}$ gives the share of a group $G=g$ among treated units that have been exposed to treatment for exactly $e$ periods (the groups that we have data for event time $e$ in Figure \ref{fig:att_gt_relative}), and is formally defined as $$w_{g,e}^{es} = \1\{g+e\leq T\} P_\omega(G=g | G+e\leq T, G\leq T).$$
Note that $ ATT_{\text{es}}(e)$ gives the average treatment effect among the units that have been exposed to treatment for exactly $e$ periods, conditional on being observed having participated in the treatment for that number of periods (the condition that $G+e\in[1,T]$) and ever-participating in the treatment by period $T$ ($G\leq T$). One can also take a simple average of all available post-treatment event times, $ATT_{\text{es}}(e),e\ge0$, and report an overall ATT measure. See \citet{Callaway2021} for a discussion of alternative aggregations based on calendar time and groups.

\begin{figure}[!ht]
\includegraphics[width=5in]{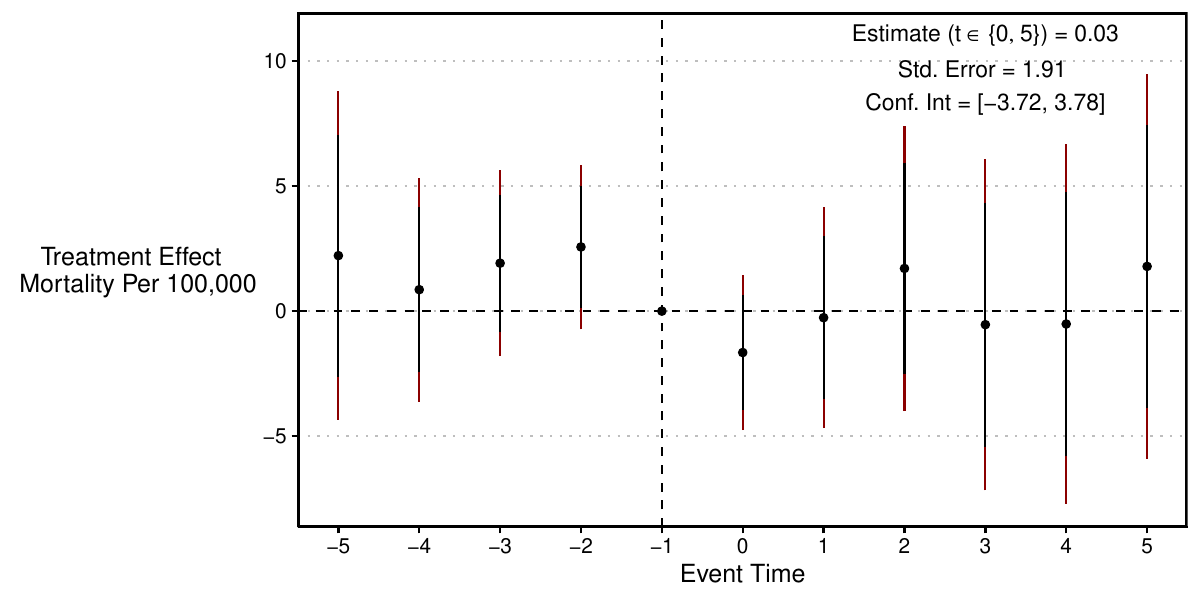}
\caption{\textbf{G $\times$ T Event Study without Covariates}}\label{fig:GXT_ES}
\justifying
\vspace{-.1cm}\noindent\scriptsize{Notes: This figure shows the event study estimates with staggered treatment timing using the doubly-robust estimation method from \citet{Callaway2021}, using 1,222 not-yet-treated units as the comparison group. The sample sizes used to estimate ATT parameters for each timing group are 2,200 for the 2014 group, 1,393 for the 2015 group, 1,315 for the 2016 group, and 1,362 for the 2019 group. The outcome variable is the crude mortality rate for adults ages 20-64. The point estimate is reported by the circles, and both 95\% point-wise (black) and simultaneous (red) confidence intervals are reported with the vertical lines. We also report the simple average of all non-negative event times as a summary of the overall ATT (together with their standard errors and 95\% confidence interval). All results use population weights.}
\end{figure}

Estimating $ATT_{\text{es}}(e)$ is straightforward and, once again, relies on the plug-in principle: we need to replace $ATT(g,g+e)$ with its sample analogs (which we already computed in Figure \ref{fig:att_gt_calendar}), and use the relative adult population share of expansion group $g$ among eventually-treated units as estimates of the event study weights. Figure \ref{fig:GXT_ES} reports our population-weighted estimates of the event-study aggregation from event times -5 to 5, which respectively correspond to 5 years before the Medicaid expansion and 5 years after the Medicaid expansion. We also report pointwise and simultaneous confidence intervals at the 95\% significance level. Overall, the results suggest that Medicaid expansion has no effect on adult mortality rates among counties that eventually experience a Medicaid expansion. The pre-trends are also fairly close to zero, suggesting that our parallel trends assumption may be reasonable. 

We conclude this section by stressing that the way we have constructed the event study parameters in Figure \ref{fig:GXT_ES} uses all available information from Figure \ref{fig:att_gt_calendar}. A potential drawback of this strategy is that we do not always use the same set of groups across all event times. Practitioners usually refer to this as ``imbalance in event time.'' For instance, the 2019 expansion group contributes only to event time $e=0$, not $e=1$ or later event times. When compositional changes are a concern, one can impose balance in event time and estimate a balanced event study aggregation:
\begin{align}
ATT_{\text{es,bal}, [\underline{e}, \overline{e}]}(e) &= \E\bigg[ATT(G,G+e) \bigg|G+\overline{e}\in[1,T], G+\underline{e}\in[1,T], G\leq T\bigg] \nonumber\\
&=\sum_{g \in \mathcal{G}_{treat}}w_{g,[\underline{e}, \overline{e}]}^{es,bal} ATT(g,g+e),\label{eqn:es_gt_balances}
\end{align} 
where balanced event-time weights $w_{g,[\underline{e}, \overline{e}]}^{es,bal}$ are given by 
$$w_{g,[\underline{e}, \overline{e}]}^{es,bal} = \1\{g+\overline{e}\leq T\} \1\{g+\underline{e}\ge 1 \} P_\omega(G=g | G+\overline{e}\in[1,T], G+\underline{e}\in[1,T], G\leq T).$$ 
\noindent Although intimidating, $w_{g,[\underline{e}, \overline{e}]}^{es,bal}$ just measures the relative size of a particular treatment group that was kept in the balanced data. We can interpret $ATT_{\text{es,bal}, [\underline{e}, \overline{e}]}(e)$ as the average group-time average treatment effect among units whose event time is equal to $e$ \textit{and is observed to participate in the treatment for at least $\overline{e}$ periods, and have at least $\underline{e}$ available pre-treatment periods} (if $\underline{e}$ is negative).\footnote{This discussion assumes that there are no ``holes'' in event-times for each treatment group. It is straightforward to adjust the interpretation to those more complicated cases, as the same logic can be applied.}

\subsubsection{Estimators for staggered designs with covariates}

As the discussions in Section \ref{sec:identification_staggered} made it clear, we can view the staggered DiD setups as a collection of simpler $2\times2$ DiD building blocks. A benefit of this interpretation is that, if parallel trends holds only after conditioning on the covariates that determine untreated potential outcome changes, we can easily leverage all the results discussed in Section \ref{sec:covariates} to identify, estimate and make inference about the $ATT(g,t)$'s, using regression adjustment, inverse probability weighting, or doubly robust methods.

Of course, to proceed in this manner, we would need to adopt conditioned on covariates versions of Assumption \ref{ass:gt-parallel-trends-never}, \ref{ass:gt-parallel-trends-nyt} or \ref{ass:gt-parallel-trends-all}, as well as impose an overlap condition. Given that all these are fairly similar to each other, here we state only an extension of Assumption \ref{ass:gt-parallel-trends-nyt} and a strong overlap condition that can be used for all three cases.

\begin{customass}{CPT-GT-NYT}[Conditional Parallel Trends based on not-yet-treated groups] \label{ass:gt-parallel-trends-nyt-cond}
For every eventually treated group $g$, not-yet-treated group $g'$, time periods $t$ such that $t\ge g$ and $g' > t$, and every covariate value $X_i$,
\begin{align*}
	\E_\omega[Y_{i,t}(\infty)-Y_{i,t-1}(\infty)|G_i=g, X_i] = \E_\omega[Y_{i,t}(\infty)-Y_{i,t-1}(\infty)|G_i=g', X_i].
\end{align*}
\end{customass}

\begin{customass}{SO-GT}[Strong overlap with staggered adoption] \label{ass:overlap-staggered}
For every group $g \in \mathcal{G}$, the conditional (weighted) probability of belonging to a treatment group $g$, given observed covariates $X_i$ that are determinants of untreated potential outcome growth, is uniformly bounded away from zero and one. That is,  for some $\epsilon>0$ and for every group $g \in \mathcal{G}$, $\epsilon < P_\omega[G_i=g|X_i] < 1-\epsilon$.
\end{customass}

Using these identifying assumptions and building on the results in Sections \ref{sec:covariates} and \ref{sec:identification_staggered}, we can follow the arguments in \citet{Callaway2021} and establish that the post-treatment $ATT(g,t)$'s are identified by the regression-adjusted, inverse-probability-weighted, and doubly-robust estimands given by\footnote{\citet{Wooldridge2021} propose alternative estimators for the $ATT(g,t)$ that incorporate covariates and their interactions with group and time dummies into \eqref{eqn:ETWFE}. Although \citet{Borusyak2023} and \citet{deChaisemartin2020} also allow for covariates in their estimation procedure, they affect the outcome levels and not their changes over time. Thus, to some extent, these procedures do not build on conditional parallel trends like Assumption \ref{ass:gt-parallel-trends-nyt-cond}.}
\begin{align*}
ATT_\text{ra}(g,t) &= \E_\omega[ Y_{i,t} -  Y_{i,t=g-1} | G_i=g] - \E_\omega\Big[ \E_\omega[ Y_{i,t} -  Y_{i,t=g-1} | X_i, G_i>t] \Big| G_i=g\Big],\\
ATT_\text{ipw}(g,t) &= \E\left[\Big(w_{\omega,G=g}(G_i) - w_{\omega,g,t}(G_i,X_i)\Big) (Y_{i,t} -  Y_{i,t=g-1}) \right], \\
ATT_\text{dr}(g,t) &= \E\left[\Big(w_{\omega,G=g}(G_i) - w_{\omega,g,t}(G_i,X_i)\Big) \Big(Y_{i,t} -  Y_{i,t=g-1} - \E_\omega[ Y_{i,t} -  Y_{i,t=g-1} | X_i, G_i>t] \Big) \right],
\end{align*}
where $(w_{\omega,G=g}(G_i)$ and $ w_{\omega,g,G>t}(G_i,X_i)$ are the analogs of the weights in \eqref{eqn:weights_ATT} and are defined as
\begin{align*}
w_{\omega,G=g}(G) &=  \omega \indicator{G=g} \Big/{\E[ \omega \indicator{G=g}]}, \\
w_{\omega,g,t}(G,X) &= \dfrac{\omega \indicator{G>t}  \indicator{G\ne g} p_{\omega,g,t}(X)}{1-p_{\omega,g,t}(X)}\Bigg/\E\left[\dfrac{\omega \indicator{G>t}  \indicator{G\ne g} p_{\omega,g,t}(X)}{1-p_{\omega,g,t}(X)}\right], 
\end{align*}
and $p_{\omega,g,t}(X) = \E_\omega[\indicator{G_i=g}|X, \indicator{G_i=g} + \indicator{G_i>t} = 1]$ denote the (weighted) probability of belonging to the group $g$ given covariates $X$ and that the unit belongs to either to group $g$---the treated group for the $ATT(g,t)$ of interest---or the not-yet-treated group $G_i>t$--- the comparison group. 

\begin{figure}[!ht]
\includegraphics[width=5in]{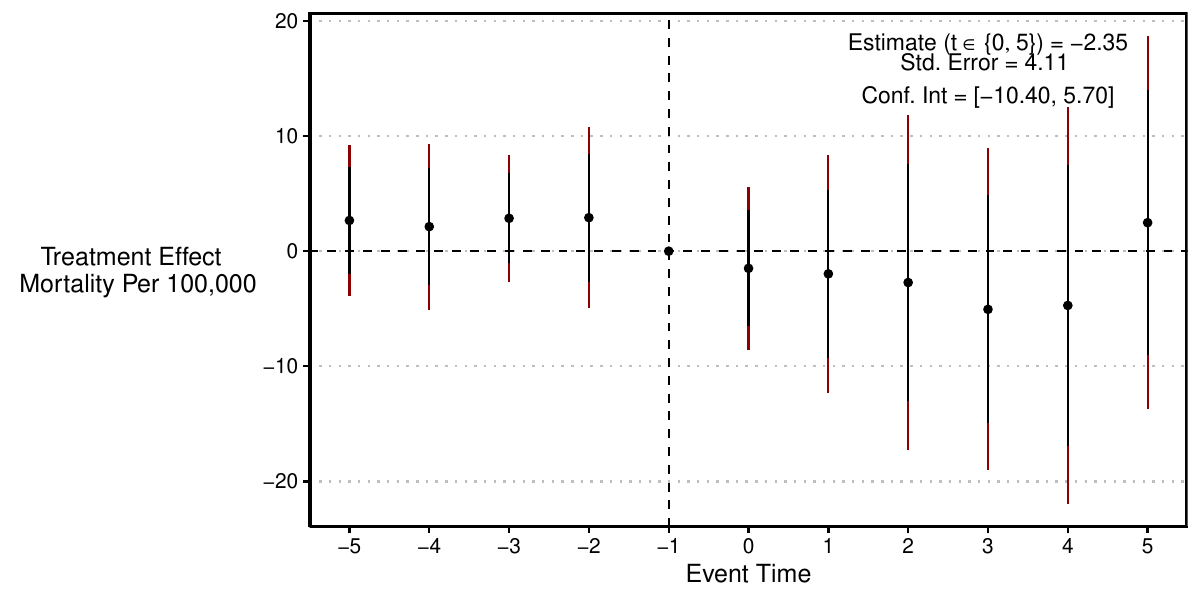}
\caption{\textbf{G $\times$ T Event Study with Covariates} }\label{fig:GXT_ES_covs}
\justifying
\vspace{-.1cm}\noindent\scriptsize{Notes: This figure shows the event-study estimates with staggered treatment timing, using the doubly-robust estimation method from \citet{Callaway2021}.  The outcome variable is the crude mortality rate for adults ages 20-64, and the covariates include the percentage of the county population that is female, the percentage of the county population that is white, the percentage of the county population that is Hispanic, the unemployment rate, the poverty rate, and county-level median income. The point estimate is reported by the circles, and both 95\% point-wise (black) and simultaneous (red) confidence intervals are reported with the vertical lines. We also report the simple average of all non-negative event times as a summary of the overall ATT (together with their standard errors and 95\% confidence interval). All results use population weights.}
\end{figure}

Estimating the $ATT(g,t)$'s follows exactly as in Section \ref{sec:covariates}, and event study aggregations follow from the arguments in Section \ref{sec:aggregation}. Figure \ref{fig:GXT_ES_covs} reports event study summary estimates incorporating covariates into the Medicaid analysis. Like Figure \ref{fig:GXT_ES}, it suggests that Medicaid expansion had no effect on adult mortality among counties that expanded Medicaid by 2019. Given the point estimates and uniform confidence interval, we can be reasonably confident that the treatment effects are not greater than 6 or less than 11 deaths per 100,000 adults for the 6 year period following Medicaid expansion.

\subsection{Limitations of TWFE regressions}\label{sec:limitations}

Our framework emphasizes building an estimator from $2\times2$ components, each of which targets a well-defined $ATT$ parameter under a specific parallel trends assumption. The most common estimator for staggered designs, a TWFE regression, comes instead from extending convenient estimation tools that work well in the $2\times2$ case. A TWFE specification that estimates a summary treatment effect parameter is:
\begin{align}
\label{eq:twfe_did} Y_{i,t} = \theta_t + \eta_i + \beta^{twfe}D_{i,t} + e_{i,t} .
\end{align}
In this section, we abstract from weights.

A major breakthrough in recent DiD research has been to demonstrate two potentially large problems with $\beta^{twfe}$ \citep{deChaisemartin2020,Goodman2021,Sun2021,Borusyak2023}. The primary issue comes from the fact that TWFE implicitly uses already-treated comparison groups. Even if PT holds for all groups and all periods, the resulting estimand can actually put negative weight on certain $ATT(g,t)$ parameters. 
The only way TWFE avoids the problem is if treatment effects do not change over time, a strong additional assumption.

To isolate this issue, consider a setting with two time periods and three groups: a group that enters treatment in the first period ($G_i=1$), a group that becomes treated in the second time period ($G_i=2$), and a never treated group ($G_i=\infty$). This is a staggered design because group 1 and group 2 are treated at different times, but because there are only two time periods, we can re-write the TWFE specification as 
\begin{align*}
\Delta Y_{i,2} = \Delta \theta_t + \beta^{twfe} \Delta D_{i,2} + \Delta e_{i,2}.
\end{align*}
Because $\Delta D_{i,2}$ takes only two values---1 for units whose treatment status increases ($G_i=2$), and 0 for units whose treatment status does not change ($G_i=1$ \textit{and} $G_i=\infty$)---the TWFE estimand is the following simple comparison of means:
\begin{small}
\begin{eqnarray}
	\beta^{twfe} &=& \E[\Delta Y_{i,2} | \Delta D_{i,2}=1] - \E[\Delta Y_{i,2} | \Delta D_{i,2} = 0 ] \nonumber \\ 
	&=& \Big(\E[\Delta Y_{i,2} | G_i=2] - \E[\Delta Y_{i,2} | G_i=\infty] \Big)(1- w_1) + \Big( \E[\Delta Y_{i,2} | G_i=2] - \E[\Delta Y_{i,2} | G_i=1] \Big) w_{1},~~~~~~~~   \label{eqn:2period_staggered_estimand}
\end{eqnarray}
\end{small}
where $w_1 = \frac{p_{1}}{p_{1} + p_\infty}$ and $p_g= P(G=g)$ is group $g$'s share of units. The TWFE coefficient, in this case, is a weighted average of two DiD terms that use the already-treated units or the never-treated units as comparisons. We have already discussed how under PT between group $G_i=2$ and never-treated units, the first term in the $\widehat{\beta}^{twfe}$ decomposition, $\E[\Delta Y_{i,2} | G_i=2] - \E[\Delta Y_{i,2} | G_i=\infty]$, equals $ATT(2,2)$. But what about the second term with the already treated comparison group? It turns out that under parallel trends and no-anticipation, if we add and subtract different terms, this type of estimand generally equals a combination of treatment effects for both groups:
\begin{small}
\begin{eqnarray*}
	\E[\Delta Y_{i,2} | G_i=2] - \E[\Delta Y_{i,2} | G_i=1] &=& \E[Y_{i,2} | G_i=2] - \E[Y_{i,1} | G_i=2] \nonumber \\
	&& - \Big(\E[Y_{i,2} | G_i=1] - \E[ Y_{i,1}| G_i=1] \Big) \nonumber \\
	&=& \E[Y_{i,2}(2) - Y_{i,2}(\infty) | G_i=2] + \E[Y_{i,2}(\infty) - Y_{i,1}(\infty) | G_i=2]  \nonumber\\
	&& - \Big(\E[Y_{i,2}(1) - Y_{i,2}(\infty) | G_i=1] - \E[ Y_{i,1}(1) - Y_{i,1}(\infty) | G_i=1] \Big)\nonumber\\
	&& - \E[Y_{i,2}(\infty) - Y_{i,1}(\infty)|G_i=1]\nonumber\\
	&=& ATT(2,2) - \Big(ATT(1,2) - ATT(1,1)\Big) \nonumber \\
	&& + \E[Y_{i,2}(\infty) - Y_{i,1}(\infty) | G_i=2] - \E[Y_{i,2}(\infty) - Y_{i,1}(\infty)|G_i=1]\nonumber\\
	&=& ATT(2,2) - \Big(ATT(1,2) - ATT(1,1)\Big) \label{eqn:2period_already_treated_bias},
\end{eqnarray*}
\end{small}
where the first equality follows from the linearity of expectations; in the second, we add and subtract averages of untreated potential outcomes and explore no-anticipation, and the observation rules that $Y_{i,t}(g)$ is observed for units with $G_i=g$; in the third equality we rearrange terms; and the last equality follows from a parallel trends for units in group $G=1$ and $G=2$.

The DiD estimand with an already treated comparison group thus equals:
\begin{equation}
\beta^{twfe} = ATT(2,2) + ATT(1,1)w_1 - ATT(1,2)w_1. \label{eqn:2period_staggered_neg_weight}
\end{equation}
The $ATT(1,2)$ building block receives \textit{negative} weight in the overall TWFE estimand unless there are no treatment effect dynamics ($ATT(1,1)=ATT(1,2)$). In this example, the problem is easy to fix: drop the always-treated units and target $ATT(2,2)$. With multiple periods, however, the problem is more complex. 

There are two primary results regarding TWFE for general staggered timing designs. One is a decomposition of the TWFE \textit{estimator}. \citet{Goodman2021} expresses the TWFE estimator as a weighted average of all possible $2\times2$ DiD comparisons between pairs of groups and time periods during which one group enters treatment and the other does not. The terms in his decomposition are not two-period $ATT(g,t)$-type estimators; they aggregate over the relevant pre- and post-periods just as $\beta^{OLS}$ from \eqref{eqn:twfe_2xT_DiD} does. They include many comparisons to already-treated units like  \eqref{eqn:2period_staggered_estimand}. This mechanical decomposition always has strictly positive weights that are larger for larger groups and for groups treated closer to the middle of the panel, which have a larger variance of $D_{i,t}$ conditional on the fixed effects. This result shows how TWFE estimators actually function and creates a clear link to the estimation approaches we outlined above for summary parameters. Both are averages of $2\times2$ DiD terms, but they differ in which comparisons they use and how they aggregate. 

The second result is a decomposition of the TWFE \textit{estimand}, which clarifies why the TWFE estimator is not guaranteed to identify a desirable parameter in staggered designs, even if parallel trends holds. The reason is their use of already-treated comparisons like the ones in  \eqref{eqn:2period_staggered_neg_weight} \citep{Goodman2021,deChaisemartin2020,Borusyak2023,Imai2021,Strezhnev2018,Sun2021}. This will tend to bias $\beta^{twfe}$ away from the sign of $ATT_{\text{avg}}$, and $\beta^{twfe}$ can even have the opposite sign of $ATT_{\text{avg}}$ \citep{baker2022much}. It may appear that a more flexible regression specification could solve this problem, but \cite{Sun2021} show that a TWFE event study specification suffers from a similar bias when the dynamics of the $ATT(g,t)$'s differ across cohorts. Moreover, the ``variance-weighting'' feature of OLS means that $\beta^{twfe}$ has non-intuitive weights even when $ATT(g,t)=ATT(g)$.

While TWFE remains common, it has well-understood, potentially serious, and easily remedied problems, and we do not recommend using it. In many cases, especially those with many untreated units or minimal treatment effect dynamics, TWFE estimates may be similar to those derived from the theoretically grounded estimators discussed above. The only way to be sure, however, is to estimate both. In that case, it is unclear why a researcher would not report the estimates from a procedure motivated by a desirable target parameter and a credible PT assumption.

\section{Conclusion}\label{sec:conclusion}

The starting point of this paper was a $2\times2$ DiD design that researchers have been using for almost 200 years. The end point was a design with five treatment groups, 11 years of data, six covariates, three types of parallel trends assumptions, and four estimation techniques. Our fundamental message is that without understanding how complex designs are built up from simpler ones, it is exceedingly difficult to navigate all the empirical tools now available for DiD designs. This lesson applies not only to the design details we considered here---weighting, covariates, and staggered designs---but to any DiD design. 

The forward-engineering philosophy we followed in this paper suggests a set of steps that researchers can follow in any DiD study:

\begin{enumerate}
\item [Step 1.] \emph{Define target parameters.} Adopt a potential outcomes notation that fits the study's specific setting and use it to define causal target parameters that answer the study's motivating question. Building block causal parameters usually aggregate across units using (conditional) weighted averages, and summary target parameters aggregate across the building blocks. This step fixes the study's goals in terms of causal quantities and facilitates comparisons with related studies.

\item [Step 2.] \emph{State (formally) the identification assumptions.} DiD studies leverage parallel trends assumptions, but they also rely on no-anticipation and, in some cases, overlap conditions, or more. Be explicit about which form of these assumptions is required for identification in the study. Engage with the theoretical arguments necessary for them to hold and generate appropriate empirical evidence, such as pre-treatment differential trends, that can falsify or (indirectly) support their plausibility. 

\item [Step 3.] \emph{Determine the appropriate estimation method.} In some DiD designs, estimation is as simple as replacing population expectations with sample means. In others, such as conditional DiD designs, estimation involves choosing econometric techniques (e.g., a regression adjustment, inverse probability weighting, or doubly robust procedure) to map theoretical quantities to estimable sample quantities. Each of these strategies relies on additional modeling restrictions that should be stated clearly.

\item [Step 4.] \emph{Discuss sources of uncertainty.} Statistical inference procedures for DiD designs stem from basic assumptions about where randomness comes from in a given design. Some researchers may adopt a sampling approach to inference, whereas others may be more comfortable with a design-based perspective. It is important to discuss what variables of the model are being treated as fixed and what variables are considered random, as well as to use inference techniques that are compatible with the model structure and assumptions.

\item [Step 5.] \emph{Estimate.} Steps 1-4 provide a specific structure for using data to estimate the causal parameters of interest.

\item [Step 6.] \emph{Conduct sensitivity analysis.} A clear statement of the identification and estimation assumptions also facilitates a clear statement of what violations of those assumptions might mean. No study is robust to all the ways its assumptions may fail, but a good study should be robust against likely violations of plausible magnitudes. Combine context-specific knowledge about how the assumptions from Step 2 might be violated, and by how much, with the structure of the estimator from Step 3 to evaluate how much the DiD estimates vary if the key identification assumptions are not exactly true.

\item [Step 7.] \emph{Conduct heterogeneity analysis.} Sometimes aggregate parameters defined in Step 1 mask important heterogeneity, in which case the forward-engineering approach simply suggests targeting sub-group parameters as well. This can include variation in parameters over time, between groups of units with different characteristics, or across different sources of treatment variation. Be clear about which types of heterogeneous effects are relevant and how they are identified and estimated.

\item [Step 8.] \emph{Keep learning.} DiD is not the only or the best research design in all settings; it is just one of many causal inference techniques. If the assumptions required for a DiD analysis appear implausible \textit{ex-ante} or are refuted by evidence or non-robustness in practice, then explore different designs. If existing DiD methods do not provide enough guidance, then use a forward engineering approach to deduce what advances would help.

\end{enumerate}

Some researchers may still prefer to use standard regression tools to conduct DiD studies. The properties and pitfalls of some popular regression specifications are now well understood, and one can easily explain how this choice fits with (and perhaps satisfies) the steps above. But using simple regressions in any DiD-type setting is an implicit choice to reverse-engineer a research design from the statistical method, rather than forward-engineer a reliable estimator from a substantive question and transparent assumptions. Ultimately, important questions and credible identification strategies should guide DiD analyses (regression-based or not), not the other way around.

Although this paper is by no means an exhaustive guide to DiD practice, the eight steps above are a rigorous framework for tackling all the DiD topics that we did not cover. The Appendix briefly discusses DiD methods for (a) treatments that turn on and off over time, (b) continuous and multi-valued treatments, (c) triple differences, (d) distributional parameters, and (e) repeated cross-section or unbalanced panel data. While each of these designs differs from what we covered in the main text, a forward-engineering approach that moves from defining parameters and assumptions, to settling on estimation and inference techniques, to probing robustness, applies equally to all of them. While the specifics of any given DiD analysis may change across research questions, treatment variables, econometric techniques, and data structures, the principles by which one can conduct reliable and transparent causal inference stay the same.\footnote{There are other DiD topics of interest that we do not cover, including fuzzy DiD and instrumented DiD designs \citep{DeChaisemartin2018, Miyaji_DiD_IV_2024}, nonlinear DiD models \citep{Wooldridge2023, TchetgenTchetgen2024_universalDiD}, issues related to few clusters \citep[Section 5]{Roth2023a}, and situations with multiple treatments \citep{deChaisemartin2023b-intertemporal-treatments,Yanagi2023}. We also do not cover some methods that address violations of parallel trends \citep{Freyaldenhoven2019, Arkhangelsky2021_SDiD,Callaway2023,Imbens_Kallus_Mao_2021}, nor do we examine setups that impose as-good-as-random treatment timing \citep{Athey2022_DiD_design_based, Roth2023c, Arkhangelsky2024_DRTWFE}.}

\pagebreak 

\renewcommand{\thetable}{A1}
\begin{table*}[h]
\centering
\caption{List of Acronyms}
\label{tab:acronym}
\begin{tabular}{ll}
	\hline
	\textbf{Acronym} & \textbf{Definition} \\
	\hline
	2 $\times$ 2 & Two-Group Two-Time-Periods DiD \\
	2 $\times$ $T$ & Two-Group $T$-Time-Periods DiD \\
	ACA   & Affordable Care Act \\
	ATT   & Average Treatment Effect on the Treated \\
	CPT   & Conditional Parallel Trends \\
	CPT-GT-NYT & Conditional Parallel Trends Based on Not-Yet-Treated Groups \\
	DiD   & Difference-in-Differences \\
	DR    & Doubly Robust \\
	ETWFE & Extended Two-Way Fixed Effects \\ 
	IPW   & Inverse Probability Weighted \\
	NA    & No Anticipation \\
	NA-S  & No Anticipation with Staggered Treatment Timing \\
	OLS   & Ordinary Least Squares \\
	PT    & Parallel Trends \\
	PT-ES & Parallel Trends Event Study \\
	PT-GT-all & Parallel Trends for Every Period and Group \\
	PT-GT-Nev & Parallel Trends Based on Never-Treated Groups \\
	PT-GT-NYT & Parallel Trends Based on Not-Yet-Treated Groups \\
	RA    & Regression Adjustment \\
	SO    & Strong Overlap \\
	SO-GT & Strong Overlap With Staggered Adoption \\
	TWFE  & Two-Way Fixed Effects \\ 
	
	\hline
\end{tabular}
\end{table*}
\renewcommand{\thetable}{\arabic{table}}
\pagebreak

\begin{appendix}
\section{Some additional DiD-related procedures}
This section discusses some important DiD-related topics that we did not cover in our main text. These discussions are short by design, and we focus on providing the main ideas related to challenges and solutions specific to the problem.  We abstract from weights and use $\E[\cdot |\cdot]$ to denote (conditional) expectations. 

\subsection{Setups with treatment turning on and off}
Our main text focuses on setups where treatment remains in place from the period it begins until the end of the sample period, but in practice, some treatments turn on and off over time. This is the setting tackled by \citet{deChaisemartin2020, deChaisemartin2023b-intertemporal-treatments}, \citet{Imai2023Matching}, and \citet{Liu2022}. 

To tackle this problem from first principles, we need to augment the potential outcomes to reflect the richer notion of treatment \textit{sequences}. Following \citet{Robins1986}, let $Y_{i,t}(\textbf{d})$ denote the potential outcome for unit $i$ at time $t$ if this unit received the $T$-dimensional treatment sequence $\textbf{d} \in \{0,1\}^T$. For simplicity, let's say that $T=3$ and that no unit is treated in the first period. In this case, we have four treatment sequences (or histories), which define four potential outcomes for each unit: $Y_{i,t} (0,0,0), Y_{i,t} (0,0,1), Y_{i,t} (0,1,0)$ and $Y_{i,t} (0,1,1)$. We then define treatment groups by treatment sequences: $G= \textbf{d}_0 \equiv (0,0,0)$ (never-treated), $G=\textbf{d}_1 \equiv (0,0,1)$ (treated in the third period), $G=\textbf{d}_2 \equiv(0,1,1)$ (treated in the second and third period), and $G = \textbf{d}_3 \equiv  (0,1,0)$ (treated only in the second period). In general, we would have as many groups as we have different (realized) treatment sequences. Recall that in a staggered timing design with an absorbing treatment, treatment timing fully characterizes a treatment sequence.

Once potential outcomes and groups are well-defined, one can move to parameters of interest. We proceed similarly to the staggered treatment setup in Section \ref{sec:staggered} and consider group-and-time specific ATTs as building blocks, except that groups are now based on more complex treatment sequences. Let $\textbf{0}$ denote a T-dimensional vector of zeros. One intuitive building block parameter on which to base a DiD analysis is $$ATT(\textbf{d}, t) = \E[Y_t(\textbf{d}) - Y_t(\textbf{0}) | G= \textbf{d}],$$ the average treatment effect at time period $t$ of being exposed to treatment sequence $\textbf{d}$  instead of never being exposed to treatment, among units that received treatment sequence $\textbf{d}$.\footnote{One could also adopt alternative building blocks not discussed here, such as the average effect of treatment lasting one period longer or a treatment spell of a given length beginning one period later.}

Next, one needs to establish identification for the parameters, and propose appropriate estimators and inference procedures. Following similar arguments to those in Section \ref{sec:staggered}, a DiD approach to this problem would involve imposing a parallel trends assumption (potentially conditional on covariates) and a no-anticipation assumption to establish that the $ATT(\textbf{d}, t)$'s are identified. If each treatment group is sufficiently large, one could proceed in a similar fashion as the staggered setup, comparing average outcome paths for a given sequence with the average outcome path for never-treated (or not-yet-treated) units. One can also aggregate these different $ATT(\textbf{d}, t)$ to form different summary parameters.

In practice, however, it is often the case that the number of treatment groups is large and each group is small. This essentially creates a ``curse of dimensionality'' problem: there are too many building block parameters defined for too-small groups to be estimated reliably. In such cases, additional assumptions that limit treatment effect dynamics (or how past treatments affect future outcomes) are often imposed, and different aggregated summary parameters are usually targeted.  We provide a brief overview of several different solutions that have been proposed to address this issue.

\citet{deChaisemartin2020} impose a ``no-carryover'' assumption that implies that past treatments do not affect future outcomes; which is to say that treatment effects in a given period last only during that period. With such an assumption (in addition to parallel trends and a no-anticipation assumption), they propose DiD estimators for an instantaneous average treatment effects parameter by comparing currently treated units with untreated units. \citet{Imai2023Matching} adopt a similar approach, though they impose a limited-carryover assumption where treatments may last for $\ell$ periods (with $\ell$ specified by the researcher). They then propose estimators for an average treatment effect of switching into treatment in period $t$ among units that experience the policy change in period $t$, and share the same treatment history over the previous $k$ periods; see \citet{Liu2022} for a related procedure. Finally, \citet{deChaisemartin2023b-intertemporal-treatments} avoid making assumptions related to carryover effects and extend the DiD framework in \citet{deChaisemartin2020} to allow for treatment effect dynamics. The way they proceed is to first ``staggerize'' treatment sequences according to first-time of treatment exposure, compute a staggered DiD procedure for this ``intention-to-treat'' type parameter, and normalize them by a DiD estimate based on the number of treated periods. A potential challenge with  \citeauthor{deChaisemartin2023b-intertemporal-treatments}'s (\citeyear{deChaisemartin2023b-intertemporal-treatments}) approach is the interpretability of their proposed summary parameter, though we should acknowledge that this is a complex setup.

One important takeaway is that comparing these DiD procedures for treatments to turn on and off may be challenging, as they target different causal parameters of interest, and practitioners should be aware of the different assumptions and limitations. We refer the reader to \citet{deChaisemartin2023-survey} and \citet{Liu2022} for additional discussions on these types of DiD estimators.

\subsection{DiD setups with continuous or multi-valued treatments}

Our paper focuses on binary treatments, but many treatments take multiple values or are even continuous. A number of recent papers have studied this particular type of treatment design. These include \citet{Callaway2021b, CallawayGoodmanBaconSantAnna2024}; \citet{deChaisemartin2022-continuous-treatment}; and \citet{deChaisemartinDHaultfoeuilleVazquezBare2024}. Here we focus on a two-period setting in which no unit is treated in period one and some units receive a treatment with varying intensities (or doses) in period two. Most of the key results that distinguish multi-valued from binary treatments are evident with two periods \citep{CallawayGoodmanBaconSantAnna2024}.

We now need to define potential outcomes that reflect varying treatment intensity. We denote $Y_{i,t}(0,d)$ as potential outcomes for unit $i$ in period $t$ if they are untreated in period one and receive treatment dosage $d$ in period two. As we focus on setups where all units are untreated in period one, we simplify notation and index potential outcomes by treatment intensity in period two; that is,  $Y_{it}(d)=Y_{i,t}(0,d)$. An important feature is that $d$ is not restricted to $\{0,1\}$ and can take on richer treatment intensities instead. We denote the treatment dosage for unit $i$ as $D_i$ in period two and stress that in this context, our notion of the treatment group is tied to units' treatment dosage: groups are defined by their treatment dosage in period 2.

A multi-valued treatment defines several different types of causal parameters that may be of interest. For instance, dose-specific average treatment effect parameters such as 
\begin{align*}
	ATT(d|d^\prime) = \E[Y_{t=2}(d) - Y_{t=2}(0) | D=d^\prime] \quad \textrm{and} \quad ATE(d) = \E[Y_{t=2}(d) - Y_{t=2}(0)],
\end{align*}
reflect the average effect of dose $d$ relative to no treatment. Here $ATT(d|d^\prime)$ is the average treatment effect for units that experienced dose $d^\prime$; when $d^\prime=d$, it is the $ATT$ among units that received dose $d$. On the right side, $ATE(d)$ is defined analogously, except that it is the effect on the overall population. Of course, one can also aggregate these dose-specific parameters to form more precisely estimable summary quantities; see, e.g., \citet{Callaway2021b}.

The two treatment effect parameters above provide average treatment effects in levels, and so one reason why they vary could be because $d$ itself varies. To account for the differences in $d$, one may be interested in ``per-dosage'' effects:
\begin{align*}
	ATT_{\text{pd}}(d|d^\prime) =\dfrac{ATT(d|d^\prime)}{d} \quad \textrm{and} \quad ATE_{\text{pd}}(d) =\dfrac{ATE(d)}{d}.
\end{align*}
One can also aggregate these parameters across dosages to analyze $\E[ATT_{\text{pd}}(D|D)| D>0]$, an average treatment effect among treated (or, more generally, among switchers). One can also consider weighted averages of these to learn about $\left.\E[ATT(D|D)| D>0]\right/ \E[D| D>0]$; see \citet{deChaisemartin2022-continuous-treatment} for a general discussion about such target parameters.

Finally, researchers are often interested in the causal effect of a marginal increment in the dose. This notion is the average causal response (ACR), similar to \citet{angrist1995}, defined as follows (when the dose is absolutely continuous):
\begin{align*}
	ACRT(d|d^\prime) = \frac{\partial ATT(l|d^\prime)}{\partial l} \bigg|_{l=d} =  \frac{\partial \E[Y_{t=2}(l)| D=d^\prime]}{\partial l} \bigg|_{l=d} \textrm{ and } ACR(d) =\frac{\partial ATE(d)}{\partial d}= \frac{\partial \E[Y_{t=2}(d)]}{\partial d}.
\end{align*}
Here $ACRT(d|d)$ equals the derivative of the average potential outcome in period two for units that received dose $d$ evaluated at $d$---this is equivalent to the derivative of $ATT(l|d)$ with respect to $l$, evaluated at $l=d$. We can interpret $ACR(d)$ analogously.\footnote{For discrete treatments, ACR's are defined in a similar way but with a slightly different notation to accommodate the discreteness of $d$: $ ACRT(d_j|d_k) = \E[Y_{t=2}(d_j) - Y_{t=2}(d_{j-1}) | D=d_k]$, and $ACR(d_j) = \E[Y_{t=2}(d_j) - Y_{t=2}(d_{j-1})]$.}

The relevant questions pertain to (a) what assumptions are needed to impose to identify these parameters, (b) how to estimate and make inferences about these parameters of interest once identification is established, (c) how to summarize treatment effect heterogeneity across doses to generate interpretable aggregated causal parameters, and (d) whether traditional regression specifications based on TWFE recover a sensible and easy-to-understand causal parameter of interest. These questions are addressed in detail by \citet{Callaway2021b} and \citet{deChaisemartin2022-continuous-treatment}.

\citet{Callaway2021b} highlight how, when no units are treated in period one, identification and estimation of $ATT(d|d)$'s (or their functionals) follows the binary case. They propose flexible nonparametric estimators for the $ATT(d|d)$ curve---the relationship between outcome changes (minus the average change for untreated units) and the dose $d$, making it possible to visualize and make inference about treatment effect heterogeneity across dosages. They also propose estimators that aggregate across dosage values and can be more precisely estimated. The identification of causal response parameters or ATE-type parameters, however, requires a stronger version of parallel trends that holds for potential outcomes at non-zero treatment doses. Under these strong parallel trends and no anticipation assumptions, they discuss estimation and inference procedures for the ACR curves and their summary measures.\footnote{Interestingly, they also show that commonly used TWFE regression specifications are too rigid to lead to easy-to-interpret causal parameters of interest. In fact, they show that one can provide several different decompositions of the TWFE treatment coefficient depending on the specific causal parameter being used as a building block for the analysis, though every decomposition considered by them has some issues related to negative-weighting, additional ``bias'' terms, or non-interpretable weights that can distort inference. They emphasize that all this can be easily resolved by adopting the forward-engineering approach.}

\citet{deChaisemartin2022-continuous-treatment} consider the setup where units are already exposed to different levels of treatment in period one. They discuss how one can identify causal quantities that generalize $ATT_{\text{pd}}(d|d^\prime)$ to this more complex setup when (a) a sizable number of units do not change treatment dosage over time (stayers), and (ii) there is no-carryover from past treatment to future outcomes. They propose estimation and inference procedures for aggregated parameters akin to $\E[ATT_{\text{pd}}(D|D)| D>0]$ and $\left.\E[ATT(D|D)| D>0]\right/ \E[D| D>0]$. 

Lastly, these papers target different causal parameters, put more emphasis on different DiD designs, and, therefore, should be viewed as complements rather than substitutes. In our view, DiD with continuous treatment is another area in which more methodological research is warranted. See \citet{Callaway2021b} and \citet{deChaisemartin2022-continuous-treatment} for a more thorough discussion of many other cases.

\subsection{Triple differences}

The causal interpretation of DiD estimates depends on the plausibility of their identification assumptions, which involve a no-anticipation and a parallel trends condition. In some applications, however, these assumptions may not hold---for example, when the trends of average untreated outcomes among men and women vary across treatment groups. In these cases, a common empirical practice is to attempt to model these violations of parallel trends directly or to conduct sensitivity analysis \citep{Freyaldenhoven_etal_2024, Rambachan2023}. In some specific treatment designs in which treatment is rolled out to different units or groups (e.g., states), but is targeted to a specific subset (partition) of the population (e.g., women), it is possible to relax DiD-type parallel trends so that partition-specific and group-specific violations of parallel trends are allowed. Such setups are often referred to as ``triple differences'' (DDD). Since its introduction by \citet{Gruber1994}, DDD has become very popular among empirical researchers---see \citet{Olden2022} for documentation. In this section, we provide a brief overview of the target parameters and identifying assumptions in DDD. We also highlight that, contrary to conventional wisdom, DDD procedures cannot generally be expressed as the difference between two DiD, especially when parallel trends assumptions only hold after conditioning on covariates or when treatment adoption is staggered. This discussion borrows heavily from \citet{Ortiz-Villavicencio_SantAnna_2025_DDD}.

We start our analysis by discussing potential outcomes and treatment design. As we focus on binary treatments (with potential staggered adoption), the potential outcome is the same as discussed in the main text, with $Y_{i,t}(g)$ denoting the potential outcome for unit $i$ in time $t$ if first exposed to treatment in period $g$. In DDD setups, a unit $i$ is \emph{exposed} to treatment in period $t$ if (i) it belongs to a group (e.g., state) that enabled treatment in period $g$ and $t$ is a post-treatment period, $t\ge g$, and (ii) it belongs to the subset of the population that qualifies (or is \emph{eligible}) for treatment (e.g., women). Let $S \in \mathcal{S} \subseteq \{2,...,T\} \cup \{\infty\}$ denote the time each group (e.g., state) enables the policy/treatment, with the notion that $S = \infty$ if the policy is not enabled in the observed time frame. We also denote the partition of the population that (eventually) qualifies for the treatment by $Q$ with $Q_i=1$ if unit $i$ is (eventually) eligible for treatment and $Q_i=0$ otherwise. With these notations, we can define the treatment groups $G_i$ according to the first time a unit $i$ is \emph{exposed} to treatment; that is, $G_i = S_i$ if $Q_i = 1$ and $G_i = \infty$ if $Q_i = 0$.\footnote{Note that when all units are eligible for treatment, we have $G_i = S_i$, which gets us back to a (staggered) DiD setup.}

Similar to standard DiD designs, DDD is interested in the $ATT(g,t)$-type parameters discussed in Section \ref{sec:building-block}. Given the particular structure of the DDD problem, we can write $ATT(g,t)$'s as
\begin{align*}
	ATT(g,t) \equiv \E[Y_{i,t}(g)-Y_{i,t}(\infty)|G_i=g] = \E[Y_{i,t}(g)-Y_{i,t}(\infty)|S_i=g, Q_i = 1],
\end{align*}
to stress that it measures the average treatment effect at time period $t$ of first being exposed to treatment in period $g$ versus not being exposed to treatment, among units that are actually exposed to treatment in period $g$, i.e., units that are in groups that the policy was first enabled in period $g$ and that qualify for treatment. One can also analyze aggregations of these $ATT(g,t)$ parameters to form causal summary parameters that can be more precisely estimated and highlight treatment effect heterogeneity in some specific directions. This would follow the exact same steps as we discussed in Section \ref{sec:aggregation}, once again highlighting the importance of our forward-engineering approach.

Identifying these causal parameters involves a no-anticipation assumption and a (conditional) parallel trends assumption. Assumption \ref{ass:NA_staggered} can be recycled here, as DDD has the same empirical content as DiD when it comes to no-anticipation. The parallel trends assumption, though, needs to be adjusted as an empirical appeal of DDD is that it can identify ATT parameters even when Assumption \ref{ass:gt-parallel-trends-all} or the other PT variations discussed in Section \ref{sec:staggered} do not hold. Here, we consider a variation of Assumption \ref{ass:gt-parallel-trends-all} that holds only after conditioning on covariates and allows for some partition-specific and group-specific non-parallel trends.

\begin{customass}{DDD-PT-GT-all}[DDD-Parallel Trends for every period and group] \label{ass:gt-parallel-trends-all-DDD}
	For every group $s$ and $s'$ and time periods $t$, with probability one,
	\begin{eqnarray*}
		\mathbb{E}\left[Y_{t}(\infty) - Y_{t-1}(\infty)| S = s, Q=1, X\right] &-& \mathbb{E}\left[Y_{t}(\infty)- Y_{t-1}(\infty) | S = s, Q=0, X \right]  \\
		&=&\\
		\mathbb{E}\left[Y_{t}(\infty) - Y_{t-1}(\infty)|S = s', Q=1, X\right] &-& \mathbb{E}\left[Y_{t}(\infty)- Y_{t-1}(\infty) | S = s', Q=0, X \right].
	\end{eqnarray*}
\end{customass}

When there are only two periods, $t=1,2$, and two groups, $S\in\{2,\infty\}$, and covariates play no role in terms of identification---that is, Assumption \ref{ass:gt-parallel-trends-all-DDD} holds without $X$ (or, equivalently, with $X=1$ for all units)---\citet{Olden2022} show that one can identify $ATT(2,2)$ as the difference of two DiD estimands:
\begin{eqnarray*} ATT(2,2) &=& \mathbb{E}\left[Y_{t=1} - Y_{t=1}| S = 2, Q=1\right] - \mathbb{E}\left[[Y_{t=1} - Y_{t=1} | S = 2, Q=0 \right] \\
	&&- \big(\mathbb{E}\left[Y_{t=1} - Y_{t=1}| S = \infty, Q=1\right] - \mathbb{E}\left[[Y_{t=1} - Y_{t=1} | S = \infty, Q=0 \right] \big).
\end{eqnarray*}
Estimation and inference would be straightforward, as one could use the analogy principle or a two-way fixed effects regression with triple interactions---see \citet{Olden2022} for details.

\citet{Ortiz-Villavicencio_SantAnna_2025_DDD} show that DDD estimands cannot be written as the difference of two DiD estimands when covariates are important for identification, or when treatment adoption is staggered over time and one wants to use not-yet-treated units as a comparison group (as is commonly done in DiD setups). They show how ignoring these considerations and proceeding as if DDD were indeed just a difference of two DiDs can lead to severely biased estimates for the $ATT(g,t)$'s. \citet{Ortiz-Villavicencio_SantAnna_2025_DDD} also show how one can avoid these issues by adopting a forward-engineering approach to the DDD problem. They propose regression-adjusted, inverse probability weighting, and doubly robust estimators for DDD setups that can reliably recover $ATT(g,t)$ and their associated summary parameters under mild assumptions. The paper discusses using multiple comparison groups to generate more precise estimates than simply using a single comparison group. Relatedly, \citet{Strezhnev_DDD_2023} discusses several limitations of common two-way fixed effects regression specifications commonly used for DDD analysis.

Sometimes researchers use the term ``triple differences'' to mean different things and often use different identification assumptions to estimate these different quantities. \citet{Caron2025} discusses using a triple difference strategy to estimate treatment effect heterogeneity. We recommend that practitioners be transparent about target parameters, research designs, and identification assumptions to allow the research community to understand the goals and the differences between DDD procedures.

\subsection{Distributional DiD procedures}

Our paper focuses on learning about \emph{average} treatment effects in various DiD setups. However, approaches that embrace heterogeneity can also target quantities that describe heterogeneity other than average treatment effect parameters. In some settings, researchers may want more information about the distributional impacts of treatment participation. For instance, if a policymaker faces two different labor market programs with very similar average effects on earnings, they may prefer the one that potentially has a higher impact on the lower tail of the income distribution. Difference-in-Differences-type strategies can also be used to identify, estimate, and make inferences about various distributional features of the outcome of interest. This area has received a substantial amount of methodological consideration by econometricians in recent years; see \citet{Athey2006}, \citet{Bonhomme2011}, \citet{Callaway2018}, \citet{Callaway2019}, \citet{Roth2023b}, \citet{Ghanem2023_CiC}, \citet{FernandezVal_2024_Dist_DiD}, and references therein. For some empirical literature using distributional DiD procedures, see \citet{Meyer_Viscusi_Durbin_1995}, \citet{Finkelstein2008}, and \citet{Cengiz2019}, among many others.

An analysis of distributional quantities does not require different potential outcomes notation from Section \ref{sec:staggered}; it just targets functionals of the potential outcome distributions other than their means. The first thing to notice is that there are several types of distributional causal parameters in the treated group that one may care about. The unique feature of them is that they are all functionals of $ F_{Y_t(g)|G=g}(y) = \mathbb{P}(Y_t(g) \leq y  | G=g)$ and $ F_{Y_t(\infty)|G=g}(y) \equiv \mathbb{P}(Y_t(\infty)\leq y | G=g)$. Examples of such functionals include distributional treatment effects in time period $t$ among units first treated in period $g$ (denominated in probability units), 
$$DTT(y|g,t) =  F_{Y_t(g)|G=g}(y) -  F_{Y_t(\infty)|G=g}(y),$$
quantile treatment effects in time period $t$ among units first treated in period $g$ (denominated in outcome units), 
$$QTT(\tau|g,t) = F_{Y_t(g)|G=g}^{-1}(\tau)- F_{Y_t(\infty)|G=g}^{-1}(\tau),$$
where $F_{Y_t(g)|G=g}^{-1}(\tau) = inf\{ y: F_{Y_t(g)|G=g}(y) \ge \tau\}$ denotes the $\tau$-quantile of $Y_t(g)$ among units in group $G=g$, and $F_{Y_t(\infty)|G=g}^{-1}(\tau)$ is defined analogously. Other functionals related to inequality measures can also be obtained; see \citet{Firpo2016} for a discussion on this topic.

To make inferences about these different causal parameters, one needs to identify $ F_{Y_t(\infty)|G=g}(y)$ and $ F_{Y_t(g)|G=g}(y)$. Identification of $ F_{Y_t(g)|G=g}(y)$ is usually non-controversial, as we can use data from units in group $G=g$ to learn about the distribution of $Y_t(g)$. The main challenge is related to how to learn the counterfactual distribution $ F_{Y_t(\infty)|G=g}(y)$ from the data. This is where different DiD-type procedures differ, as each paper in this literature relies on different and often non-nested identification assumptions that, if true, identify $ F_{Y_t(\infty)|G=g}(y)$. Given the space constraints, we do not provide explicit and detailed discussion about how these different DiD-related distributional procedures function.  However, all distributional DiD estimators share our forward-engineering approach; they clearly state their identification assumptions and target parameters and then provide estimators that recover well-defined causal quantities. We also note that most distributional DiD methodological papers focus on two-period and two-group setups. However, it is straightforward to build similar arguments to those in Section \ref{sec:staggered} to extend the designs to more general settings, which is again another benefit of the forward-engineering approach to causal inference.

We close this section by noting that there exist other types of distributional parameters of interest related to the distribution \emph{of} the treatment effects in period $t$ among the units in group $g$, $ \mathbb{P}(Y_t(g) - Y_t(\infty)\leq y | G=g)$. In general, such causal quantities cannot be point identified, as discussed in \citet{Heckman1997b}, \citet{FanYu2012} and \citet{Callaway2021c}. However, often one can still partially identify such policy-relevant parameters under different restrictions. We refer the reader to \citet{Callaway2021c} for a more detailed discussion of this topic.

\subsection{Repeated cross-sections and unbalanced panel data}
An appealing feature of DiD procedures is that, although helpful, a balanced panel is not a requirement for DiD analyses, which can also be deployed with repeated cross-sectional data or unbalanced panels. Indeed, as discussed in Section \ref{sec:estimation} and made explicit in equation \ref{eqn:2x2_ATT}, the $2\times 2$ building block in unconditional DiD analyses involves only averages that are group and time specific, and does not require the same unit to be observed in all periods. As discussed in \citet{Callaway2021}, the same applies to unconditional staggered adoption setups, and one need not enforce a balanced panel even within each subset of the data used to estimate the $ATT(g,t)$ building blocks. One caveat is that the interpretation of the parameter of interest may change, which we discuss more below.

When covariates are available and play an important role in the plausibility of the identification assumptions, the differences between DiD with a balanced panel and repeated cross-sections (or unbalanced panel) are subtle, can be practically important, and are often not discussed in methodological papers. The gist of the problem relates to potential compositional changes over time. Most DiD papers that rigorously discuss repeated cross-section setups, including \citet{Abadie2005}, \citet{SantAnna2020}, and \citet{Callaway2021}, rule out compositional changes by assuming that the joint distribution of covariates and treatment groups is invariant over time, a stationarity-type assumption. However, this may not be warranted in empirical applications, and erroneously imposing this additional assumption can lead to biases \citep{Hong2013, Santanna_Qi_DiD_compositional_2023}. On the other hand, when this stationarity assumption is justified and correctly used, the gains in power when conducting inference for DiD parameters can be noticeable \citep{Santanna_Qi_DiD_compositional_2023}. In what follows, we use the $2 \times 2$ setup to explain how compositional changes can complicate the analysis and why ruling it out leads to a gain in precision. 

To see how issues related to compositional changes affect the analysis, let us first assume that there are no compositional changes and that the stationarity assumption is valid. In this case, the average treatment effect on the treated in period two (post-treatment) can be written as 
\begin{align}
	ATT(2) &\equiv \E[Y_{i,t=2}(1)|D_i=1]  - \E[Y_{i,t=2}(0)|D_i=1]  \nonumber\\
	&=  \E[Y_{i,t=2}(1)|D_i=1, T_{i,t=2}=1]  - \E[Y_{i,t=2}(0)|D_i=1, T_{i,t=2}=1] \nonumber 
	\\
	&=  \E[Y_{i}(1)|D_i=1, T_{i,t=2}=1]  - \E[Y_{i}(0)|D_i=1, T_{i,t=2}=1] \label{eqn:RCS_ATT_no_comp_change},
\end{align}
where $T_{i,t}$ is an indicator if unit $i$ is observed in period $t$, $Y_{i}(d)=T_{i,t=2}~Y_{i,t=2}(d)+ T_{i,t=1}~ Y_{i,t=1}(d)$ is the potential outcome for unit $i$, $D_i = 1\{G_i = 2\}$ is a treatment group dummy that equals one if a unit is first treated in period two and zero if it is untreated in both periods. We also set $X_i$ to be a vector of (pre-treatment) covariates.  Note that even here, we already use the stationarity condition that the joint distribution of $(D_i, X_i)$ is invariant to $T_{i,t=2}$ to move from the first to the second line and establish \eqref{eqn:RCS_ATT_no_comp_change}.

To identify $ATT(2)$ it is often constructive to first establish the identification of its conditional-on-covariates analog; that is, the conditional ATT in period two among units with covariates $X_i$, $ATT_{X_i}(2)$. This is exactly how we proceeded in Section \ref{sec:2x2_DiD_with_X}. Under the stationarity condition, and similarly to \eqref{eqn:RCS_ATT_no_comp_change}, we can express this quantity as\footnote{To guarantee that all the conditional expectations in \eqref{eqn:RCS_ATT_no_comp_change_X} are well-defined, we need an overlap condition that guarantees that $P(T_{i,t=2}=1, D_i=1 |X_i)>0$.  We discuss this below.}
\begin{align}
	ATT_{X_i}(2) &\equiv \E[Y_{i,t=2}(1)|D_i=1, X_i]  - \E[Y_{i,t=2}(0)|D_i=1, X_i]  \nonumber\\
	&=  \E[Y_{i}(1)|D_i=1, X_i, T_{i,t=2}=1]  - \E[Y_{i}(0)|D_i=1, X_i, T_{i,t=2}=1] \label{eqn:RCS_ATT_no_comp_change_X}.
\end{align}

Next, we have to establish the identification of this quantity. As expected, we will again use conditional parallel trends, no-anticipation, and overlap assumptions. The no-anticipation condition used here is the same as the one in the main text. The conditional parallel trends and overlap assumptions need to be modified, as we now work with multiple partitions of the data depending on treatment status and the period a unit is observed. In this sense, we modify Assumptions \ref{ass:cond-parallel-trends} and \ref{ass:overlap} to the following related, but different, assumptions. These modifications are warranted regardless of whether compositional changes are present; this step is instead tied to data structure.\footnote{In setups where we rule out compositional changes and impose the stationarity condition that the joint distribution of $(D_i, X_i)$ is invariant to $T_{i,t=2}$, we may not need to modify Assumption \ref{ass:cond-parallel-trends}. We do it here for transparency purposes.}

\begin{customass}{CPT-RCS}[$2\times2$ Conditional Parallel Trends with repeated cross-sections] \label{ass:cond-parallel-trends-rcs}
	We assume that, with probability one,
	\begin{eqnarray}
		\E[Y_{i,t=2}(0)| X_i, D_i=1, T_{i,t=2}=1] &-& \E_\omega[Y_{i,t=1}(0) | X_i, D_i=1,  T_{i,t=1}=1] \nonumber  \\
		&=&\label{eqn:cond-parallel-trends-rcs}\\
		\E[Y_{i,t=2}(0) | X_i, D_i=0,  T_{i,t=2}=1] &-& \E[ Y_{i,t=1}(0) | X_i, D_i=0,  T_{i,t=1}=1]. \nonumber
	\end{eqnarray}
\end{customass}

\begin{customass}{SO-RCS}[Strong overlap with repeated cross-sections] \label{ass:overlap-rcs}
	For some $\epsilon>0$ and every $(d,s) \in \{0,1\} \times \{0,1\}$, $\epsilon < P[D_i=d, T_{i,t=2}=s|X_i] < 1-\epsilon$.
\end{customass}

With this modification, we can now show that when Assumptions \ref{ass:NA}, \ref{ass:cond-parallel-trends-rcs}, and \ref{ass:overlap} hold, the conditional ATT parameter $ATT_{X_i}(2)$ is identified by\footnote{We also require the assumption that the pooled repeated cross-section data $\left\{ Y_{i},D_{i},X_{i}, T_{i,t=2},  T_{i,t=1}\right\}_{i=1}^{n}$ is $iid$, though this is fairly standard and uncontroversial; see, for instance, \citet{Abadie2005} and \citet[Assumption 1]{SantAnna2020}. We maintain this condition as an assumption throughout this section.}
\begin{align}
	ATT_{X_i}(2) =& (\E[Y_{i}|D_i=1, T_{i,t=2}=1, X_i] - \E[Y_{i}|D_i=1, T_{i,t=1}=1, X_i]) \nonumber \\
	& - (\E[Y_{i}|D_i=0, T_{i,t=2}=1, X_i] - \E[Y_{i}|D_i=0, T_{i, t=1}=1, X_i]).\label{eqn:RCS_CATT}
\end{align}
This step provides a methodological justification to estimate the $ATT_{X_i}(2)$'s using four conditional expectations that use only the available data. In addition, it highlights that, under the stationarity assumption, once we learn the $ATT_{X_i}(2)$'s, we can aggregate them using the covariate distribution of treated units \emph{available from both time periods} to get the $ATT(2)$. More formally, $ATT(2)$ is identified by 
\begin{align*}
	ATT(2) =& \E[ ATT_{X_i}(2) | D_i = 1] \nonumber \\
	=& \E[ ATT_{X_i}(2) | D_i = 1, T_{i,t=2}=1] P(T_{i,t=2}=1|D_i=1) \nonumber \\
	&+ \E[ ATT_{X_i}(2) | D_i = 1, T_{i,t=1}=1] P(T_{i,t=1}=1|D_i=1), \label{eqn:att_rcs_id}
\end{align*}
where $ATT_{X_i}(2)$ is given by \eqref{eqn:RCS_CATT}. This is the second point at which the stationarity assumption and the absence of compositional changes are necessary: under these conditions, covariates from treated units across the entire dataset can be used to identify $ATT(2)$. The fact that you can pool data across all periods to learn about $ATT(2)$ translates to gains in power, as formally discussed by \citet{SantAnna2020} and \citet{Santanna_Qi_DiD_compositional_2023}. The third place where the stationarity condition affects the analysis is in the characterization of how ``the most precise'' (regular and asymptotically linear) estimator for the $ATT(2)$ should look. This point relates to the semi-parametric efficiency bound and the construction of efficient (and doubly robust) estimators. As these points are slightly more technical, we refer the readers to \citet{SantAnna2020} and \citet{Santanna_Qi_DiD_compositional_2023} for more details. 

Overall, when group composition does not change over time, one can pool information across the entire dataset, which has an impact on the definition of target parameters and leads to more precise inference procedures. But what happens when this condition fails? How does this affect the analysis? 

First, this matters for the definition of the treatment effect of interest. In setups where the sampling varies across periods, we do not have a single notion of $ATT(2)$. Instead, we need to accommodate the fact that the $ATT(2)$ may vary across units sampled from different periods. Thus, when we do not rule out compositional changes, we must be explicit about the treated subpopulation that we are interested in. It is common to focus on the average treatment effect in period two among treated units \emph{that are also sampled in period two}, that is,\footnote{One may also be interested in the ATT in period two among treated units that are sampled in period one. The arguments required to establish (point) identification of this parameter differ from those we use here. A main challenge is that we do not observe $\E[Y_{i,t=2}(1)|D_i=1, T_{i,t=1}=1]$, and the parallel trends assumption we leverage does not involve treated potential outcomes.} 
\begin{eqnarray}
	ATT(2|T_{t=2} = 1) &\equiv& \E[Y_{i,t=2}(1)|D_i=1, T_{i,t=2}=1]  - \E[Y_{i,t=2}(0)|D_i=1, T_{i,t=2}=1] \nonumber  \\
	&=& \E[Y_{i}(1)|D_i=1, T_{i,t=2}=1]  - \E[Y_{i}(0)|D_i=1, T_{i,t=2}=1] . \label{eqn:cc}
\end{eqnarray}

Although \eqref{eqn:cc} has the same statistical estimand as \eqref{eqn:RCS_ATT_no_comp_change}---that is, the formulas on the right-hand side of the equation coincide---it has a very different interpretation. It is the $ATT(2)$ among units sampled in period $2$, and is not an ``overall'' $ATT(2)$. One may think this difference is merely cosmetic, but as discussed below, this has implications for constructing estimands when covariates are important for identification. Where covariates do not play an important role, it is simply a matter of changing the interpretation of your reported estimates (which also applies to unconditional staggered setups, to be clear).

When covariates do play an important role, however, and when Assumptions \ref{ass:cond-parallel-trends-rcs}, \ref{ass:NA} and \ref{ass:overlap-rcs} hold, the conditional ATT parameter among units sampled in period two, $ATT_{X_i}(2| T_{t=2})$ is identified by
\begin{eqnarray}
	ATT_{X_i}(2| T_{t=2}) &=& (\E[Y_{i}|D_i=1, T_{i,t=2}=1, X_i] - \E[Y_{i}|D_i=1, T_{i,t=1}=1, X_i]) \nonumber \\
	& &- (\E[Y_{i}|D_i=0, T_{i,t=2}=1, X_i] - \E[Y_{i}|D_i=0, T_{i, t=1}=1, X_i]),\label{eqn:CATT_comp_change}
\end{eqnarray}
which, in turn, implies that $ATT(2|T_{t=2} = 1) $ is identified by
\begin{eqnarray}
	ATT(2|T_{t=2} = 1) = \E[ ATT_{X_i}(2| T_{t=2}) | D_i = 1, T_{i,t=2}=1].\label{eqn:att_rcs_comp_chage}
\end{eqnarray}
Several remarks are worth making. First, the statistical estimand in \eqref{eqn:CATT_comp_change} is the same as when one rules out compositional changes as in \eqref{eqn:RCS_CATT}, suggesting that, once again, what changes in this step is the interpretation. However, these interpretative issues have direct consequences on the appropriate method for aggregating across covariate values. As clearly stated in \eqref{eqn:att_rcs_comp_chage}, in the presence of potential compositional changes, one is not allowed to pool information across periods to identify (and also estimate and make inference) about $ATT(2|T_{t=2} = 1)$. As discussed in \citet{Santanna_Qi_DiD_compositional_2023}, ignoring these issues and pooling data from all periods in the presence of compositional changes leads to a bias that is important to be aware of. We refer the reader to \citet{Santanna_Qi_DiD_compositional_2023} for a discussion related to unbalanced panels and also on a discussion about doubly robust and semiparametric efficient DiD estimators under compositional changes. We are not aware of any papers that formally extend the discussion in \citet{Santanna_Qi_DiD_compositional_2023} to staggered DiD designs. Still, this extension is surely possible by following our forward-engineering approach to DiD. 

We close this section by highlighting that, in practice, it is possible to test for compositional changes by comparing the estimates from estimators that impose it and those that do not. \citet{Santanna_Qi_DiD_compositional_2023} discuss Hausman-type tests in the two-period setting, though one can extend those to more general setups. We also highlight that when it comes to DiD setups with staggered adoption, some equivalence results discussed in Section \ref{sec:staggered} no longer hold with repeated cross-sections or unbalanced panel data. For instance, the \citet{Sun2021} regression-based strategy to estimate $ATT(g,t)$'s using \eqref{eqn:SA_reg} no longer coincides with \citeauthor{Callaway2021}'s (\citeyear{{Callaway2021}}) estimators using the analog of \eqref{eqn:att_gt_never_hat}:
\begin{align*}
	\widehat{ATT}_{\text{never}}(g,t) = (\overline{Y}_{G=g, t} - \overline{Y}_{G=g,t=g-1}) - (\overline{Y}_{G=\infty, t} - \overline{Y}_{G=\infty, t=g-1}),
\end{align*}
where $\overline{Y}_{G=a, t=s}$ is the sample mean of $Y$ among units that belong in group $G=a$ and are observed in period $t=s$. In fact, it is unclear exactly what estimand is being recovered when one uses \eqref{eqn:SA_reg} with an unbalanced panel. If one replaces unit fixed effects with treatment group dummies in \eqref{eqn:SA_reg}, such equivalence is restored, though we suspect that many practitioners do not use this alternative specification. In general, we caution against extrapolating from a well-motivated regression specification that was studied under one specific setup to another related but inherited different framework. This practice has led to many issues in DiD, which can be fully avoided by adopting the forward-engineering approach discussed in this paper.

\end{appendix}
\small{
\setlength{\bibsep}{1pt plus 0.3ex}
\putbib
}

\end{bibunit}

\end{document}

%% file: Preambles/header.tex
\usepackage{changepage}

\usepackage[svgnames]{xcolor}
\usepackage[english]{babel}
\usepackage{amsfonts}
\usepackage{amssymb}
\usepackage{bm}
\usepackage{amsmath}
\usepackage{adjustbox}
\usepackage{threeparttable}
\usepackage{booktabs}	
\usepackage[normalem]{ulem}
\usepackage{enumerate}
\usepackage{bbm}
\usepackage{mathtools}
\usepackage{subfig}
\usepackage{booktabs}	
\usepackage{setspace}
\usepackage{graphicx}
\usepackage{fullpage}

\usepackage{soul}
\usepackage{xspace}
\usepackage{csquotes}
\usepackage[left=2cm,right=2cm,top=2cm,bottom=2cm]{geometry}
\usepackage{placeins}
\usepackage[footfont=normalsize,font=normalsize]{floatrow}
\usepackage{lscape}
\usepackage{longtable}
\usepackage{mathabx}
\usepackage{accents}
\usepackage{float}
\usepackage{fullwidth}
\usepackage{verbatim}
\usepackage{autobreak}
\usepackage[bottom]{footmisc} 

\usepackage{tikz}
\usetikzlibrary{calc,intersections}

	\usepackage[longnamesfirst]{natbib}
	\usepackage{bibunits}
	\defaultbibliography{Bibliography.bib}  
	\defaultbibliographystyle{aer}

\definecolor{darkcandyapplered}{rgb}{0.64, 0.0, 0.0}
\definecolor{cite_blue}{rgb}{0.0, 0.1, 0.3}
\definecolor{BrickRed}{rgb}{0.8, 0.25, 0.33}
	\usepackage{hyperref}
	\hypersetup{
    colorlinks=true,
    linkcolor=cite_blue,
    filecolor=cite_blue,      
    urlcolor=cite_blue,
    citecolor = cite_blue
}
	\usepackage{cleveref}

    \usepackage{rotating}
	\usepackage[framemethod=tikz]{mdframed}

    \usepackage{tcolorbox}
    \newtcbox{\feedback}{nobeforeafter,colframe=black,colback=white,boxrule=0.5pt,arc=2pt,
      boxsep=0pt,left=2pt,right=2pt,top=2pt,bottom=2pt,tcbox raise base}
      
    \usepackage{amsthm}

    \theoremstyle{definition}

\numberwithin{equation}{section}

\usepackage{array}
\newcolumntype{L}[1]{>{\raggedright\let\newline\\\arraybackslash}m{#1}}
\newcolumntype{C}[1]{>{\centering\let\newline\\\arraybackslash\hspace{0pt}}m{#1}}
\newcolumntype{R}[1]{>{\raggedleft\let\newline\\\arraybackslash\hspace{0pt}}m{#1}}

\usepackage{ragged2e}
\newlength\ubwidth


%% file: Preambles/notation.tex
	\newcommand{\E}{\mathop{}\!\mathbb{E}}

	\newcommand{\1}{\mathbf{1}}

\newcommand{\indicator}[1]{\mathbf{1}\{#1\}}

%% file: Tables/adoptions.tex
\begin{table}[H]

\caption{\label{tab:adoptions}Medicaid Expansion under the Affordable Care Act}
\centering
\resizebox{\linewidth}{!}{
\begin{threeparttable}
\begin{tabular}[t]{c>{\centering\arraybackslash}p{20em}ccc}
\toprule
Expansion 
 Year & States & Share of States & Share of Counties & Share of Adults (2013)\\
\midrule
Pre-2014 & DE, MA, NY, VT & 0.08 & 0.03 & 0.09\\
2014 & AR, AZ, CA, CO, CT, HI, IA, IL, KY, MD, MI, MN, ND, NH, NJ, NM, NV, OH, OR, RI, WA, WV & 0.44 & 0.36 & 0.45\\
2015 & AK, IN, PA & 0.06 & 0.06 & 0.06\\
2016 & LA, MT & 0.04 & 0.04 & 0.02\\
2019 & ME, VA & 0.04 & 0.05 & 0.03\\
2020 & ID, NE, UT & 0.06 & 0.04 & 0.02\\
2021 & MO, OK & 0.04 & 0.06 & 0.03\\
2023 & NC, SD & 0.04 & 0.05 & 0.03\\
Non-Expansion & AL, FL, GA, KS, MS, SC, TN, TX, WI, WY & 0.20 & 0.31 & 0.26\\
\bottomrule
\end{tabular}
\begin{tablenotes}[para]
\item \vspace{-4ex} \singlespacing \footnotesize{The table shows which states adopted 
                  the ACA's Medicaid expansion in each year as well as the share of all states, counties, and adults in 
                  each expansion year.}
\end{tablenotes}
\end{threeparttable}}
\end{table}

%% file: Tables/two_by_two_ex.tex
\begin{table}[!h]

\caption{\label{tab:two_by_two_ex}Simple 2 $\times$ 2 DiD}
\centering
\fontsize{12}{14}\selectfont
\begin{threeparttable}
\begin{tabular}[t]{>{}c>{}c>{}c>{}c>{}c>{}c>{}c}
\toprule
\multicolumn{1}{c}{ } & \multicolumn{3}{c}{Unweighted Averages} & \multicolumn{3}{c}{Weighted Averages} \\
\cmidrule(l{3pt}r{3pt}){2-4} \cmidrule(l{3pt}r{3pt}){5-7}
 & Expansion & No Expansion & Gap/DiD & Expansion & No Expansion & Gap/DiD\\
\midrule
\textcolor{black}{2013} & \textcolor{black}{419.2} & \textcolor{black}{474.0} & \em{-54.8} & \textcolor{black}{322.7} & \textcolor{black}{376.4} & \em{-53.7}\\
\textcolor{black}{2014} & \textcolor{black}{428.5} & \textcolor{black}{483.1} & \em{-54.7} & \textcolor{black}{326.5} & \textcolor{black}{382.7} & \em{-56.2}\\
\em{\textcolor{black}{Trend/DiD}} & \em{\textcolor{black}{9.3}} & \em{\textcolor{black}{9.1}} & \em{\em{\textcolor{BrickRed}{0.1}}} & \em{\textcolor{black}{3.7}} & \em{\textcolor{black}{6.3}} & \em{\em{\textcolor{BrickRed}{-2.6}}}\\
\bottomrule
\end{tabular}
\begin{tablenotes}[para]
\item \vspace{-4ex} \singlespacing \footnotesize{This table reports average county-level 
                  mortality rates (deaths among adults aged 20-64 per 100,000 adults) in 2013 (top row) and 2014 (middle row) 
                  in states that expanded adult Medicaid eligibility in 2014 (columns 1 and 4, 978 counties) and states that have not expanded 
                  by 2019 (columns 2 and 5, 1,222 counties). The first three columns present unweighted averages and the second three columns 
                  present population-weighted averages. Columns 1, 2, 4, and 5 in the third row show time trends in mortality 
                  between 2013 and 2014 for each group of states. The first two rows of columns 3 and 6 show the cross-sectional 
                  gap in mortality between expansion and non-expansion states in 2013 and 2014. The entries in red text in 
                  the bottom row show the simple 2 $\times$ 2 difference-in-differences estimates without weights (column 3) and with them 
                  (column 6)}
\end{tablenotes}
\end{threeparttable}
\end{table}

%% file: Tables/regdid_2x2.tex
\begin{table}[!ht]
   \centering
   \begin{adjustbox}{width = 0.8\textwidth, center}
      \begin{threeparttable}[b]
         \caption{\label{tab:regdid_2x2} Regression 2 $\times$ 2 DiD}
         \bigskip
         \renewcommand*{\arraystretch}{0.8}
         \begin{tabular*}{\textwidth}{@{\extracolsep{\fill}}lcccccc}
            \toprule
             & \multicolumn{3}{c}{Unweighted} & \multicolumn{3}{c}{Weighted} \\ \cmidrule(lr){2-4} \cmidrule(lr){5-7}
             & \multicolumn{2}{c}{Crude Mortality Rate} & $\Delta$ & \multicolumn{2}{c}{Crude Mortality Rate} & $\Delta$ \\ \cmidrule(lr){2-3} \cmidrule(lr){4-4} \cmidrule(lr){5-6} \cmidrule(lr){7-7}
                                              & (1)           & (2)   & (3)         & (4)           & (5)        & (6)\\  
            \midrule 
            Constant                          & 474.0$^{***}$ &       & 9.1$^{***}$ & 376.4$^{***}$ &            & 6.3$^{***}$\\   
                                              & (4.3)         &       & (2.6)       & (7.6)         &            & (1.1)\\   
            Medicaid Expansion                & -54.8$^{***}$ &       &             & -53.7$^{***}$ &            &   \\   
                                              & (6.3)         &       &             & (11.5)        &            &   \\   
            Post                              & 9.1$^{***}$   &       &             & 6.3$^{***}$   &            &   \\   
                                              & (2.6)         &       &             & (1.1)         &            &   \\   
            Medicaid Expansion $\times$ Post  & 0.1           & 0.1   & 0.1         & -2.6$^{*}$    & -2.6$^{*}$ & -2.6$^{*}$\\   
                                              & (3.7)         & (3.7) & (3.7)       & (1.5)         & (1.5)      & (1.5)\\   
            \midrule 
            County fixed effects              & No            & Yes   & No          & No            & Yes        & No\\  
            Year fixed effects                & No            & Yes   & No          & No            & Yes        & No\\  
            \bottomrule
         \end{tabular*}
         
         \begin{tablenotes}\item This table reports the 
                              regression 2 $\times$ 2 DiD estimate comparing 978 counties in states that expanded Medicaid in 2014 with 1,222 counties in states that did 
                              not expand Medicaid by 2019, using only data for the years 2013 and 2014. Columns 1-3 report unweighted regression 
                              results, while columns 4-6 weight by county population aged 20-64 in 2013. Columns 1 and 4 report results from 
                              regressing the crude mortality rate for adults ages 20-64 on indicators for expansion states (Treat) and post-expansion year (Post); 
                              the DiD estimate is the coefficient on the interaction term. Columns 2 and 5 report the corresponding 
                              results for the interaction term using county and year fixed effects. Finally, Columns 3 and 6 report the results
                              of the long difference in county mortality rates on a treatment indicator. Standard errors (in parentheses) are clustered at the county level.
         \end{tablenotes}
      \end{threeparttable}
   \end{adjustbox}
\end{table}

%% file: Tables/cov_balance.tex
\begin{table}[!h]

\caption{\label{tab:cov_balance}Covariate Balance Statistics}
\centering
\resizebox{\linewidth}{!}{
\begin{threeparttable}
\begin{tabular}[t]{lcccccc}
\toprule
\multicolumn{1}{c}{ } & \multicolumn{3}{c}{Unweighted} & \multicolumn{3}{c}{Weighted} \\
\cmidrule(l{3pt}r{3pt}){2-4} \cmidrule(l{3pt}r{3pt}){5-7}
Variable & Non-Adopt & Adopt & Norm. Diff. & Non-Adopt & Adopt & Norm. Diff.\\
\midrule
\addlinespace[0.3em]
\multicolumn{7}{l}{\textbf{2013 Covariate Levels}}\\
\hspace{1em}\% Female & 49.43 & 49.33 & -0.03 & 50.48 & 50.07 & -0.24\\
\hspace{1em}\% White & 81.64 & 90.48 & 0.59 & 77.91 & 79.54 & 0.11\\
\hspace{1em}\% Hispanic & 9.64 & 8.23 & -0.10 & 17.01 & 18.86 & 0.11\\
\hspace{1em}Unemployment Rate & 7.61 & 8.01 & 0.16 & 7.00 & 8.01 & 0.50\\
\hspace{1em}Poverty Rate & 19.28 & 16.53 & -0.42 & 17.24 & 15.29 & -0.37\\
\hspace{1em}Median Income & 43.04 & 47.97 & 0.43 & 49.31 & 57.86 & 0.68\\
\addlinespace[0.3em]
\multicolumn{7}{l}{\textbf{2014 - 2013 Covariate Differences}}\\
\hspace{1em}\% Female & -0.02 & -0.02 & 0.00 & 0.02 & 0.01 & -0.09\\
\hspace{1em}\% White & -0.21 & -0.21 & 0.01 & -0.32 & -0.33 & -0.04\\
\hspace{1em}\% Hispanic & 0.20 & 0.21 & 0.04 & 0.25 & 0.33 & 0.29\\
\hspace{1em}Unemployment Rate & -1.16 & -1.30 & -0.21 & -1.08 & -1.36 & -0.55\\
\hspace{1em}Poverty Rate & -0.55 & -0.28 & 0.14 & -0.41 & -0.35 & 0.05\\
\hspace{1em}Median Income & 0.98 & 1.11 & 0.06 & 1.10 & 1.74 & 0.32\\
\bottomrule
\end{tabular}
\begin{tablenotes}[para]
\item \vspace{-4ex} \singlespacing \footnotesize{This table reports the 
                  covariate balance between 978 counties in states that expanded Medicaid in 2014 and 1,222 counties in states that did not expand by 2019.  In the top panel, we report the averages 
                  and standardized differences of each variable, measured in 2013, by adoption status. All variables are measured in 
                  percentage values, except for median household income, which is measured in thousands of U.S. dollars. In the bottom panel
                  we report the average and standardized differences of the county-level long differences between 2014 and 
                  2013 of each variable. We report both weighted and unweighted measures of the averages to correspond to the different
                  estimation methods of including covariates in a 2 $\times$ 2 setting.}
\end{tablenotes}
\end{threeparttable}}
\end{table}

%% file: Tables/regdid_2x2_covs.tex
\begin{table}[H]
   \centering
   \begin{adjustbox}{width = 0.8\textwidth, center}
      \begin{threeparttable}[b]
         \caption{\label{tab:regdid_2x2_covs} Regression 2 $\times$ 2 DiD with Covariates}
         \bigskip
         \renewcommand*{\arraystretch}{0.8}
         \begin{tabular*}{\textwidth}{@{\extracolsep{\fill}}lcccccc}
            \toprule
             & \multicolumn{3}{c}{Unweighted} & \multicolumn{3}{c}{Weighted} \\ \cmidrule(lr){2-4} \cmidrule(lr){5-7}
             & No Covs & $X_{i, t = 2013}$ & $X_{i, t}$ & No Covs & $X_{i, t = 2013}$ & $X_{i, t}$ \\ \cmidrule(lr){2-2} \cmidrule(lr){3-3} \cmidrule(lr){4-4} \cmidrule(lr){5-5} \cmidrule(lr){6-6} \cmidrule(lr){7-7}
                               & (1)    & (2)    & (3)    & (4)         & (5)    & (6)\\  
            \midrule 
            Medicaid Expansion & 0.12   & -2.35  & -0.49  & -2.56$^{*}$ & -2.56  & -1.37\\   
                               & (3.75) & (4.29) & (3.83) & (1.49)      & (1.78) & (1.62)\\   
            \bottomrule
         \end{tabular*}
         
         \begin{tablenotes}\item This table reports the 
                              regression 2 $\times$ 2 DiD estimate comparing 978 counties that expanded Medicaid in 2014 with 1,222 counties that did not expand Medicaid by 2019, adjusting for covariates (percent female, percent white, percent hispanic,
                              the unemployment rate, the poverty rate, and median household income). Columns 1-3 report unweighted regression 
                              results, while columns 4-6 weight by county population aged 20-64 in 2013. Columns 1 and 4 report results for 
                              expansion states without covariates, columns 2 and 5 adjust for the baseline levels of the covariates in 2013, 
                              and columns 3 and 6 control for the time-varying covariate values in 2014 and 2013. 
                   Standard errors (in parentheses) are clustered at the county level.
         \end{tablenotes}
      \end{threeparttable}
   \end{adjustbox}
\end{table}

%% file: Tables/reg_pscore_cs.tex
\begin{table}[H]
   \centering
   \begin{adjustbox}{width = 0.8\textwidth, center}
      \begin{threeparttable}[b]
         \caption{\label{tab:reg_pscore_cs} Outcome Regression and Propensity Score Models}
         \bigskip
         \renewcommand*{\arraystretch}{0.8}
         \begin{tabular*}{\textwidth}{@{\extracolsep{\fill}}lcccc}
            \toprule
             & \multicolumn{2}{c}{Unweighted} & \multicolumn{2}{c}{Weighted} \\ \cmidrule(lr){2-3} \cmidrule(lr){4-5}
             & Regression & Propensity Score & Regression & Propensity Score \\ \cmidrule(lr){2-2} \cmidrule(lr){3-3} \cmidrule(lr){4-4} \cmidrule(lr){5-5}
                              & (1)     & (2)            & (3)        & (4)\\  
                              &  OLS    & Logit          & OLS        & Logit\\  
            \midrule 
            Constant          & -20.91  & -10.00$^{***}$ & -4.62      & -8.17$^{***}$\\   
                              & (69.85) & (1.35)         & (44.98)    & (0.01)\\   
            \% Female         & 0.04    & -0.04$^{**}$   & -0.09      & -0.19$^{***}$\\   
                              & (0.82)  & (0.02)         & (0.68)     & (0.00)\\   
            \% White          & 0.15    & 0.06$^{***}$   & 0.20$^{*}$ & 0.04$^{***}$\\   
                              & (0.22)  & (0.00)         & (0.11)     & (0.00)\\   
            \% Hispanic       & -0.08   & -0.02$^{***}$  & -0.08      & -0.02$^{***}$\\   
                              & (0.20)  & (0.00)         & (0.08)     & (0.00)\\   
            Unemployment Rate & 1.14    & 0.32$^{***}$   & 0.88       & 0.68$^{***}$\\   
                              & (1.56)  & (0.03)         & (0.99)     & (0.00)\\   
            Poverty Rate      & 0.21    & 0.03$^{*}$     & -0.13      & 0.11$^{***}$\\   
                              & (0.98)  & (0.02)         & (0.53)     & (0.00)\\   
            Median Income     & 0.09    & 0.08$^{***}$   & -0.05      & 0.15$^{***}$\\   
                              & (0.52)  & (0.01)         & (0.24)     & (0.00)\\   
            \bottomrule
         \end{tabular*}
         
         \begin{tablenotes}\item This table reports the outcome regression propensity score models that enter into the estimator 
                   from \citet{SantAnna2020} and \citet{Callaway2021}. The first two columns report the results for unweighted regressions and the 
                   second two report results from weighted regression models. The regression model predicts changes in the 
                   outcome variable (mortality rates) as a function of 2013 covariate values for the 1,222 counties that do not expand 
                   Medicaid in 2014. The propensity score model uses data on all 2,200 counties for 2013 and estimates a logit model of an expansion
                   indicator variable on the 2013 covariate levels. Standard errors (in parentheses) are clustered at the county level.
         \end{tablenotes}
      \end{threeparttable}
   \end{adjustbox}
\end{table}

%% file: Tables/2x2_csdid.tex
\begin{table}[!h]

\caption{\label{tab:2x2_csdid}DiD Estimates with Covariates}
\centering
\begin{threeparttable}
\begin{tabular}[t]{lcccccc}
\toprule
\multicolumn{1}{c}{ } & \multicolumn{3}{c}{Unweighted} & \multicolumn{3}{c}{Weighted} \\
\cmidrule(l{3pt}r{3pt}){2-4} \cmidrule(l{3pt}r{3pt}){5-7}
  & Regression & IPW & Doubly Robust & Regression & IPW & Doubly Robust\\
\midrule
Medicaid Expansion & -1.62 & -0.86 & -1.23 & -3.46 & -3.84 & -3.76\\
 & (4.73) & (4.76) & (4.61) & (2.29) & (3.22) & (3.59)\\
\bottomrule
\end{tabular}
\begin{tablenotes}[para]
\item \vspace{-4ex} \singlespacing \footnotesize{This table reports the 2 $\times$ 2 DiD 
                        estimate comparing 978 counties in states that expand Medicaid in 2014 to 1,222 counties in states that did not expand Medicaid by 2019, adjusting for
                         2013 covariate values using the methodologies discussed in \citet{SantAnna2020} and \citet{Callaway2021}. The first column
                        reports results using regression adjustment, the second column uses inverse probability weighting based on a 
                        propensity score model using the included covariates, and the third column uses the doubly robust combination of 
                        the two approaches. Standard errors (in parentheses) are clustered at the county level.}
\end{tablenotes}
\end{threeparttable}
\end{table}